\documentclass[final,3p,fleqn,11pt]{elsarticle}

\usepackage{amssymb}
\usepackage{multicol}
\usepackage[svgnames]{xcolor}

\usepackage{booktabs}
\usepackage{array}
\newcommand{\PreserveBackslash}[1]{\let\temp=\\#1\let\\=\temp}
\newcolumntype{C}[1]{>{\PreserveBackslash\centering}p{#1}}
\newcolumntype{R}[1]{>{\PreserveBackslash\raggedleft}p{#1}}
\newcolumntype{L}[1]{>{\PreserveBackslash\raggedright}p{#1}}

\usepackage{amsmath}
\usepackage{amsfonts}
\usepackage{amsthm}
\usepackage{mathtools}
\usepackage{subfig}
\usepackage{lineno}
\usepackage[hidelinks]{hyperref}
\usepackage{textcomp}
\usepackage{algorithm,algpseudocode,algorithmicx}
\usepackage{pifont}
\usepackage{tabularx}

\usepackage{physics}
\usepackage{MnSymbol}

\usepackage{float}
\usepackage{array,multirow}
\usepackage{color}
\usepackage{tikz}
\usetikzlibrary{patterns}
\usetikzlibrary{spy}
\usetikzlibrary{spy,decorations.fractals,shapes.misc}

\usetikzlibrary{backgrounds}
\usepackage{pgfplots}
\usepgfplotslibrary{patchplots}
\usepgfplotslibrary{fillbetween}
\usepgfplotslibrary{groupplots}
\usepgfplotslibrary{polar}
\usepgfplotslibrary{smithchart}
\usepgfplotslibrary{statistics}
\usepgfplotslibrary{dateplot}
\usepgfplotslibrary{ternary}
\usetikzlibrary{arrows.meta, positioning, shapes.geometric}

\pgfplotsset{compat=1.16}
\newsavebox\Axis

\definecolor{lightgray}{gray}{0.80}

\usetikzlibrary{decorations.pathreplacing}
\usepackage[listings,theorems,skins,breakable]{tcolorbox}
\tcbset{skin=enhanced}
\newtcolorbox{lbracebox}[1][Word]{%
   frame hidden,enlarge left by=2cm,width=\linewidth-2cm,%
  overlay unbroken = {\draw [decorate,decoration={brace,amplitude=10pt},]%
                     (frame.south west)-- (frame.north west)
                    node [black,midway,left,xshift=-.6cm] {#1};},%
}

\definecolor{grey1}{rgb}{0.5, 0.5, 0.5}
\definecolor{green1}{rgb}{0.4660, 0.6740, 0.1880} 
\definecolor{blue1}{rgb}{0.26, 0.41, 0.88} 
\definecolor{red1}{rgb}{0.8600, 0.0800, 0.2400}
\definecolor{yellow1}{rgb}{1.0, 0.76, 0.15}
\definecolor{purple1}{rgb}{0.4940, 0.1840, 0.5560}
\definecolor{lightblue1}{rgb}{0.3010, 0.7450, 0.9330}
\definecolor{bordeaux1}{rgb}{0.6350, 0.0780, 0.1840}
\definecolor{brown1}{rgb}{0.65, 0.16, 0.16}
\definecolor{pink1}{rgb}{1.0, 0.08, 0.58}
\definecolor{green2}{rgb}{0.4, 1, 0}

\definecolor{burntorange}{rgb}{0.74902,0.341176,0}

\usepackage{color, soul}

\DeclareRobustCommand{\reviewerI}[1]{{\sethlcolor{pink}\hl{#1}}}
\DeclareRobustCommand{\reviewerII}[1]{{\sethlcolor{green2}\hl{#1}}}
\DeclareRobustCommand{\reviewerIII}[1]{{\sethlcolor{yellow}\hl{#1}}}

\soulregister\reviewerI{1}
\soulregister\reviewerII{1}
\soulregister\reviewerIII{1}

\soulregister\ref7
\soulregister\pageref7
\soulregister\eqref7
\soulregister\cite7

\newcommand{\mathcolorbox}[2]{\colorbox{#1}{$\displaystyle #2$}}

\makeatletter
\renewcommand*\env@matrix[1][*\c@MaxMatrixCols c]{%
  \hskip -\arraycolsep
  \let\@ifnextchar\new@ifnextchar
  \array{#1}}
\makeatother

\theoremstyle{plain}
\newtheorem{theorem}{Theorem}[section]

\newtheorem{remark}[theorem]{Remark}

\theoremstyle{definition}

\newcommand{\vect}[1]{\boldsymbol{#1}} 									
\newcommand{\mat}[1]{\mathbf{#1}} 											
\newcommand{\domain}{\Omega}														

\newcommand{\map}{{\Theta}} 
\newcommand{\refconfig}{\Phi}
\newcommand{\curconfig}{\varphi}

\newcommand{\datafieldset}{\mathcal{D}}
\newcommand{\dataset}{\mathfrak{D}}

\newcommand{\linB}{\mathcal{B}_1}

\newcommand{\ndofs}{N}																	
\newcommand{\mdofs}{M}																	
\newcommand{\numDataPts}{n_D}

\newcommand{\arclen}{\xi}

\newcommand{\xF}{x}																	
\newcommand{\yF}{y}																	
\newcommand{\lF}{z}																	

\newcommand{\xSpace}{X}																	
\newcommand{\ySpace}{Y}																	
\newcommand{\lSpace}{Z}																	

\newcommand{\xh}{x_h}																	

\newcommand{\ytilde}{\tilde{y}}																	
\newcommand{\yhtilde}{\tilde{y}_h}
\newcommand{\ystar}{\tilde{y}^*}																	

\newcommand{\yhat}{\hat{y}}

\newcommand{\qV}{\vect{q}}                            
\newcommand{\qVh}{\vect{q}_h}                            
\newcommand{\qhat}{\hat{\vect{q}}}

\newcommand{\idset}{\Xi}               

\newcommand{\globObjFunc}{\text{dist}_G}      
\newcommand{\eleObjFunc}{\text{dist}_E}      

\newcommand{\scaleFac}{\beta_s}     
\newcommand{\scaleFacHi}{\textcolor{blue1}{\beta_s}}

\DeclareMathOperator*{\argmin}{arg\,min}

\journal{Computers \& Structures}

\allowdisplaybreaks

\modulolinenumbers[5]

\begin{document}

\begin{frontmatter}

\title{Solving strategies for data-driven one-dimensional elasticity exhibiting nonlinear strains}

\author[address1]{Thi-Hoa Nguyen \corref{cor1}}
\ead{hoa.nguyen@uib.no}

\author[address1,address2]{Viljar H. Gjerde}
\ead{viljargjerde@gmail.com}

\author[address1]{Bruno A. Roccia}
\ead{bruno.roccia@uib.no}

\author[address1]{Cristian G. Gebhardt}
\ead{cristian.gebhardt@uib.no}

\cortext[cor1]{Corresponding author}

\address[address1]{Geophysical Institute and Bergen Offshore Wind Centre, University of Bergen, Norway}

\address[address2]{Eviny Fornybar AS, Norway}

\begin{abstract}
In this work, we extend and generalize our solving strategy, first introduced in \cite{viljar2025}, based on a greedy optimization algorithm and the alternating direction method (ADM) for nonlinear systems computed with multiple load steps.  
In particular, we combine the greedy optimization algorithm with the direct data-driven solver based on ADM which is firstly introduced in \cite{ortiz_ddcm_2016} and combined with the Newton-Raphson method for nonlinear elasticity in \cite{Keip_ddcm_nonlinearbar}. 
We numerically illustrate via one- and two-dimensional bar and truss structures exhibiting nonlinear strain measures and different constitutive datasets that our solving strategy generally achieves a better approximation of the globally optimal solution. 
This, however, comes at the expense of higher computational cost which is scaled by the number of “greedy” searches. 
Using this solving strategy, we reproduce the first cycle of the cyclic testing for a nylon rope that was performed at industrial testing facilities for mooring lines manufacturers. 
We also numerically illustrate for a truss structure that our solving strategy generally improves the accuracy and robustness in cases of an unsymmetrical data distribution and noisy data. 
\end{abstract}

\begin{highlights}
\item We extend and generalize our solving strategy based on a greedy optimization algorithm and the alternating direction method for nonlinear systems computed with multiple load steps.
\item We numerically illustrate that our solving strategy leads to a better approximation of the global optima for both one- and two-dimensional structures exhibiting nonlinear strains and different constitutive datasets. 
\item We apply our solving strategy to a real experimental dataset of a cyclic testing of a nylon rope performed in industrial testing facilities for mooring line manufacturers.
\item We numerically illustrate that our solving strategy generally improves the accuracy and robustness in cases of an unsymmetrical data distribution and noisy data.
\end{highlights}


\end{frontmatter}



\section{Introduction}
  
Data-driven computational mechanics (DDCM) emerged nearly a decade ago as a novel numerical approach in the field of computational mechanics. 
Firstly introduced in \cite{ortiz_ddcm_2016}, the direct DDCM approach enables the 
use of constitutive data (i.e. stress-strain pairs) obtained from experiments, bypassing the need for ad-hoc material models and avoiding information loss. 
The main idea is to formulate the boundary-value or initial-boundary-value problem as an optimization problem, seeking the stress-strain pairs from a given dataset that are closest to those pairs satisfying the equilibrium, compatibility, boundary, and initial conditions \cite{ortiz_ddcm_2016,Kirchdoerfer2017ddcm,Kirchdoerfer2018ddcmDyn}. 
Minimizing the distance between these two pairs is obtained by finding the global minima of an objective function that is the weighted sum of the square $L^2$-norm of this distance. 
Hence, the resulting optimization problem is often considered as a distance-minimizing problem, or 
a discrete–continuous quadratic optimization problem due to the discrete and continuous nature of the dataset and the variable fields, respectively. 
In \cite{ortiz_ddcm_2016}, the authors proposed a solving strategy that begins with randomly chosen but fixed initial stress-strain pairs and solves for solution that satisfies the constraint set and are simultaneously closest to these initial pairs. 
It then searches for new stress-strain pairs from the dataset that are closest to those from the obtained variable fields, i.e. the obtained structural solution, and repeats these steps until the stress-strain data converges to the same value. 
This idea is the same as the alternating direction method (ADM) without initialization \cite{Douglas1956adm,Gabay1976adm} 
(see also \cite{Glowinski2014adm} for a historical overview of ADM). 
In general, ADM 
is applicable to both convex and certain non-convex formulations, although its basic form often suffers from slow or unreliable convergence because coupling constraints are enforced only implicitly. 
To tackle this issue, particularly for convex problems, an extension of ADM is the alternating direction method of multipliers (ADMM) which employs augmented Lagrangian method and hence dual variables to speed up convergence \cite{Fortin1983admm,Boyd2011admm}. 
These enhancement stabilizes the iteration and provides convergence guarantee for convex problems. 
For non-convex problems, ADM remains the chosen method in practice, although it generally does not guarantee local or global optimal solution despite globally optimal solutions of each subproblem (see e.g. discussions in \cite{Gebhardtddcmsolution2025}).

There exist alternatives to the direct DDCM approach that include 
the inverse DDCM approach \cite{Ibanez2017ddcm,Ibanez2018ddcm,Ibanez2019ddcm}, which reconstructs the material model based on the given dataset as a traditional energy functional for the computations, 
and the hybrid DDCM approach \cite{Gebhardtddcmstatic2020,Gebhardtddcmdynamic2020,Kanno2021hybridddcm,Gebhardtddcmcontacts2024}, which combines both the direct and inverse approaches. 
The direct DDCM approach has been 
extended by numerous other works to 
noisy data in combination with the so-called max-ent approaches \cite{Kirchdoerfer2017ddcm} or locally convex recontruction (LCR) approaches \cite{He2020ddcmnoise}, 
structural dynamics \cite{Kirchdoerfer2018ddcmDyn,Gebhardtddcmdynamic2020}, 
large deformations \cite{Keip_ddcm_nonlinearbar,Conti2020nonlinDDCM,Platzer2021admnonlin}, 
snap-through problems \cite{Kuang2023ddcmsnapthrough}, 
composite structures \cite{Kuang2023composites,Xu2020}, 
inelasticity \cite{Eggersmann2019ddcminelastic}, 
fracture \cite{Carrara2020fracture}, 
uncertainty quantification \cite{Zschocke2022ddcmuncertainty,Prume2023ddcmuncertainty}, 
and multiscale mechanics \cite{Korzeniowski2021ddcmmultiscale,Gorgogianni2023ddcmmultiscale,Prume2025ddcm}, 
and also coupled with game theory 
\cite{Weinberg2023ddcm}. 
Another field of application is material modeling where the direct DDCM approach is reformulated to identify the stress field based on the given strain data \cite{Leygue2018datamodel,Dalemat2019datamodel,Stainier2019datamodel,Flaschel2022datamodel}. 
In other fields than mechanics, the direct DDCM approach also finds its applications, for instance, in 
magnetism \cite{Gersem2020ddcmmagnetic}, 
electro-mechanical coupled problems \cite{Marenic2022ddcmelectro}, or 
simulation of electrical circuits \cite{Gebhardtelectric2025}.

Different properties of the direct DDCM approach have been investigated and various approaches developed for their improvement. 
One of the first aspects addressed, e.g., in \cite{kanno_data_driven_2019,Galetzka2021ddcm,Nguyen2022ddcm}, is to achieve globally optimal solution. 
This 
plays a crucial role, particularly in cases of sparse datasets, three-dimensional systems, and nonlinear constitutive relations, 
however, 
is not guaranteed by the direct solving strategy. 
To tackle this, Kanno 
reformulated the optimization problem associated with the direct DDCM approach as 
a mixed-integer programming problem that can be solved globally using 
a standard mixed-integer programming solver to obtain global optima in \cite{kanno_data_driven_2019}. 
Such a solver, however, requires high computational cost that rapidly increases with increasing number of data points. 
Recently, another approach that guarantees global optima for linear systems in 
certain symmetric cases 
is introduced in \cite{Gebhardtddcmsolution2025} where the authors proposed and integrated a structure-specific initialization for the stress-strain pairs in the original solving strategy \cite{ortiz_ddcm_2016}. 
Other attempts to achieve a better approximation of the global optima, improving the accuracy, include 
the development of local or adaptive weighting parameters of the objective function, i.e. the tangent of the constitutive manifold \cite{Galetzka2021ddcm,Nguyen2022ddcm}, 
a solving algorithm integrating the material tangent information from or into the dataset, based on the tensor voting method, in the case of sparse datasets \cite{Eggersmann2021ddcmaccuracy,Ciftci2022ddcm}, 
and a modified fixed-point algorithm to escape local minima \cite{Prume2025ddcm}. 
In our previous work \cite{viljar2025}, we propose to combine a greedy optimization algorithm \cite{Temlyakov2008,TEMLYAKOV2014greedy} and the standard solving strategy \cite{ortiz_ddcm_2016,Keip_ddcm_nonlinearbar} for one-dimensional bars exhibiting nonlinear strains. 
In particular, our solving strategy reduces the value of the global objective function by searching for alternatives to the initial stress-strain pairs in a given dataset. 
Based on a similar idea, the authors of \cite{Rocha2025} proposed the so-called defective restarting approach to avoid the standard solver based on the alternating direction method getting trapped in local optima. 
Both approaches introduced in \cite{viljar2025} and \cite{Rocha2025} search the nearest-neighbourhood and consider the second-nearest data point as alternative, which is an idea inspired by the Dropout regularisation used by the Deep Neural Networks community \cite{Rocha2025}.
However, while the former searches this alternative for the stress and strain states across all elements, the latter performs this search for some randomly chosen parts of the states only.  
Another important aspect of the direct DDCM approach that has been addressed and investigated is the computational cost, mainly due to the search of the optimal stress-strain pairs from a given dataset. 
In \cite{Eggersmann2021ddcm}, the authors proposed 
approximate nearest-neighbor algorithms to reduce the effort of this search through the dataset. 
Another approach to accelerate this search is the tree-based (or $k$-d-tree-based) nearest-neighbor search algorithms \cite{Bentley1975,Zheng2020}, which can be coupled with neural networks for multi-fidelity data \cite{Bahmani2021}, and is particularly beneficial in cases of high-dimensional datasets. 
In \cite{Nguyen2022ddcm}, the authors illustrated that using adaptive hyperparameters of the objective that represent the tangent of the constitutive manifold also reduces the computational cost of the employed distance-minimizing method for the data-driven solving strategy. 
Furthermore, the mathematical structure and proofs of solution existence have been addressed and investigated in \cite{Gebhardtddcmsolution2025,Conti2020nonlinDDCM,Conti2018ddcm,Gebhardtddcmhilbert2025}.

In this work, we extend and generalize the solving strategy, introduced in our previous work \cite{viljar2025}, for nonlinear systems computed with multiple load steps. 
The main idea is to integrate a greedy optimization algorithm \cite{Temlyakov2008,TEMLYAKOV2014greedy} into the standard solving strategy of the direct DDCM approach \cite{ortiz_ddcm_2016,Keip_ddcm_nonlinearbar} to find alternatives to the stress-strain pairs in a given dataset, which reduce the value of the global objective function. 
We initialize these stress-strain data pairs at the first load step using one of the three initialization approaches: a random initialization \cite{ortiz_ddcm_2016}, an initialization with the stress-free state, i.e. zero stresses and strains, \cite{Gebhardtddcmstatic2020}, and the structure-specific initialization \cite{Gebhardtddcmsolution2025}. 
Starting from the second load step, we initialize these pairs with the those obtained from the preceding load step. 
We extend our numerical studies in our previous work \cite{viljar2025} to two-dimensional truss structures exhibiting nonlinear strains and large deformations. 
We numerically illustrate that our solving strategy achieves a better approximation of the globally optimal solution, however, at the expense of higher computational cost, compared to the standard solving strategy. 
We apply our solving strategy to the simulation of a cyclic test of nylon ropes that are performed at industrial testing facilities for mooring lines. 
We also briefly discuss and numerically show via a truss structure that our solving strategy generally improves the accuracy and robustness in cases of an unsymmetrical data distribution and noisy data.

The outline of the paper is as follows: 
in Section \ref{sec:formulations}, we state the optimization problem of one-dimensional elasticity with nonlinear strain measures in the continuous and discrete setting. 
Here, we also reformulate the resulting formulation as a mixed-integer programming problem, which is solved to obtain the global optima as a reference for our computations. 
In Section \ref{sec:solvingstrategy}, we describe the algorithm of our solving strategy and three initialization approaches considered in this work. 
In Section \ref{sec:results}, we numerically illustrate via bar and truss structures favorable properties of our solving strategy. 
In Section \ref{sec:conclusions}, we summarize our results and the main conclusions.






\section{Nonlinear data-driven one-dimensional elasticity}\label{sec:formulations}

In this section, we state the formulation of the boundary-value problem for one-dimensional elasticity with nonlinear strain measures studied in this work. 
We start with the formulation 
in the spatially continuous setting, which is the formulation of a discrete-continuous nonlinear optimization problem (DCNLP). 
We then employ the finite element method and obtain the associated spatially discrete formulation. 
We also derive the linearized system of equations in a matrix form that is the backbone of the solving strategy discussed in the next section. 
We close this section with the discussion and reformulation of the DCNLP as a mixed-integer quadratic programming problem which yields global optima when solved with several existing software packages.

\subsection{Continuous setting}\label{sec:continuous_form}

Consider a bounded domain 
$\domain \subset \mathbb{R}$ 
and a physical body that occupies the closure $\bar{\domain}$, which is represented as a one-parameter curve $\vect{\curconfig} = \vect{\curconfig}(\arclen) \in \mathbb{R}^2$, where $\arclen \in [0,L_0]$ is the arc-length coordinate. 
Assume that the physical body 
is subjected to body forces $\vect{f} \in L^2(\domain)$ and exhibits only axial deformations. 
Let $\vect{u} \in H_0^1(\domain)$, $e \in L^2(\domain)$, $s \in L^2(\domain)$ denote the displacement, axial strain, and stress fields, respectively, 
$\vect{\lambda} \in H_0^1(\domain)$, $\mu \in L^2(\domain)$ the dual fields of Lagrange multipliers. 
While the indicated function spaces fit perfectly for linear analysis, the nonlinear strain measure employed in this work, defined in the following Equation \eqref{eq:strainmeasures}, might require spaces with increased regularity. This study does not pursue a dedicated analysis of these regularity requirements, leaving such considerations for future work. 
Furthermore, for the sake of notation, we define: 
\begin{equation*}
    \begin{aligned}
        & \xF := (\vect{u},\,e,\,s) \in \xSpace := H_0^1(\domain) \, \times \, L^2(\domain) \, \times \, L^2(\domain) \,, \\
        & \yF := (\,e,\,s) \in \ySpace := L^2(\domain) \, \times \, L^2(\domain) \,, \\
        & \lF := (\vect{\lambda},\,\mu) \in \lSpace := H_0^1(\domain) \, \times \, L^2(\domain) \,.
    \end{aligned}
\end{equation*}
Let $\map$ define the enforcement of the equilibrium and the compatibility conditions using Lagrange multipliers:
\begin{equation}\label{eq:equilibriumNcompatibilityConds}
    \map\left( \lF,\, \xF; \, \vect{f} \right) = 0 \quad \forall \,\lF \in \lSpace \,,
\end{equation}
with
\begin{equation}
    \map\left( \lF,\, \xF; \,\vect{f} \right) := \langle \vect{\lambda}, \,\mathcal{B}^T s - \vect{f}\rangle_{L^2(\domain)} + \langle\mu,\, \epsilon(\vect{u}) - e\rangle_{L^2(\domain)} \,.
\end{equation}

\noindent
Here, 
$\vect{\lambda}$ and $\mu$ correspond to the enforcement of the equilibrium and compatibility conditions, respectively, 
$\epsilon(\vect{u})$ is the displacement-based strain, and   
$\langle\cdot,\cdot\rangle$ denotes the $L^2$ scalar product. 
In this work, we consider the following displacement-based strain:
\begin{equation}\label{eq:strainmeasures}    
    \epsilon(\vect{u}) = \vect{\refconfig}^\prime \cdot \vect{u}^\prime + \frac{1}{2} \alpha \, \vect{u}^\prime \cdot \vect{u}^\prime \,, \quad \alpha \in \{0,1\} \,,
\end{equation}
which corresponds to the Green-Lagrange strain measure simplified for a one-dimensional bar formulation. To conveniently switch between linear and nonlinear strain representations in the numerical studies, the quadratic term is scaled by a binary parameter, $\alpha \in {0,1}$. Choosing $\alpha=0$ recovers the linear strain measure, whereas $\alpha=1$ activates the nonlinear contribution. 
Here, $\vect{\refconfig} = \vect{\refconfig}(\arclen)$ denotes the reference configuration of the physical body, which relates to the current configuration as $\vect{\curconfig} = \vect{\refconfig} + \vect{u}$. 
The corresponding strain-displacement operator is then: 
\begin{equation}
    \mathcal{B}(\cdot) = \vect{\refconfig}^\prime \cdot (\cdot)^\prime + \alpha \, \vect{u}^\prime \cdot (\cdot)^\prime \,, 
\end{equation}
where 
$(\cdot)^\prime$ denotes the weak differentiation with respect to the arc-length coordinate $\arclen$, $(\cdot)^\prime := \partial(\cdot) / \partial \, \arclen$.

Consider a given closed data set of stress-strain pairs $\dataset$. One can construct the corresponding strain and stress fields from $\dataset$ as follows:
\begin{equation}
    \datafieldset := \{ \ytilde := (\tilde{e},\, \tilde{s}) \in \mathcolorbox{pink}{L^2(\domain) \cross L^2(\domain)}: \left(\tilde{e}(\arclen),\, \tilde{s}(\arclen) \right) \in \dataset \; \forall \, \arclen \in \domain \} \,.
\end{equation}
We note that in general, the dataset consists of available experimental measurements, i.e. discrete data points. 
The case when a constitutive manifold exists is a special case (see also \cite{Gebhardtddcmhilbert2025}). 
The static structural analysis of the physical body can be formulated as a discrete-continuous quadratic optimization problem \cite{ortiz_ddcm_2016,Gebhardtddcmsolution2025} as follows:

\noindent
Find $\xF \in \xSpace$ such that:
\begin{equation}\label{eq:dcnlp}
    \begin{aligned}
        & \inf_{\xF,\,\ytilde} \; \sup_{\lF \in \lSpace} \;
        \globObjFunc \left(\yF ,\, \ytilde\right)
        \, + \, \map\left( \lF,\, \xF; \, \vect{f} \right)
        \text{ s.t. } 
        \ytilde \in \datafieldset \,, \\
        \text{with} \quad & \globObjFunc \left(\yF ,\, \ytilde\right) := 
        \frac{c}{2} ||e-\tilde{e}||^2_{L^2(\domain)} + \frac{1}{2c} ||s-\tilde{s}||^2_{L^2(\domain)} \,,
    \end{aligned}
\end{equation}
where $c$ is a constant weighting scalar to ensure unit consistency. 
Here, $\globObjFunc(\cdot,\cdot)$ denotes the global objective function. 
We note that \eqref{eq:dcnlp} is a \textit{discrete-continuous nonlinear optimization problem} 
since the objective function $\globObjFunc(\cdot,\cdot)$ is a quadratic function 
and \eqref{eq:dcnlp} combines the data field $\datafieldset$ of a discrete nature and continuous primal and dual fields, $\xF$ and $\lF$, respectively, that subject to the equilibrium and compatibility conditions in $\map$.

The first-order necessary optimality conditions (Karush-Kuhn-Tucker (KKT) conditions) state 
the resulting stationary problem as follows: 

\noindent
Given a fixed $\ytilde = (\tilde{e},\,\tilde{s}) \in \datafieldset$:
\begin{equation}\label{eq:1stOptCondEq}
    \begin{aligned}
        0 = \, \delta\left( \globObjFunc \left(\yF ,\, \ytilde\right)
        + \map\left( \lF,\, \xF; \, \vect{f} \right)
        \right) 
        = \, & \langle \delta \vect{u}, \, \mathcal{B}^T \mu + \linB \vect{\lambda} \, s \rangle_{L^2(\domain)} \\
        + \, & \langle \delta e, \, c (e - \tilde{e}) - \mu \rangle_{L^2(\domain)} \\
        + \, & \langle \delta s, \, \frac{1}{c} (s - \tilde{s}) + \mathcal{B} \vect{\lambda} \rangle_{L^2(\domain)} \\
        + \, & \langle \delta \mu, \, \epsilon(\vect{u}) - e \rangle_{L^2(\domain)} \\
        + \, & \langle \delta \vect{\lambda}, \, \mathcal{B}^T s - \vect{f} \rangle_{L^2(\domain)}
        = \, \vect{g}(\qV) \,,
    \end{aligned}
\end{equation}
where $\qV^T = \left[\vect{u}^T \; e \; s \; \mu \; \vect{\lambda}^T \right]^T$ and 
$\linB = \alpha \,(\cdot)^\prime (\cdot)^\prime \mat{I}$, with $\mat{I}$ is the identity matrix. 
For the derivation of $\linB$, we refer to \ref{sec:linearization}. 
Linearizing Equation \eqref{eq:1stOptCondEq}, using e.g. the Taylor expansion, leads to:
\begin{equation}\label{eq:linearization}
    \begin{aligned}
        0 = \, & \vect{g}(\qV) \, \approx \, \vect{g}(\qV) + \frac{\partial}{\partial \qV} \vect{g}(\qV) \, \Delta \qV 
        = \vect{g}(\qV) \,+ \\                        
        +\, & \langle \delta \vect{u}, \, 
            \mu \, \linB \Delta \vect{u} + 
            \linB \vect{\lambda} \, \Delta s + 
            \mathcal{B}^T \Delta \mu + 
            s \, \linB \Delta \vect{\lambda} 
            \rangle_{L^2(\domain)} \\
        +\, & \langle \delta e, \, c \Delta e - \Delta \mu \rangle_{L^2(\domain)} \\
        +\, & \langle \delta s, \, 
            \vect{\lambda}^T \linB \Delta \vect{u} + 
            \frac{1}{c} \Delta s + 
            \mathcal{B} \Delta \vect{\lambda} \rangle_{L^2(\domain)} \\
        +\, & \langle \delta \mu, \, \mathcal{B} \Delta \vect{u} - \Delta e \rangle_{L^2(\domain)} \\
        +\, & \langle \delta \vect{\lambda}, \, 
            s \, \linB \Delta \vect{u} +
            \mathcal{B}^T \Delta s \rangle_{L^2(\domain)} \,.
    \end{aligned}
\end{equation}    
For more detail on the intermediate steps of the linearization of $\vect{g}(\qV) = 0$, we refer to \ref{sec:linearization}. 
The solution of \eqref{eq:1stOptCondEq} (or \eqref{eq:linearization}) consists of the structural solution, $\qV$, and stress-strain pairs in $\ytilde$ which is then the minimizer of \eqref{eq:dcnlp}. 
In this work, we denote the obtained minimizer with the superscript $(\cdot)^*$, i.e. $\ystar := (\tilde{e}^*,\,\tilde{s}^*)$. 

\begin{remark}
    We note that one can strongly enforce the compatibility conditions by directly replacing the strain field $e$ with the kinematically admissible strain
    $\epsilon(\vect{u})$. 
    The resulting formulation is a mixed formulation involving only displacement and stress as primary variable fields. 
    In the present work, however, we intentionally treat the strain as an independent variable field. This choice allows us to simultaneously achieve high accuracy for all variable fields and to retain full flexibility in selecting suitable function spaces for displacement, strain, and stress. Furthermore, the independent treatment of the strain field follows naturally from the objective function, which explicitly minimizes the distance between the stress-strain variable fields and data.
\end{remark}

\subsection{Spatial discretization}\label{sec:discretization}

In this work, we deal with Equations \eqref{eq:linearization} derived in the previous subsection using the finite element method. 
In particular, 
we spatially discretize the primal and dual fields with $\ndofs$ linear Lagrange polynomials, $N_i(\arclen)$, $i=1,\ldots,\ndofs$, 
and $\mdofs$ constant basis function, $R_j(\arclen)$, $j=1,\ldots,\mdofs$,
as follows:
\begin{align}\label{eq:discretizing}
    & \vect{u} \approx \vect{u}_h = \sum_{i=1}^\ndofs \, N_i \, \hat{\vect{u}}_i = \mat{N} \, \hat{\vect{u}} \,, &\vect{u}_h \in \mathcal{V}^1_h \subset H_0^1(\domain) \\
    & e \approx e_h = \sum_{i=1}^\mdofs \, R_i \, \hat{e}_i = \mat{R} \, \vect{\hat{e}} \,, &e_h \in \mathcal{V}^0_h \subset L^2(\domain) \,, \\
    & s \approx s_h = \sum_{i=1}^\mdofs \, R_i \, \hat{s}_i = \mat{R} \, \vect{\hat{s}} \,, &s_h \in \mathcal{V}^0_h \subset L^2(\domain) \,,\\
    & \mu \approx \mu_h = \sum_{i=1}^\mdofs \, R_i \, \hat{\mu}_i = \mat{R} \, \vect{\hat{\mu}} \,, &\mu_h \in \mathcal{V}^0_h \subset L^2(\domain) \,,\\
    & \vect{\lambda} \approx \vect{\lambda}_h = \sum_{i=1}^\ndofs \, N_i \, \hat{\vect{\lambda}}_i = \mat{N} \, \hat{\vect{\lambda}} \,, &\vect{\lambda}_h \in \mathcal{V}^1_h \subset H_0^1(\domain) \,,
\end{align}
where $\mat{N}$ and $\mat{R}$ denote the matrix of the basis functions $N_i$ and $R_i$, respectively, 
$\hat{\vect{u}}$, $\hat{\vect{e}}$, $\hat{\vect{s}}$, $\hat{\vect{\mu}}$, and $\hat{\vect{\lambda}}$ the vector of the unknown coefficients of the corresponding variable fields. 
Here, the spaces $\mathcal{V}^0_h$ and $\mathcal{V}^1_h$ are:
\begin{equation*}
    \begin{aligned}
        & \mathcal{V}^0_h := \left\{v_h \in L^2(\domain): v_h|_{\domain_{ei}} \in \mathbb{P}_0(\domain_{ei}) \, \forall i \in \{1,2,\ldots,\mdofs\}\right\} \,, \\
        & \mathcal{V}^1_h := \left\{v_h \in H^1_0(\domain): v_h|_{\domain_{ei}} \in \mathbb{P}_1(\domain_{ei}) \, \forall i \in \{1,2,\ldots,\mdofs\}\right\} \,, \\
    \end{aligned}
\end{equation*}
respectively, where $\domain_{ei}$ denotes the domain of the $i$-th element and $\mdofs$ also the number of elements. 
For the sake of notation employed later on, we define:
\begin{equation*}
    \begin{aligned}
        & \yhat := \left(\hat{\vect{e}},\, \hat{\vect{s}}\right)\,, \\
        & \yhat_i := \left(\hat{e}_i,\, \hat{s}_i\right)\,, \; i=1,\ldots,\mdofs\,,
    \end{aligned}
\end{equation*}
and the element objective function $\eleObjFunc\left(\cdot,\cdot\right)$ corresponding to the global one in \eqref{eq:dcnlp} as follows:
\begin{equation}\label{eq:eleObjectiveFunc}
    \eleObjFunc\left(\yhat_i ,\, \ytilde_j\right) := 
        \frac{c}{2} ||\hat{e}_i-\tilde{e}_j||^2_{L^2(\domain_{ei})} + \frac{1}{2c} ||\hat{s}_i-\tilde{s}_j||^2_{L^2(\domain_{ei})}
\end{equation}
where $\ytilde_j := (\tilde{e}_j,\,\tilde{s}_j) \in \dataset$ denotes the arbitrary $j$-th stress-strain pair in $\dataset$, $j=1,\ldots,\numDataPts$ with $\numDataPts$ is the number of available stress-strain pairs in $\dataset$. 
We also spatially discretize the reference configuration with $\ndofs$ functions $N_i(\arclen)$ as follows:
\begin{equation}
    \vect{\refconfig} \approx \vect{\refconfig}_h = \mat{N} \, \hat{\vect{\refconfig}} \,,
\end{equation}
where $\hat{\vect{\refconfig}}$ is the vector of nodal positions in the reference configuration. 
The discrete current configuration is then:
\begin{equation}
    \vect{\curconfig}_h = \vect{\refconfig}_h + \vect{u}_h\,.
\end{equation}

Inserting the discretization above into the 
the first-order necessary optimality conditions \eqref{eq:1stOptCondEq}, we obtain the following spatially discrete formulation of these conditions:
\begin{equation}\label{eq:1stOptCondEq_discrete}
    \begin{aligned}
        0 \,=\, \vect{g}(\qVh) \,=\, & 
            \langle \delta \vect{u}_h, \, \mathcal{B}^T \mu_h + \linB \vect{\lambda}_h \, s_h \rangle_{L^2(\domain)} \\
            + \, & \langle \delta e_h, \, c (e_h - \tilde{e}_h) - \mu_h \rangle_{L^2(\domain)} \\
            + \, & \langle \delta s_h, \, \frac{1}{c} (s_h - \tilde{s}_h) + \mathcal{B} \vect{\lambda}_h \rangle_{L^2(\domain)} \\
            + \, & \langle \delta \mu_h, \, \epsilon_h(\vect{u}_h) - e_h \rangle_{L^2(\domain)} \\
            + \, & \langle \delta \vect{\lambda}_h, \, \mathcal{B}^T s_h - \vect{f} \rangle_{L^2(\domain)} \,,
    \end{aligned}
\end{equation}
where $\qVh^T = \left[\vect{u}_h^T \; e_h \; s_h \; \mu_h \; \vect{\lambda}_h^T \right]^T$ and 
$\epsilon_h(\vect{u}_h)$ is the discrete displacement-based axial strain:
\begin{equation}\label{eq:discretizeESstar}
    \epsilon_h(\vect{u}_h) = \vect{\refconfig}_h^\prime \cdot \vect{u}_h^\prime + \frac{1}{2} \alpha \, \vect{u}_h^\prime \cdot \vect{u}_h^\prime \,.
\end{equation}
Here, $\tilde{e}_h$ and $\tilde{s}_h$ are:
\begin{equation}
    \tilde{e}_h = \sum_{i=1}^\mdofs \, R_i \, \tilde{e}_i \,, \qquad 
    \tilde{s}_h = \sum_{i=1}^\mdofs \, R_i \, \tilde{s}_i \,,
\end{equation}
respectively. 
For the sake of notation, we define $\yhtilde := \left( \tilde{e}_h,\, \tilde{s}_h\right)$, 
and 
$\tilde{\vect{e}} = \left[\tilde{e}_1 \, \ldots \, \tilde{e}_\mdofs \right]^T$ and 
$\tilde{\vect{s}} = \left[\tilde{s}_1 \, \ldots \, \tilde{s}_\mdofs \right]^T$ 
the vector of the strain and stress data in $\ytilde \in \dataset$, respectively.

The linearization of the conditions \eqref{eq:1stOptCondEq_discrete} can be obtained by inserting the spatial discretization into \eqref{eq:linearization} or by directly linearizing \eqref{eq:1stOptCondEq_discrete}, that is:
\begin{equation}\label{eq:linearization_discrete}
    \begin{aligned}
        0 = \, \vect{g}(\qVh) \,+\,                      
        & \langle \delta \vect{u}_h, \, 
            \mu_h \, \linB \Delta \vect{u}_h + 
            \linB \vect{\lambda}_h \, \Delta s_h + 
            \mathcal{B}^T \Delta \mu_h + 
            s_h \, \linB \Delta \vect{\lambda}_h 
            \rangle_{L^2(\domain)} \\
        +\, & \langle \delta e_h, \, c \Delta e_h - \Delta \mu_h \rangle_{L^2(\domain)} \\
        +\, & \langle \delta s_h, \, 
            \vect{\lambda}_h^T \linB \Delta \vect{u}_h + 
            \frac{1}{c} \Delta s_h + 
            \mathcal{B} \Delta \vect{\lambda}_h \rangle_{L^2(\domain)} \\
        +\, & \langle \delta \mu_h, \, \mathcal{B} \Delta \vect{u}_h - \Delta e_h \rangle_{L^2(\domain)} \\
        +\, & \langle \delta \vect{\lambda}_h, \, 
            s_h \, \linB \Delta \vect{u}_h +
            \mathcal{B}^T \Delta s_h \rangle_{L^2(\domain)} \,,
    \end{aligned}
\end{equation}
which can be expressed in the following matrix form (KKT system):
\begin{equation}\label{eq:kkt_matrix_sys}
    \mat{A} \, \Delta \qhat \,=\, \vect{b}\,,
\end{equation}
where 
\begin{equation}\label{eq:kkt_matrices}
    \mat{A} = \begin{bmatrix}
        \mat{A}_{11} & \mat{H}^T \\
        \mat{H} & \mathbf{0}
    \end{bmatrix} \,, \quad \text{and} \quad 
    \vect{b} = - \int_\domain \, \begin{bmatrix}
        \mat{N}^{\prime\,T} \left(\mu_h \, \vect{\refconfig}_h^\prime + \alpha \, \mu_h \, \vect{u}_h^\prime + \alpha \, s_h \, \vect{\lambda}_h^\prime \right) \\
        c \, \mat{R}^T \mat{R} \, \left(\vect{\hat{e}} - \vect{\tilde{e}}\right) - \mat{R}^T \mu_h \\
        \frac{1}{c} \, \mat{R}^T \mat{R} \, \left(\vect{\hat{s}} - \vect{\tilde{s}}\right) + \mat{R}^T \left(\vect{\refconfig}_h^\prime + \alpha \, \vect{u}_h^\prime \right) \cdot \, \vect{\lambda}_h \\
        \mat{R}^T \left( \epsilon_h(\vect{u}_h) - e_h \right) \\
        \mat{N}^{\prime\,T} \left(\vect{\refconfig}_h^\prime + \alpha \, \vect{u}_h^\prime \right) \, s_h - \mat{N}^T \vect{f}
        \end{bmatrix} \, \mathrm{d} \arclen \,,
\end{equation}
with    
\begin{equation}
    \begin{aligned}
        & \mat{A}_{11} = \int_\domain \, \begin{bmatrix}
                \alpha \, \mu_h \, \mat{N}^{\prime\,T} \mat{N}^\prime & \mathbf{0} & \alpha \, \mat{N}^{\prime\,T} \vect{\lambda}_h^\prime \mat{R} \\
                \mathbf{0} & c\,\mat{R}^T \mat{R} & \mathbf{0} \\
                \alpha \,\mat{R}^T \, \vect{\lambda}_h^{\prime \, T} \mat{N}^\prime & \mathbf{0} & \frac{1}{c} \,\mat{R}^T \mat{R}
            \end{bmatrix} \, \mathrm{d} \arclen \,, \vspace{0.3cm}\\
        & \mat{H} = \int_\domain \, \begin{bmatrix}
                \mat{R}^T \left(\vect{\refconfig}_h^\prime + \alpha \, \vect{u}_h^\prime \right)^T \mat{N}^\prime & -\mat{R}^T \mat{R} & \mathbf{0}  \\
                \alpha \, s_h \, \mat{N}^{\prime\,T} \mat{N}^\prime & \mathbf{0} & \mat{N}^{\prime\,T} \left(\vect{\refconfig}_h^\prime + \alpha \, \vect{u}_h^\prime \right) \mat{R}
            \end{bmatrix} \, \mathrm{d} \arclen \,, \quad \text{and} \vspace{0.3cm} \\
        & \qhat^T = \left[\vect{\hat{u}}^T \; \vect{\hat{e}}^T \; \vect{\hat{s}}^T \; \vect{\hat{\mu}}^T \; \vect{\hat{\lambda}}^T \right]^T \,.
    \end{aligned}
\end{equation}
We note that the system matrix $\mat{A}$ is symmetric and has the expected form of a saddle point problem.

\begin{remark}\label{rmk:scalingEq}
    Based on numerical experiments, we observe that the system matrix $\mat{A}$ has a large condition number. To reduce this, we selectively scale certain equations of the matrix equations \eqref{eq:kkt_matrix_sys} and describe the intermediate steps and further details in \ref{sec:scalingEq}. 
    For our computations in Section \ref{sec:results}, we employ a scaling factor of $10^{-5}$ for all cases.
\end{remark}

\subsection{Mixed-integer nonlinear optimization}\label{sec:minlpForm}

An alternative to the data-driven solver introduced in \cite{ortiz_ddcm_2016} is the mixed-integer programming solver, as Kanno introduced in \cite{kanno_data_driven_2019}. 
The main idea is to formulate the data selection behind the methodology of the data-driven solver introduced in \cite{ortiz_ddcm_2016}, i.e. the selection of 
stress-strain pairs in $\dataset$ that are closest to the stress-strain fields satisfying the equilibrium and compatibility conditions, 
in terms of binary constraints. 
Since the objective function is a quadratic function, the resulting problem is then a mixed-integer quadratic programming (MIQP) problem which can be solved globally with a standard mixed-integer programming solver provided by several software packages such as Gurobi\footnote{\url{https://www.gurobi.com/}} or CPLEX\footnote{\url{https://www.ibm.com/products/ilog-cplex-optimization-studio}}. 
Such solvers guarantee optimal solution and hence can be employed to obtain reference solutions for benchmarking of other algorithms and solvers. 
However, the required computational cost increases rapidly when the size of the problem increases \cite{viljar2025,kanno_data_driven_2019}. 
In this work, to investigate the global optimality of our solving strategy discussed in the next section, 
we reformulate the optimization problem \eqref{eq:dcnlp} as a MIQP problem based on \cite{kanno_data_driven_2019}. 
For our computations in Section \ref{sec:results}, we employ the \texttt{JuMP.jl}\footnote{\url{https://jump.dev/JuMP.jl/stable/}} optimization framework with Gurobi as the backend solver, provided as a package in the \texttt{Julia} programming language\footnote{\url{https://julialang.org/}}. 
To reformulate the optimization problem \eqref{eq:dcnlp} as a MIQP problem, we adapt the representation of the data selection introduced by Kanno in \cite{kanno_data_driven_2019}. 
We note that Kanno formulated this selection for each member of the structure, i.e. each element, which necessarily means that the resulting MIQP problem corresponds to the spatially discretized formulation of the optimization problem \eqref{eq:dcnlp} (see also Equation \eqref{eq:1stOptCondEq_discrete}). 
In the following, we employ the same notation of the discrete variable fields introduced in the previous subsection.

Let $t_{ij} \in \{0,1\}$ be a binary variable, $i=1,\ldots,\mdofs$ and $j=1,\ldots,\numDataPts$, where $\numDataPts$ is the number of available stress-strain pairs in $\dataset$ \cite{kanno_data_driven_2019}. 
We recall from the previous subsections that $(\tilde{e}_i^*,\,\tilde{s}_i^*) \in \dataset$ denotes the $i$-th minimizer, i.e. the stress-strain pair in $\dataset$ closest to $(\hat{e}_i,\,\hat{s}_i)$ of the $i$-th element that satisfies the equilibrium and compatibility conditions \eqref{eq:equilibriumNcompatibilityConds}, 
where $i=1,\ldots,\mdofs$ (see also Equations \eqref{eq:discretizing} and \eqref{eq:discretizeESstar}), and 
$\ytilde_j := (\tilde{e}_j,\,\tilde{s}_j)$ the arbitrary $j$-th stress-strain pair in $\dataset$, $j=1,\ldots,\numDataPts$.  
Let $(\check{e}_i,\,\check{s}_i)$ 
denote the unknown stress-strain pair that we want to select from $\dataset$ for the $i$-th element.  
One can formulate the selection of $(\check{e}_i,\,\check{s}_i)$ as follows \cite{kanno_data_driven_2019}: 
\begin{equation}
    \begin{aligned}
        & (\check{e}_i,\,\check{s}_i) = \sum_{j=1}^{\numDataPts} \, (\tilde{e}_j,\,\tilde{s}_j) \, t_{ij}\,,\\
        & \text{with } \sum_{j=1}^{\numDataPts} \, t_{ij} = 1\,, \quad \text{and } t_{ij} \in \{0,1\}\,, \quad \text{where } i=1,\ldots,\mdofs, \; j = 1,\ldots, \numDataPts \,.
    \end{aligned}
\end{equation}
We note that $t_{ij} = 1$ 
if $(\tilde{e}_j,\,\tilde{s}_j) = (\tilde{e}_i^*,\,\tilde{s}_i^*)$, 
and $t_{ij} = 0$ otherwise. 
Let $\xh := \left(\vect{u}_h,e_h,s_h\right) \in \mathcal{V}^1_h \,\times \, \mathcal{V}^0_h \, \times \, \mathcal{V}^0_h \, \subset \, \xSpace$, 
the spatially discretized MIQP formulation of the optimization problem \eqref{eq:dcnlp} is then:
\begin{equation}\label{eq:MIQPform}
    \begin{aligned}
        \min_{\xh} \; & \frac{c}{2} ||\hat{\vect{e}} - \check{\vect{e}}||^2_{L^2(\domain)} + \frac{1}{2c} ||\hat{\vect{s}}-\check{\vect{s}}||^2_{L^2(\domain)}\,, \\
        \text{s.t.} \; & (\check{e}_i,\,\check{s}_i) = \sum_{j=1}^{\numDataPts} \, (\tilde{e}_j,\,\tilde{s}_j) \, t_{ij}\,,\\
        & \sum_{j=1}^{\numDataPts} \, t_{ij} = 1\,, \quad t_{ij} \in \{0,1\}\,, \; i=1,\ldots,\mdofs, \; j = 1,\ldots, \numDataPts \,, \\
        & \int_\domain \,\mat{R}^T \left( \epsilon_h(\vect{u}_h) - e_h \right) = 0 \,, \\
        & \int_\domain \, \mat{N}^{\prime\,T} \left(\vect{\refconfig}_h^\prime + \alpha \, \vect{u}_h^\prime \right) \, s_h - \mat{N}^T \vect{f} = 0 \,,
    \end{aligned}
\end{equation}
where 
$\check{\vect{e}} = \left[\check{e}_1 \, \ldots \check{e}_\mdofs\right]^T$, 
$\check{\vect{s}} = \left[\check{s}_1 \, \ldots \check{s}_\mdofs\right]^T$. 
Here, the last two constraints are the spatially discrete formulation of the compatibility and equilibrium conditions (see also the first-order necessary optimality conditions \eqref{eq:1stOptCondEq_discrete}). 
We note that the optimization problem \eqref{eq:dcnlp} is now formulated as a minimization problem of a convex quadratic function subjected to linear equality and binary constraints. 
An alternative to the $L^2-$norm employed in the objective function is the $L^1-$norm, which enables the objective function to be implemented as a linear function, using auxiliary variables. Hence, this alternative can greatly reduce the computational cost when using mixed-integer programming solver \cite{viljar2025}. 
The resulting optimization problem is categorized as a mixed-integer nonlinear optimization problem (MINLP).

\section{Solving strategy}\label{sec:solvingstrategy}

In this section, we discuss  
the direct data-driven solver based on the alternating direction method (ADM) introduced in \cite{ortiz_ddcm_2016} 
and the combination with the Newton-Raphson method \cite{Keip_ddcm_nonlinearbar} for nonlinear elasticity. 
We then 
extend and generalize our solving strategy introduced in \cite{viljar2025} for nonlinear systems. 
In particular, we combine the ADM-based solving algorithm with the greedy optimization algorithm, aiming to achieve a globally optimal solution. 
We close this section with an overview and a brief discussion on three different data initialization approaches considered in this work: a random, stress-free, and structure-specific initialization.

\subsection{Alternating direction method}

\begin{algorithm}[h]
\textbf{Input}:
	dataset $\dataset$,
	external force vector $\vect{f}$ \\
\textbf{Output}: $\qhat$, $\ystar$
\begin{algorithmic}[1]
    \State $\qhat^{(0)} = \vect{0}$      \Comment{Initial solution guess for the 1st load step}
    \State Inititalize $\ytilde^{(0)}$ with Algorithm \ref{alg:datainitialize}    
\For{$j$ in $1,\ldots,$number of load steps}
    \State $\vect{f}^{(j)} = \gamma_j \, \vect{f}$      \Comment{$j$-th load factor $\gamma_j$}
    \State $\ytilde^{(j)} = \ytilde^{(j-1)}$
    \State $\qhat^{(j)}, \, \ytilde^{*(j)}$ = ADM-solver $\left(\qhat^{(j-1)}, \, \ytilde^{(j)}, \, \vect{f}^{(j)}, \, \dataset\right)$      \Comment{See Algorithm \ref{alg:admsolver}}
\EndFor
\caption{Solving strategy for geometrically nonlinear data-driven problems based on the alternating direction method (ADM).}\label{alg:admstrategy}
\end{algorithmic}
\end{algorithm}

\begin{algorithm}[h]
\textbf{Input}:
	solution guess $\qhat_0$, 
    initial selected data $\ytilde_0$,
    dataset $\dataset$, 
	external force vector $\vect{f}$ \\
\textbf{Output}: $\qhat$, $\ystar$
\begin{algorithmic}[1]
    \State $i=0$     \Comment{Number of iterations}
    \State $\ytilde_{i+1}$ = $\ytilde_{i}$ + $\delta$     \Comment{Initial error}
    \While{$\ytilde_{i+1}$ $\neq$ $\ytilde_{i}$}
        \State $\qhat$ = Newton-Raphson scheme $\left(\qhat_0,\, \ytilde_{i},\, \vect{f}\right)$        \Comment{See Algorithm \ref{alg:NRscheme}}
    	\State $\ytilde_{i+1}$ = Local state assignment $\left(\qhat,\,\dataset \right)$        \Comment{See Algorithm \ref{alg:localstateassigment}}
        \State $i\,+=\,1$
	\EndWhile
    \State $\ystar$ = $\ytilde$     \Comment{Converged $\ytilde$ is the obtained minimizer}
\caption{ADM-solver: Direct data-driven solver based on the Alternating direction method (ADM), combined with the Newton-Raphson iteration scheme.}\label{alg:admsolver}
\end{algorithmic}
\end{algorithm}

\begin{algorithm}[h]
\textbf{Input}:
	solution $\qhat$, 
    dataset $\dataset$ \\
\textbf{Output}: $\ytilde$
\begin{algorithmic}[1]
    \State Collect $\yhat = \left(\hat{\vect{e}},\, \hat{\vect{s}}\right)$ from $\qhat$
    \For{$i$ in $1,\ldots,\mdofs$}
        \State Choose $\ytilde_i \in \dataset$ s.t. 
        $\eleObjFunc \left(\yhat,\,\ytilde_i\right)$ = min.       \Comment{See also Equation \eqref{eq:eleObjectiveFunc}}
    \EndFor
    \State Return $\ytilde$
\caption{Local state assignment: selection of the stress-strain pairs $\ytilde \in \dataset$ that are closest to $\yhat$.}\label{alg:localstateassigment}
\end{algorithmic}
\end{algorithm}

To tackle the data-driven structural analysis formulated as a 
discrete-continuous nonlinear optimization problem, 
the authors of \cite{ortiz_ddcm_2016} introduced the direct data-driven solver that seeks the stress-strain pairs in a given dataset, $\ytilde \in \dataset$, closest to the stress-strain pairs that is parts of the structural solution, $\yF$, satisfying the equilibrium and compatibility conditions. 
This solver is an iterative scheme where $\ytilde$ are firstly fixed with initial selected data in $\dataset$ to solve for the variable fields that satisfy the equilibrium and compatibility conditions. The solver then seeks new values of $\ytilde$ that are closest to the obtained solution of the variable fields by minimizing the objective function at the element level and repeats until $\ytilde$ converge to a minimizer $\ystar$. 
This necessarily means that the direct data-driven solver introduced in \cite{ortiz_ddcm_2016} alternates between two minimization subproblems: 
i) minimizing the objective function with fixed discrete variables $\ytilde \in \dataset$ to yield optimal continuous variables $\xF$, and 
ii) minimizing the objective function with fixed continuous variables $\xF$ to obtain optimal discrete variables $\ytilde \in \dataset$. 
This solving algorithm is the main idea behind the alternating direction method (ADM) \cite{Douglas1956adm,Gabay1976adm}, which  
also iteratively alternates between minimizing separate blocks of variables while holding the others fixed in solving each of the subproblems.

In this work, we consider the direct data-driven solver introduced in \cite{ortiz_ddcm_2016} and its extension to nonlinear elasticity by combining with the Newton-Raphson method \cite{Keip_ddcm_nonlinearbar}. 
Algorithm \ref{alg:admstrategy} describes this solving strategy for the spatially discretized optimization problem \eqref{eq:1stOptCondEq_discrete} discussed in the previous section. 
Here, we explicitly describe the computation at each load step including the initialization of the stress-strain data pairs $\ytilde$. 
For the first load step, we initialize $\ytilde$ considering three different approaches, including the random initialization \cite{ortiz_ddcm_2016,Keip_ddcm_nonlinearbar} 
(see line 2 in Algorithm \ref{alg:admstrategy}), which  
we discuss more in details later in Section \ref{sec:datainitialization}.  
Starting from the second load step, we initialize $\ytilde$ with the minimizer obtained from the preceding load step  
(see line 5 in Algorithm \ref{alg:admstrategy}). 
In each load step, we employ the direct data-driven solver, to which we refer to as ADM-solver, to obtain the solution $\qhat$ and update $\ytilde$. 
Algorithm \ref{alg:admsolver} describes the details of the solving steps, including the Newton-Raphson scheme for solving the underlying geometrically nonlinear structural problem (see line 4 in Algorithm \ref{alg:admsolver}). 
We note that we employ the same solution guess for the Newton-Raphson scheme at each alternating iteration and update only the selected data pairs $\ytilde$. 
For completeness, we describe the details of the Newton-Raphson scheme employed in Algorithm \ref{alg:admsolver} in \ref{sec:Newtonraphsonalgo}. 
In Algorithm \ref{alg:localstateassigment}, we describe the details of the local state assignment step which seeks the stress-strain pair in the dataset that is closest to the obtained structural solution of each element via minimizing the element objective function \eqref{eq:eleObjectiveFunc}.

\begin{remark}\label{rmk:admcompcost}
  We note that given a fixed value of the stress-strain data pairs $\ytilde$, the computational effort for each update of $\ytilde$ 
  primarily lies in solving the linearized system \eqref{eq:kkt_matrix_sys} within the Newton-Raphson scheme. Therefore it only depends on the dimension of the resulting system of equations. 
  Moreover, the quadratic convergence of the Newton-Raphson method is well-known. 
  The number of iterations required for updating $\ytilde$ and consequently the convergence of the ADM-solver, however, has not yet been analyzed or established in the current state of the art. 
  For more discussions in this regard, we refer to \cite{Gebhardtelectric2025}.
\end{remark}

  \begin{algorithm}[h]
\textbf{Input}:
	dataset $\dataset$,
	external force vector $\vect{f}$ \\
\textbf{Output}: $\qhat$, $\ystar$
\begin{algorithmic}[1]
    \State $\qhat^{(0)} = \vect{0}$      \Comment{Initial solution guess for the 1st load step}
    \State Inititalize $\ytilde^{(0)}$ with Algorithm \ref{alg:datainitialize}    
\For{$j$ in $1,\ldots,$number of load steps}
    \State $\vect{f}^{(j)} = \gamma_j \, \vect{f}$      \Comment{$j$-th load factor $\gamma_j$}
    \State $\ytilde^{(j)} = \ytilde^{(j-1)}$
    \State $\qhat^{(j)}, \, \ytilde^{*(j)}$ = GO-ADM-solver $\left(\qhat^{(j-1)}, \, \ytilde^{(j)}, \, \vect{f}^{(j)}, \, \dataset\right)$      \Comment{See Algorithm \ref{alg:goadmsolver}}
\EndFor
\caption{Solving strategy for geometrically nonlinear data-driven problems based on the \textbf{greedy optimization algorithm} and the alternating direction method (ADM).}\label{alg:goadmstrategy}
\end{algorithmic}
\end{algorithm}
  \begin{algorithm}[ht]
\textbf{Input}:
	solution guess $\qhat_0$, 
    initial selected data $\ytilde_0$,
    dataset $\dataset$, 
	external force vector $\vect{f}$ \\
\textbf{Output}: $\qhat$, $\ystar$
\begin{algorithmic}[1]
    \State $\yhat \gets \qhat, \, \ytilde$ = ADM-solver $\left(\qhat_0, \, \ytilde_0, \, \vect{f}, \, \dataset\right)$      \Comment{First results. See Algorithm \ref{alg:admsolver}}
    \State dist$^{(0)}$ = dist$_G\left(\yhat,\, \ytilde\right)$      \Comment{Evaluate global objective function}
    \State $k=0$     \Comment{Number of ``greedy'' searches}
    \While{$k \leq k_{\text{max}}$}
        \State $\idset = [i]_{1,\ldots,\mdofs}$ s.t. $\left[\text{dist}_E\left(\yhat_i,\, \ytilde_i\right) \right]_{i \in \idset}$ is descending
        \For{$i$ in $\idset$}
            \State $k+=1$, \textcolor{blue1}{$\ytilde^{n}$} = $\ytilde$
            \State $q$, $p$ = $\argmin\left( \, \text{dist}_E\left(\yhat_i,\, \ytilde_j\right) \, \right)$, $j=1,\ldots,\numDataPts$     \Comment{Find 2 best alternatives to $\ytilde_i$}
            \If{\textcolor{blue1}{$\ytilde^{n}_i$} $\neq \, \ytilde_{q}$}
                \State \textcolor{blue1}{$\ytilde^{n}_i$} = $\ytilde_{q}$      \Comment{Get the best alternative.}
            \Else
                \State \textcolor{blue1}{$\ytilde^{n}_i$} = $\ytilde_{p}$      \Comment{Get the 2nd-best alternative.}
            \EndIf
            \State $\yhat \gets \qhat$, \textcolor{blue1}{$\ytilde^{n}$} = ADM-solver $\left(\qhat_0, \, \textcolor{blue1}{\ytilde^{n}}, \, \vect{f}, \, \dataset\right)$      \Comment{Recompute with new $\textcolor{blue1}{\ytilde^{n}_i}$ in $\textcolor{blue1}{\ytilde^{n}}$}
            \State dist$^{(k)}$ = dist$_G\left(\yhat,\, \textcolor{blue1}{\ytilde^{n}}\right)$
            \If{dist$^{(k)}$ $<$ dist$^{(k-1)}$ or dist$^{(k)} \leq \delta$}
                \State $\ytilde_i = \textcolor{blue1}{\ytilde^{n}_i}$
                \If{dist$^{(k)} \leq \delta$}: $k=k_{\text{max}}+1$
                \EndIf
                \State break
            \EndIf
        \EndFor
	\EndWhile
\caption{GO-ADM-solver: Direct data-driven solver based on the \textbf{greedy optimization algorithm} and the Alternating direction method (ADM).}\label{alg:goadmsolver}
\end{algorithmic}
\end{algorithm}

  \subsection{Greedy-optimization based alternating direction method}\label{sec:goadmalgorithm}

  It is well known that  
  the direct data-driven solver based on the alternating direction method (ADM-solver) does not guarantee globally optimal solution 
  \cite{Gebhardtddcmsolution2025,kanno_data_driven_2019}, as also discussed in the previous section. 
  To achieve a better approximation of the global optimal via reducing the value of the global objective function, 
  we combine this ADM-solver with a greedy optimization algorithm, 
  searching for alternative data points in the nearest-neighbourhood \cite{viljar2025}. 
  Based on this similar idea, the so-called defective restarting approach has been introduced in \cite{Rocha2025}, where the second-nearest data points are employed as alternative for some randomly chosen parts of the stress and strain states. 
  The key motivation to use alternative data points in the nearest-neighbourhood is to avoid the alternating solving strategies getting trapped in local optima \cite{Rocha2025}. 
  In this work, based on a greedy optimization algorithm, we systematically search for alternative data points corresponding to the stress and strain states across all elements. 
  Moreover, our solving strategy generally allows to consider more than one alternative point in the nearest-neighbourhood and to choose   
  a desired upper bound of the global objective function as the convergence criterion, 
  rather than a fixed maximal number of repeated computations as the defective restarting approach. 
  Regarding the greedy optimization algorithms, firstly introduced in \cite{Temlyakov2008}, 
  they belong to a class of approximation methods that
  iteratively improves the local optima with the aim of approximating the global optima. 
  This type of algorithms has been applied in various fields, for instance in computational mechanics for model order reduction \cite{Lappano2016greedy,Jelich2021greedy,Babbepalli2025greedy}, 
  data and signal compression \cite{Blanchard2015greedy}, and approximation problems \cite{TEMLYAKOV2014greedy}. 
  For an overview of its applications, we refer to \cite{Garcia2025greedyOpt}. 
  In this work, we attempt to employ this for structural analysis in the context of data-driven computational mechanics. 
  In general, greedy optimization algorithms 
  do not always guarantee the absolute best outcome, they are highly practical in situations where computing an exact optimal solution would be prohibitively expensive or even not possible.

  The ADM-solver, described in Algorithm \ref{alg:localstateassigment}, 
  seeks the stress-strain pairs from the given data set, $\ytilde \in \dataset$, at the element level and hence yields at best local optima. 
  Employing a greedy optimization algorithm, we now iteratively search for alternatives to $\ytilde$ in its nearest-neighbourhood in $\dataset$ for each element, which possibly reduce the value of the global objective function. 
  We describe the details and each step of the proposed solving strategy in Algorithms \ref{alg:goadmstrategy} and \ref{alg:goadmsolver}. 
  The resulting solver is a direct data-driven solver based on alternating direction method (ADM) combined with a greedy optimization algorithm, and hence we refer to this solver as GO-ADM solver. 
  Algorithm \ref{alg:goadmstrategy} shows the solving strategy for each load step, including data initialization and solving with the GO-ADM solver described in Algorithm \ref{alg:goadmsolver}. 
  We note that the Algorithm \ref{alg:goadmstrategy} is the same as the solving strategy based on ADM described in Algorithm \ref{alg:admstrategy} except for the employed solver that is now the GO-ADM- instead of the ADM-solver.

  Within the GO-ADM-solver, 
  we compute first results using the ADM-solver and evaluate the global objective function using these results, $\globObjFunc(\cdot,\cdot)$ (see lines 1-2 in Algorithm \ref{alg:goadmsolver}). 
  We then systematically search for alternative of the stress and strain state of each element, which reduces $\globObjFunc(\cdot,\cdot)$. 
  We start this search for the element that has the largest contribution to $\globObjFunc(\cdot,\cdot)$, i.e. with element index $i \in \idset$, $i = 1,\,\ldots,\,\mdofs$, such that: 
  \begin{equation}
    \mathcolorbox{pink}{\left[\text{dist}_E\left(\yhat_i,\, \ytilde_i\right) \right]_{i \in \idset}} \; \text{is descending (see line 5 of Algorithm \ref{alg:goadmsolver}).} \nonumber
  \end{equation}

  \noindent
  For each $i$-th element in $\idset$, we find two closest data points to its stress and strain states:
  \begin{equation}
    \mathcolorbox{pink}{
      q,\, p = \argmin\left( \, \text{dist}_E\left(\yhat_i,\, \ytilde_j\right) \, \right), \, j=1,\ldots,\numDataPts
    } \,. \nonumber
  \end{equation}

  \noindent
  If the data point that is already found for this $i$-th element, $\ytilde^{n}_i$, is \textit{not} the closest one, i.e. $\ytilde^{n}_i \, \neq\, \ytilde_{q}$, we overwrite $\ytilde^{n}_i$ with $\ytilde_{q}$. If $\ytilde^{n}_i$ is the closest data point, we overwrite $\ytilde^{n}_i$ with the second-nearest data point, $\ytilde_{p}$, (see lines 9-13 of Algorithm \ref{alg:goadmsolver}). 
  We then employ $\ytilde^{n}$ with the new data point for the $i$-th element as initial selected data points and resolve the optimization problem using the standard ADM-solver (see lines 14-15 of Algorithm \ref{alg:goadmsolver}). 
  If the global objective function, $\globObjFunc(\cdot,\cdot)$, is now reduced, we keep the new data point $\ytilde^{n}_i$ and continue the same search for the next element in $\idset$. 
  Our GO-ADM solver keeps searching until $\globObjFunc(\cdot,\cdot)$ is smaller than a desired upper bound, $\delta$, or until the number of searches meets the maximum value. 
  We note that the latter criterion is purely for numerical stability, for instance, to maintain a practical computing time, particularly when $\delta$ is chosen to be very small. 
  Moreover, we see in Algorithm \ref{alg:goadmsolver} that it allows to consider more than two closest data points for the stress and strain states of each element. In general, one can resolve the optimization problem with $n_{Al} \geq 1$ nearest-neighbours and select the data point that leads to minimal value of $\globObjFunc(\cdot,\cdot)$ among these neighbours.

  \begin{remark}
      Our solver presented in Algorithm \ref{alg:goadmsolver} searches
      for the nearest neighbours as alternative data points
      over the complete dataset for each element. 
      One can reduce this searching region to the neighbourhood of the current stress-strain state within a defined radius.
      This possibly reduces the computational effort, particularly, in cases of very dense datasets.
  \end{remark}

  Regarding the computational cost of the proposed GO-ADM-solver,
  the number of ``greedy'' searches is unpredictable, as well as the number of iterations for each call of the ADM-solver (see also Remark \ref{rmk:admcompcost}). 
  In other words, the total number of searches and iterations, and hence also the total required computational effort is unknown. 
  Compared to the ADM-solver, the GO-ADM-solver linearly scales the computational cost by the number of ``greedy'' searches since it calls the former once for each search to resolve the optimization problem using alternative initial data pairs. 
  When considering more than one nearest neighbours as alternative data points, the computational cost is additionally scaled with this number of considered alternatives. 
  Moreover, the new initial data pairs may not associate with any structural solution 
  and hence require a higher number of iterations within the ADM-solver at each search.

  We note that, compared to the defective restarting approach introduced in \cite{Rocha2025}, our GO-ADM solving strategy is computationally more expensive since it searches for alternative data of stress-strain states across all elements, rather than a subset of randomly selected states.
  As an alternative to random state selection, one could restrict the search to elements that make a significant contribution to the global objective function,
  for example those whose element objective function values are at least 10\% (or another defined percentage) of the global objective function.
  Another possible strategy is to apply GO-ADM only up to or at selective load steps instead of at every step. 
  This approach, however, does not guarantee the improvement of the solution and global objective function for sequential steps.
  In the present work, we focus on evaluating the performance of our GO-ADM solving strategy in terms of solution accuracy, reduction of the global objective function, and computational cost, in comparison with the standard ADM-solver. 
  Given the very competitive performance of the defective restarting approach \cite{Rocha2025}, we plan to carry out a comprehensive comparison between GO-ADM and this method in future work. 
  This will include a statistical analysis to account for the randomness inherent in the state selection of the defective restarting approach and will serve as a basis for further refinement and development of the GO-ADM solving strategy.

  \begin{remark}
      Since the GO-ADM solver builds on the standard ADM framework, it inherits the same limitations when applied to non-convex optimization problems, such as snap-through or buckling phenomena, where multiple equilibrium solutions may exist. 
      In these cases, the data-driven solver may switch between different admissible solutions across iterations, preventing convergence. 
      Addressing such problems typically requires extensions of the standard data-driven approach, as proposed for example in \cite{Kuang2023ddcmsnapthrough,Romero2026ddcm}. 
      We plan to study and integrate these enhancements 
      with our GO-ADM solving strategy in future work.
  \end{remark}

  \subsection{Data initialization approaches}\label{sec:datainitialization}

    \begin{algorithm}[h]
\textbf{Input}:
    dataset $\dataset$ \\
\textbf{Output}: $\ytilde$
\begin{algorithmic}[1]
    \If{random initialization}
        \State $\idset$ = random selected $\mdofs$-indices $i \in \{1,\ldots,\numDataPts\}$
        \State Select $\ytilde_i \in \dataset$, with $i \in \idset$ for $\mdofs$ elements
    \ElsIf{stress-free initialization}
        \State Select $\ytilde_i = (0,\,0) \in \dataset$, $\forall \, i=1,\ldots,\mdofs$
    \Else           \Comment{Default with structure-specific initialization}
        \State Solve Equation \eqref{eq:equilibriumCondMatrixEq}
        \State Select $\ytilde_i \in \dataset$ s.t. $\frac{1}{2c} ||\hat{s}_i-\tilde{s}_i||^2_{L^2(\domain_{ei})}$ = min, $i=1,\ldots,\mdofs$
    \EndIf
    \State Return $\ytilde$
\caption{Data initialization scheme.}\label{alg:datainitialize}
\end{algorithmic}
\end{algorithm}

    We now discuss and give an overview of three common approaches for the initialization of the stress-strain data pairs $\ytilde$ in the solving strategies discussed in the previous subsections (see also Algorithms \ref{alg:admsolver} and \ref{alg:goadmsolver}) considered in this work: 
    i) a random initialization \cite{ortiz_ddcm_2016}, 
    ii) a stress-free initialization \cite{Gebhardtddcmstatic2020}, 
    iii) and a structure-specific initialization \cite{Gebhardtddcmsolution2025}. 
    We describe the initialization using these approaches in details in Algorithm \ref{alg:datainitialize}. 
    The first approach is to randomly select $\ytilde$ as firstly introduced together with the direct data-driven solver in \cite{ortiz_ddcm_2016}. 
    In general, this random selection is not related to any physical state of the underlying structure and loading scenario, hence, can lead to diverging Newton-Raphson scheme. 
    For geometrically linear structural problems, we can expect convergence, however, a larger number of iterations for both the Newton-Raphson scheme and the ADM-solver. 
    Although the initialized data $\ytilde$ is employed only for the first load step, the inaccurate solution might affect the accuracy and convergence at following load steps. 
    Moreover, noises and pollutions in the dataset may amplify this effect over the load steps.

    Alternatively, one can select zeros values for the data pairs $\ytilde$, which necessary correspond to the stress-free state of the underlying structure. 
    This initialization approach reduces the chance but theoretically does not eliminate the case of diverging Newton-Raphson scheme for geometrically nonlinear structural problems. 
    The third approach considered in this work is a structure-specific initialization first introduced in \cite{Gebhardtddcmsolution2025}. 
    The main idea is to select the initial data pairs for $\ytilde$ that consist of stresses being closest to the     
    solution of the equilibrium condition of the underdetermined linear system. 
    This initialization contains structure-specific information and guarantees globally optimal solution for geometrically linear structures in certain symmetric cases \cite{Gebhardtddcmsolution2025}. 
    For geometrically nonlinear cases, this is no longer guaranteed. Nevertheless, the structure-specific initialization avoids the divergence of the Newton-Raphson scheme since the resulting values of $\ytilde$ relate to the solution of the underlying linear system, i.e. the solution obtained in the first Newton-Raphson iteration at the first load step. 
    Hence, this initialization approach has the most favorable properties among the three considered approaches in this work. Therefore, we choose this as the default choice in Algorithm \ref{alg:datainitialize}.

    For completeness, we state here the equations of the equilibrium conditions of the linear system (i.e. linear strains or $\alpha = 0$) corresponding to the optimization problem \eqref{eq:1stOptCondEq_discrete} studied in this work:
    \begin{equation}\label{eq:equilibriumCondMatrixEq}
      \int_\domain \, \mat{N}^{\prime\,T} \vect{\refconfig}_h^\prime \, \mat{R} \, \vect{\hat{s}} = \int_\domain \, \mat{N}^T \vect{f} = 0\,.
    \end{equation}
    Solving this system of equations for $\vect{\hat{s}}$, one can then seek the stress-strain pairs in the given dataset $\dataset$ with stresses being closest to $\vect{\hat{s}}$ (see also line 8 in Algorithm \ref{alg:datainitialize}). 
    For more theoretical details and discussion on solving this system of equations, we refer to \cite{Gebhardtddcmsolution2025}.
    We note that one can also employ $\vect{\hat{s}}$ as 
    initial stresses in the solution guess of $\qhat$ for the first load step (see also Algorithms \ref{alg:admsolver} and \ref{alg:goadmsolver}). 
    Given an approximated constitutive manifold, one can further obtain the corresponding strains $\vect{\hat{e}}$ to improve the solution guess of $\qhat$. 
    Another possible extension is to solve the underlying linear structural problem with an approximated constitutive manifold to obtain the complete solution $\qhat$ for the first load step. 
    This approach is out of scope of this work and is considered in future work.

\section{Numerical studies}\label{sec:results}

In this section, we numerically illustrate the favorable properties of the introduced solving strategy based on the greedy optimization algorithm and the alternating direction method (ADM), discussed in the previous section (see also Algorithm \ref{alg:goadmstrategy}). 
We start with the benchmarking of our implementation for a one-dimensional bar and a truss structure in two-dimensional space. 
To illustrate the application of the introduced solving strategy to a practical scenario with a real experimental dataset, we consider a cyclic test of a mooring line provided by an anonymous industry partner. 
We then numerically illustrate via a truss structure that our solving strategy generally leads to better approximation of the global optima. 
This, however, requires higher computational cost compared to the standard direct data-driven solving strategy. 
Lastly, we briefly discuss the effect of the quality of the dataset on the solution.

\subsection{Benchmarking}

  We first benchmark our implementation of the solving strategies discussed in the previous section. 
  To this end, we consider the  
  linear constitutive law and generate the dataset based on this relation, which enables the computation of the reference solution for the considered benchmark examples, using standard tools and formulations. 
  We consider the structure-specific initialization approach since this also requires its own implementation compared to the other two approaches, which we benchmark at the same time as the solving strategies.

    \subsubsection{One-dimensional bar structure}

\begin{figure}
    \centering
    \subfloat[][Sketch of 1D bar]{\includegraphics[width=0.45\textwidth]{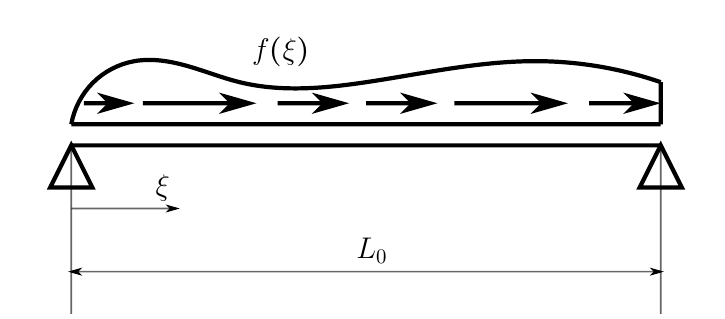}}
    \subfloat[][Axial force function]{\includegraphics[width=0.48\textwidth]{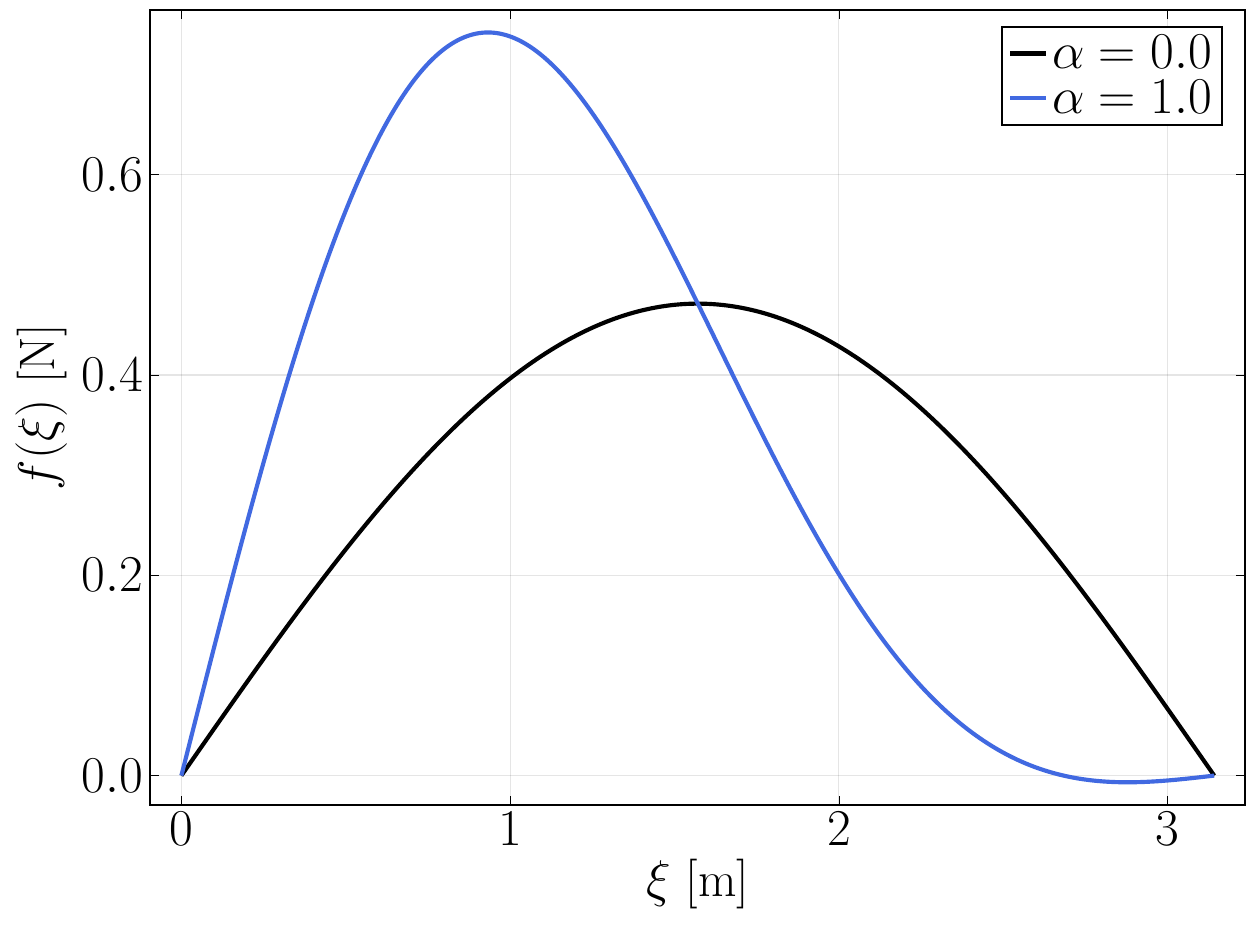}}
    \vspace{0.2cm}
    \caption{Sketch of 1D bar (a) subjected to axial force (b) resulting from a manufactured axial displacement in the case of linear ($\alpha=0.0$) and nonlinear ($\alpha=1.0$) strain measures.}\label{fig:1dbarGeometry}
\end{figure}

\begin{figure}
    \centering
    \subfloat[][Axial displacement $u_{h,x}$, $\alpha=0.0$]{\includegraphics[width=0.48\textwidth]{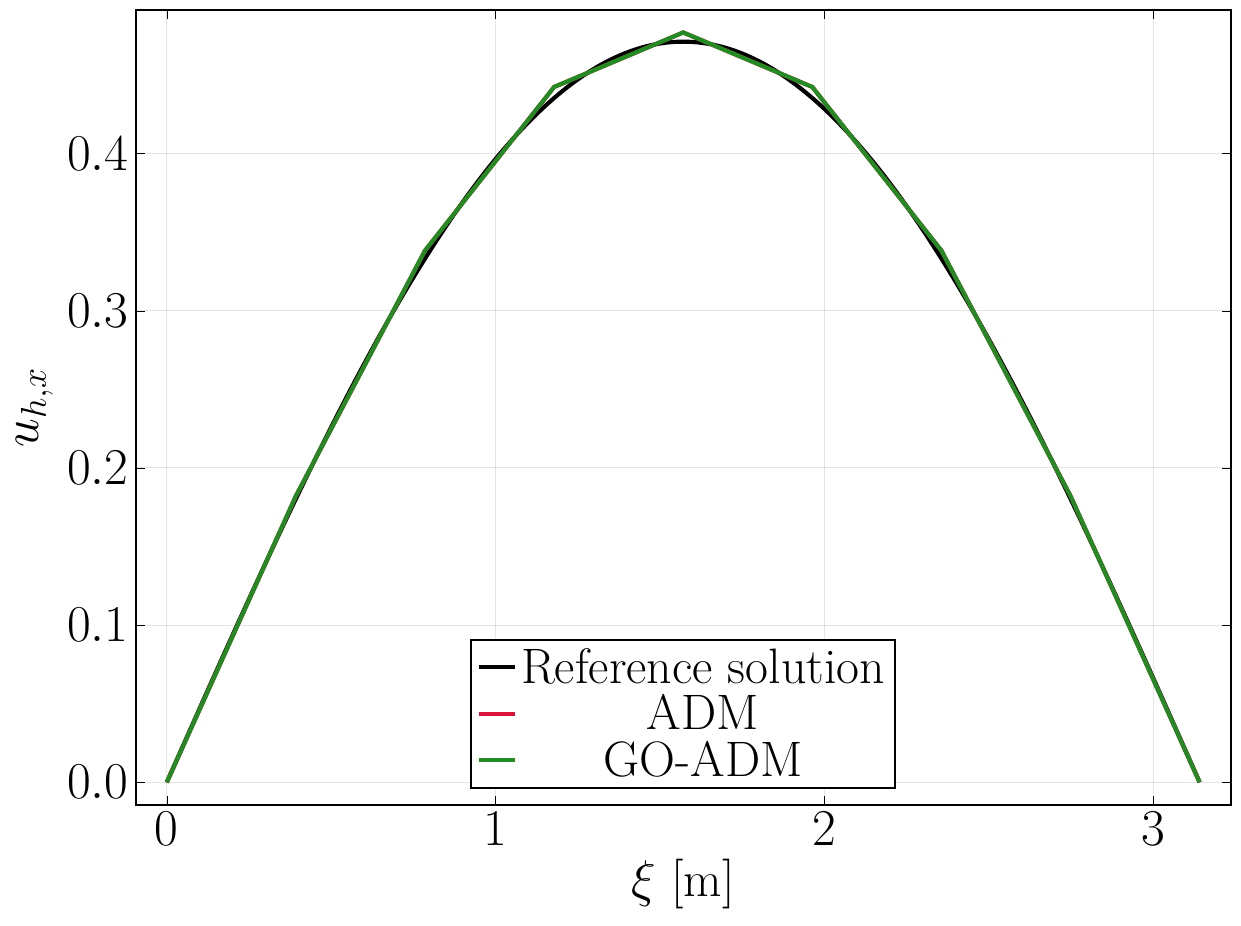}}
    \subfloat[][Axial displacement $u_{h,x}$, $\alpha=1.0$]{\includegraphics[width=0.48\textwidth]{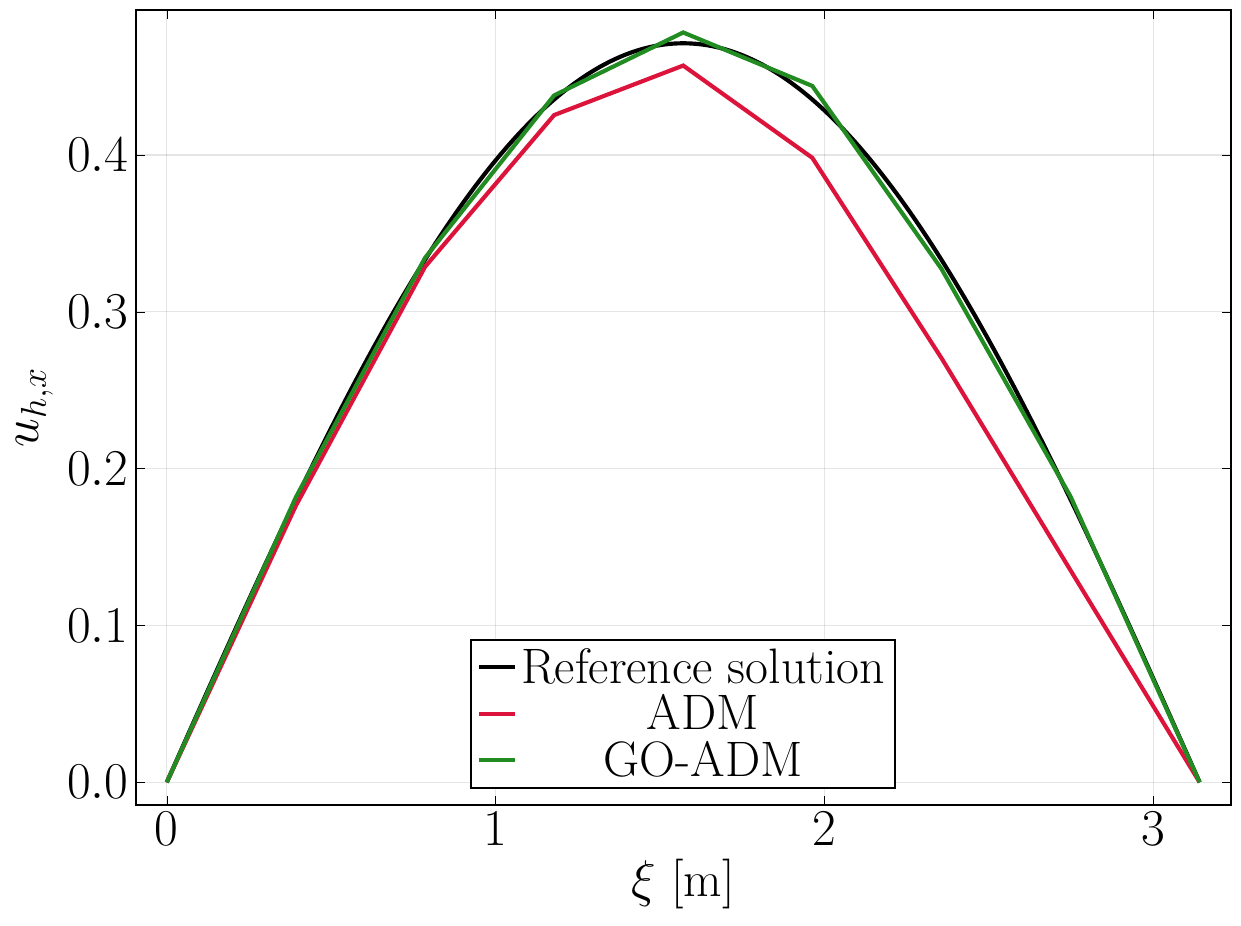}}

    \subfloat[][$e_h$, $\alpha=0.0$]{\includegraphics[width=0.48\textwidth]{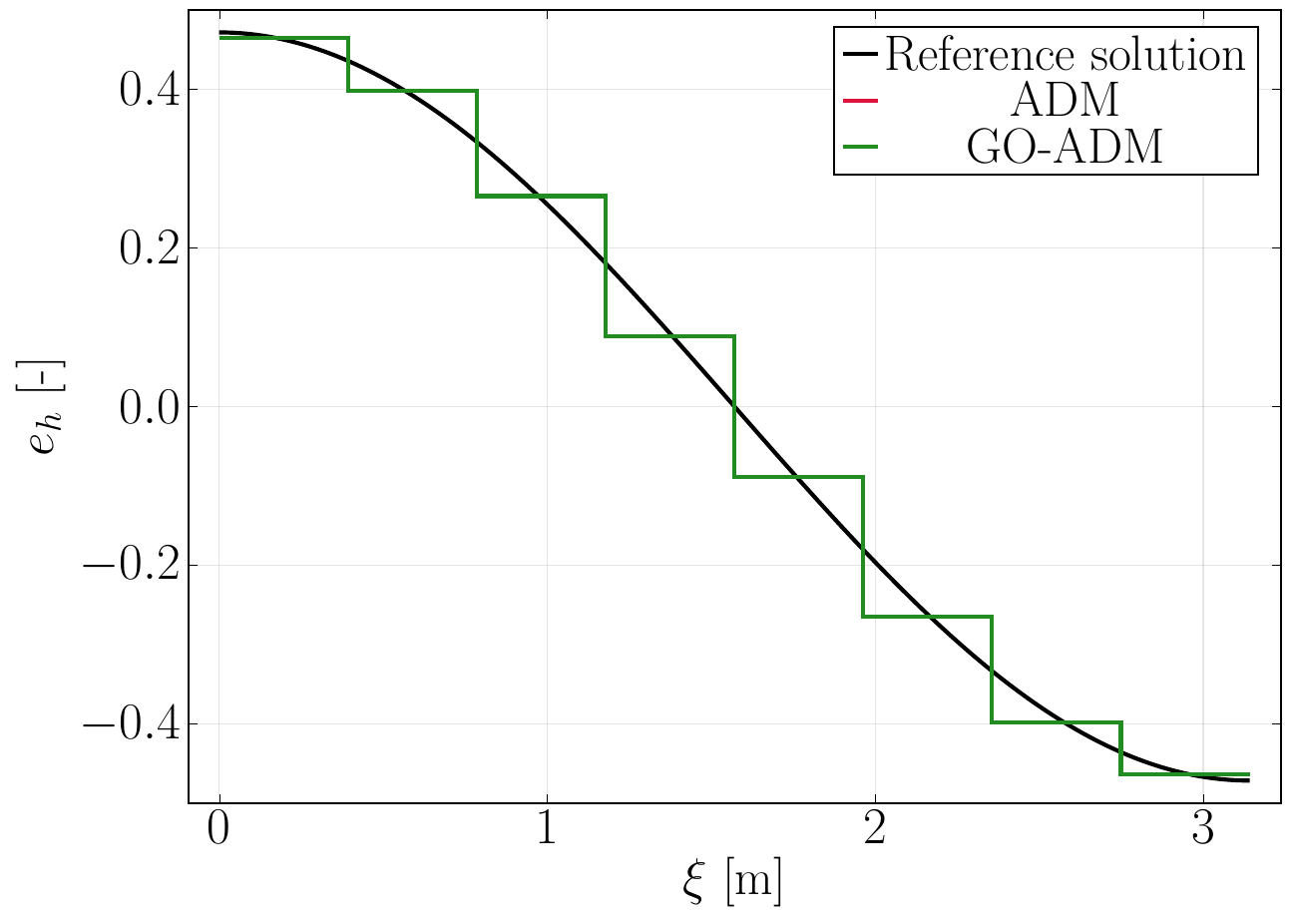}}
    \subfloat[][$e_h$, $\alpha=1.0$]{\includegraphics[width=0.48\textwidth]{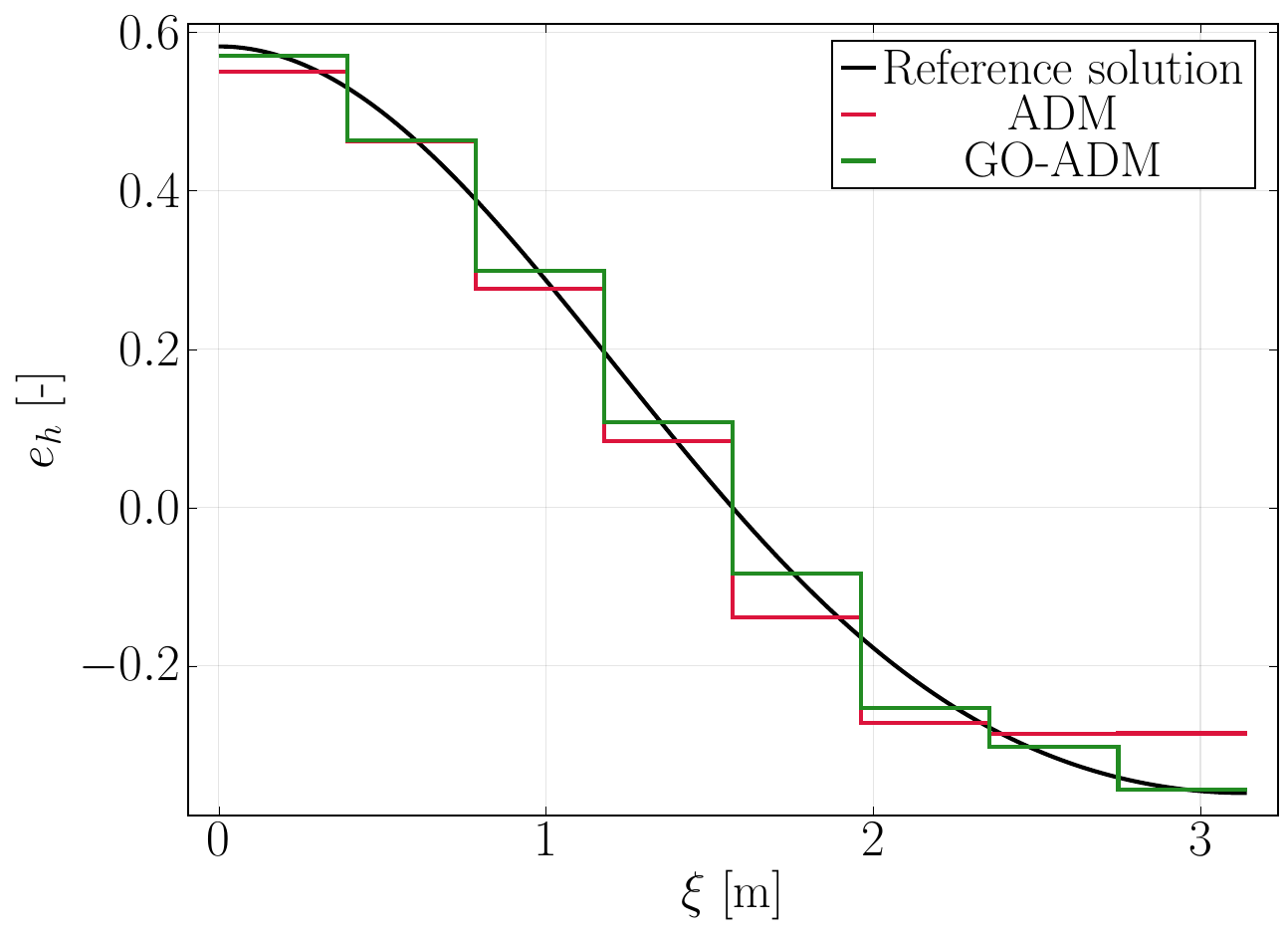}}

    \subfloat[][$\dataset$, $\alpha=0.0$]{\includegraphics[width=0.48\textwidth]{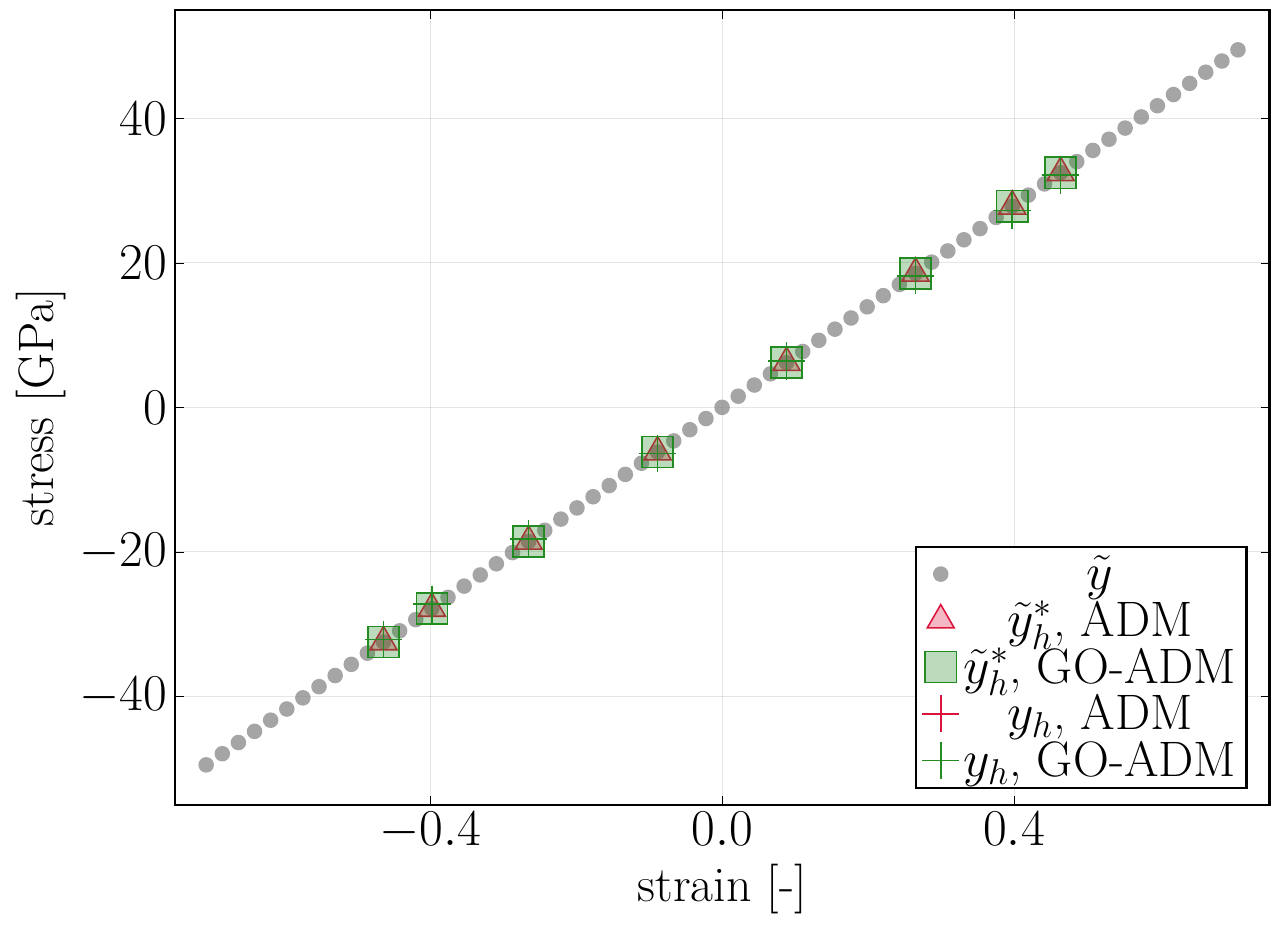}}
    \subfloat[][$\dataset$, $\alpha=1.0$]{\includegraphics[width=0.48\textwidth]{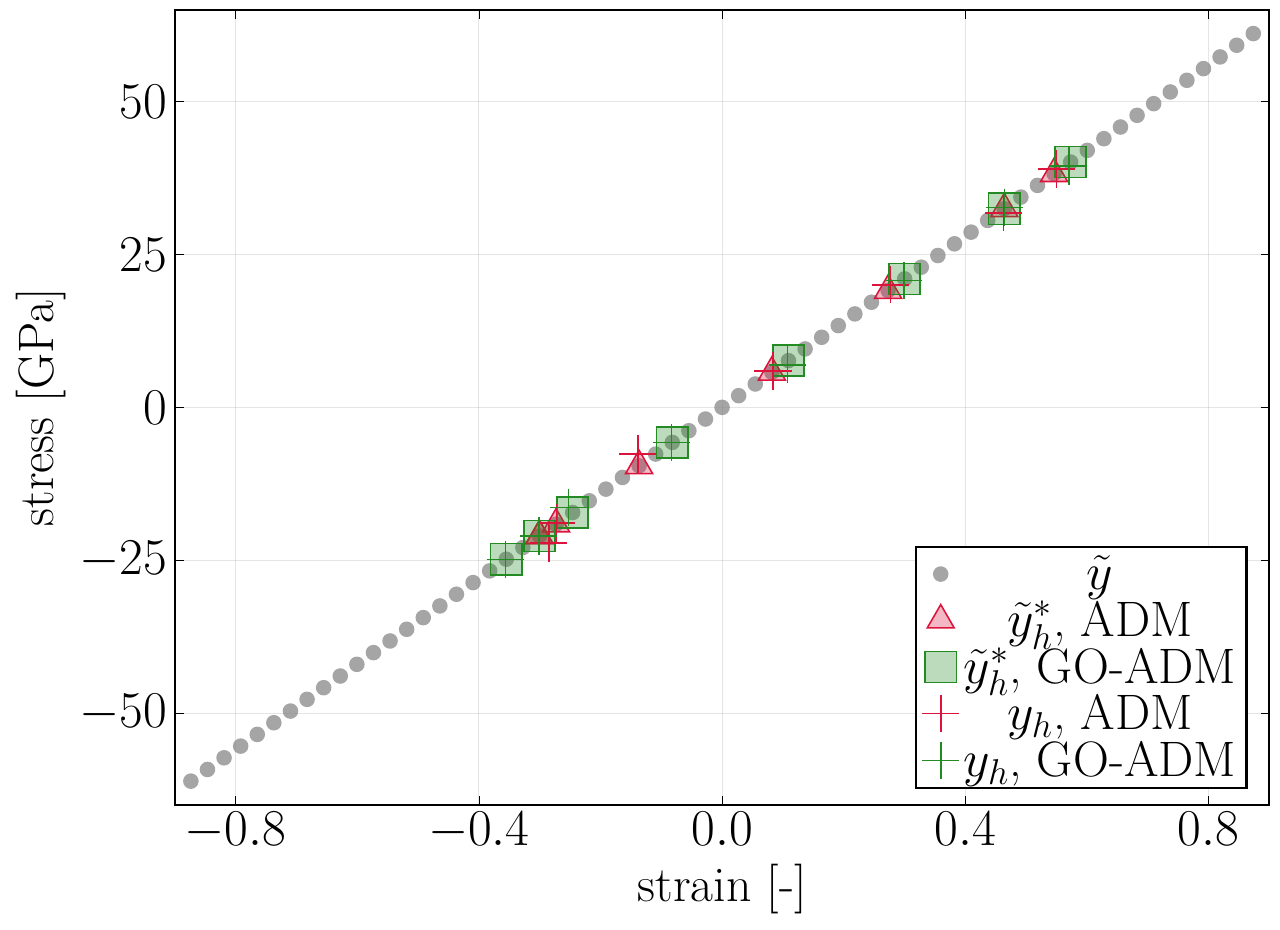}}

    \caption{Discrete axial displacements, strains, and the dataset including the solved stress-strain pairs and minimizer, computed with different solvers and strain measures, using \textbf{8 elements and 65 data points}.}\label{fig:1dbarResults}
\end{figure}

    The first benchmark is a fixed-fixed aluminum bar with the following manufactured axial displacement field:
    \begin{equation}
        \mathcolorbox{pink}{u(\arclen) = \beta \, \sin( \frac{\pi}{L_0}\, \arclen)}\,,
    \end{equation}
    that satisfies the boundary conditions. Here, 
    $\beta$ is the amplitude of $u(\arclen)$ and $L_0$ the initial length of the bar. 
    We obtain the corresponding axial external force function by inserting $u(\arclen)$ in the equilibrium equation of the underlying structure with a linear constitutive relation and the studied strain measures \eqref{eq:strainmeasures}. For the derivation of the axial force function in this case, we refer to \cite{viljar2025} and repeat in the following the expression of the resulting axial force:
    \begin{equation}
        f(\arclen) = - E u^{\prime\prime} \, \left(1 + 3 \alpha u^{\prime} + \frac{3\alpha^2}{2} u^{\prime} u^{\prime} \right) \,,
    \end{equation}
    where $E$ is the Young's modulus of the bar associated with the linear constitutive law. 
    Figure \ref{fig:1dbarGeometry}a illustrates the studied bar subjected to the axial load distributed over its length and Figure \ref{fig:1dbarGeometry}b shows the load function over the bar length for the case of linear ($\alpha=0.0$) and nonlinear ($\alpha=1.0$) strain measures. 
    The corresponding manufactured axial displacement is then the reference solution for our benchmarking. 
    To this end, we choose an initial length of $L_0=\pi$ m, a circular cross-section with a radius of $0.02$ m, a Young's modulus of $70$ GPa, and 
    a maximum axial displacement of $\beta=0.15L_0$.

    \begin{table}[htbp]
  \centering  
  \begin{tabular}{|l|c|c|}
    \hline
     & ADM-solver & GO-ADM-solver \\
    \hline
    $\alpha = 0.0$ & 0.004273811626 & 0.004273811626 \\
    $\alpha = 1.0$ & 0.038841703897 & 0.007188298872 \\
    \hline
  \end{tabular}
  \caption{Value of the global objective function obtained with the ADM- and GO-ADM-solvers for the aluminum bar illustrated in Figure \ref{fig:1dbarGeometry}a.}
  \label{tab:adm-go-adm-alpha-lines}
\end{table}

    \begin{figure}
      \centering
      \subfloat[][ADM]{\includegraphics[width=0.48\textwidth]{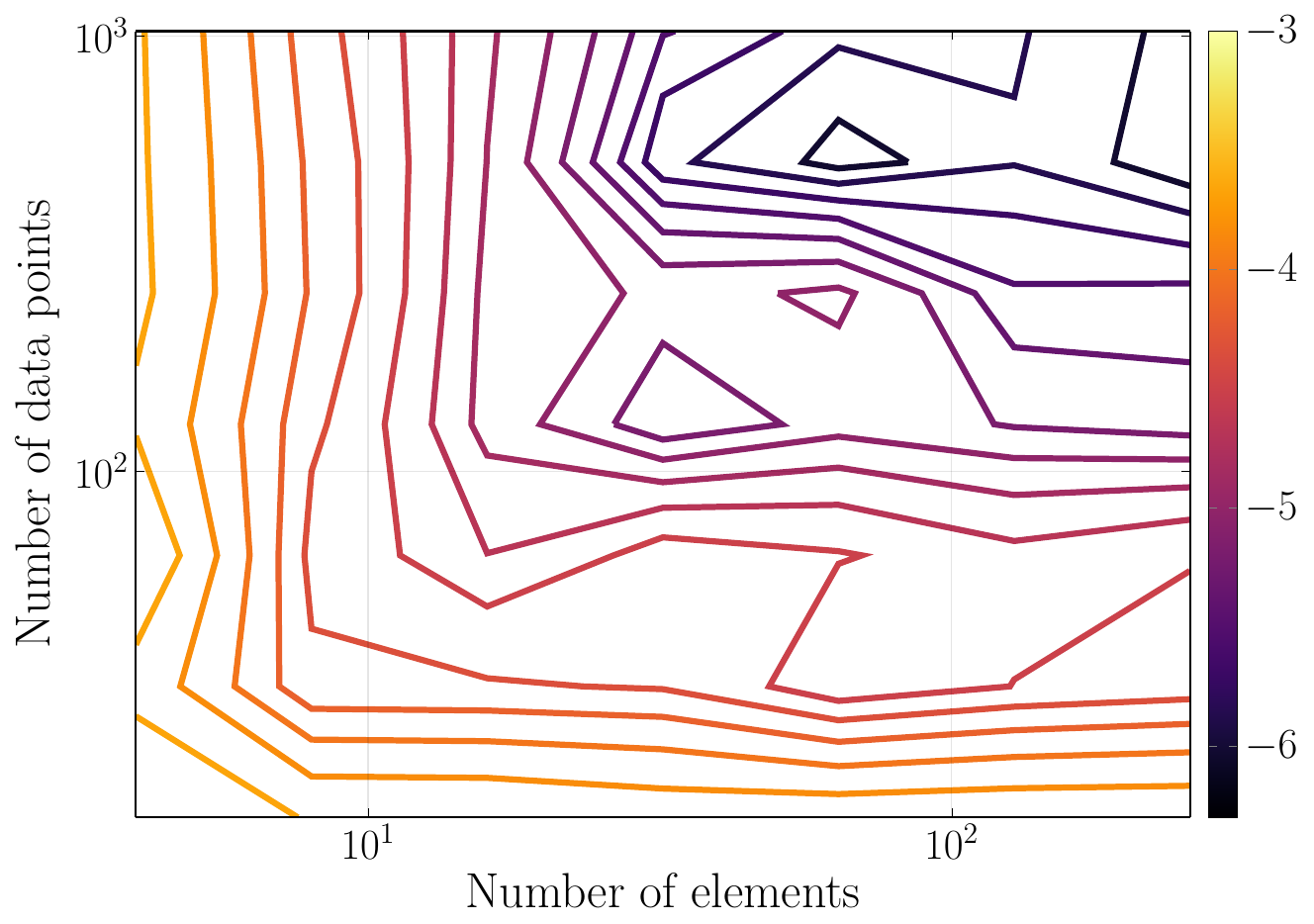}}
      \subfloat[][GO-ADM]{\includegraphics[width=0.48\textwidth]{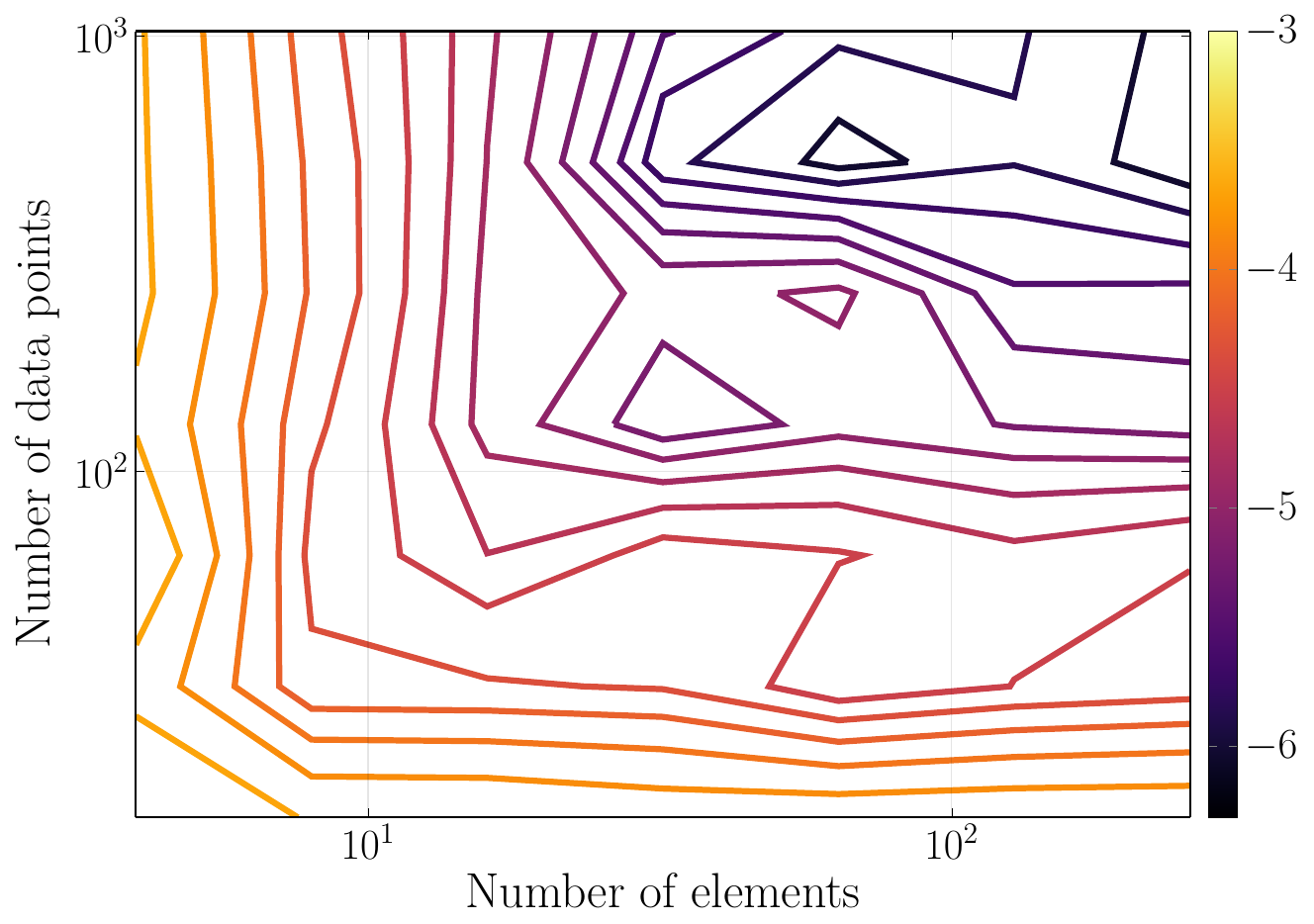}}
      \caption{Convergence of the relative $L^2$ error in the axial displacement of the bar, illustrated in Figure \ref{fig:1dbarGeometry}a, with \textbf{linear strain measures ($\alpha=0.0$)}, using different solvers. The colors correspond to $\log_{10}(\cdot)$ of the errors and the external load is applied in 1 load step.}\label{fig:1dbarConvLin}
    \end{figure}

    Figure \ref{fig:1dbarResults} illustrates the discrete axial displacement, axial strain, the computed stress-strain pairs $\yhat$, and the converged minimizer $\ystar$ of the studied bar, using the ADM- and GO-ADM-solver for both linear and nonlinear strain measures. 
    We discretize the bar with 8 elements and generate the dataset with 65 data points including the origin corresponding to the stress-free state, symmetrically distributed with respect to the origin. 
    We apply the external force in one load step in all cases. 
    For the case of linear strain measures, we observe in Figures \ref{fig:1dbarResults}a and c that both the ADM- and GO-ADM solvers lead to the same axial displacement and strains that are a very good approximation of the reference solution for the employed discretization. 
    Figure \ref{fig:1dbarResults}e shows the computed $\yhat$ and converged minimizer $\ystar$ obtained with both solvers. 
    We see that they lead to the same results and hence the same value of the global objective function, as showed in Table \ref{tab:adm-go-adm-alpha-lines}. 
    This necessarily means that both solvers lead to 
    the same optima for this linear case. 
    For the case with nonlinear strain measures, we observe in Figures \ref{fig:1dbarResults}b and d that the GO-ADM-solver leads to a better approximation of the reference axial displacement and strain than the ADM-solver. 
    These results confirm the improved approximation capability of the former and that the structure-specific initialization does not guarantee global optimality for nonlinear systems, as discussed in the previous section.  
    Table \ref{tab:adm-go-adm-alpha-lines} shows the value of the global objective function obtained with both solvers. 
    We can see that the GO-ADM-solver leads to a value of almost one order of magnitude smaller than the ADM-solver. 
    This is reflected in Figure \ref{fig:1dbarResults}f where we see that the stress-strain pairs $\yhat$ obtained with the former is closer to the corresponding converged minimizer. 
    Moreover, following the discussions in \cite{Gebhardtddcmhilbert2025}, we also check that the dataset and results are thermomechanically consistent. 
    As expected for the case of linear constitutive relation, all data points and elements are thermomechanically consistent. 
    We conclude that for the studied one-dimensional benchmark, the ADM solving strategy leads to globally optimal solution for the case of linear strain measures, however, not for nonlinear strains. 
    Using the GO-ADM solving strategy improves the solution and the value of the global objective function. 
    The expected results for this benchmark confirms our implementation for this one-dimensional structure.

    \begin{figure}
      \centering
      \subfloat[][ADM]{\includegraphics[width=0.48\textwidth]{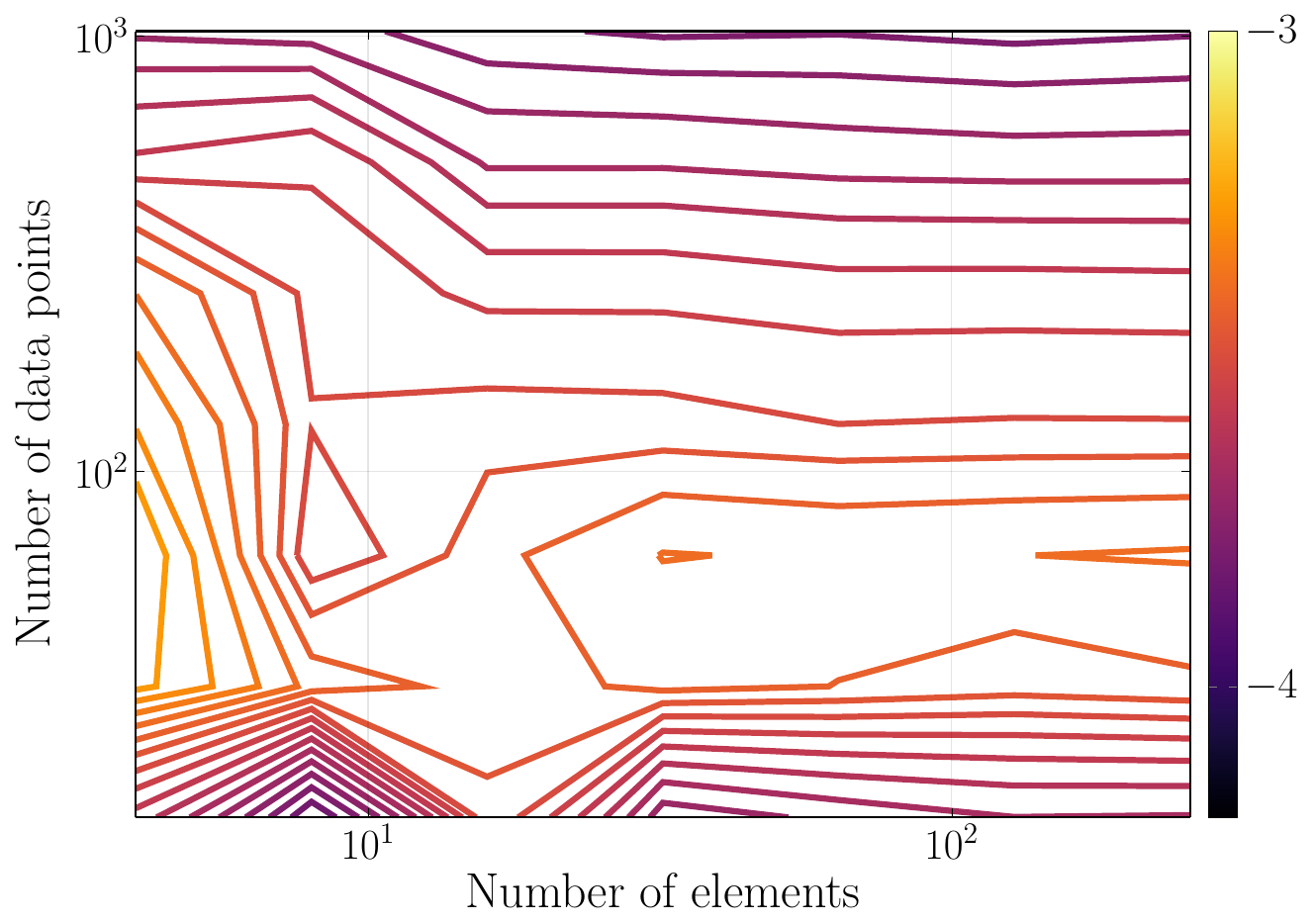}}
      \subfloat[][GO-ADM]{\includegraphics[width=0.48\textwidth]{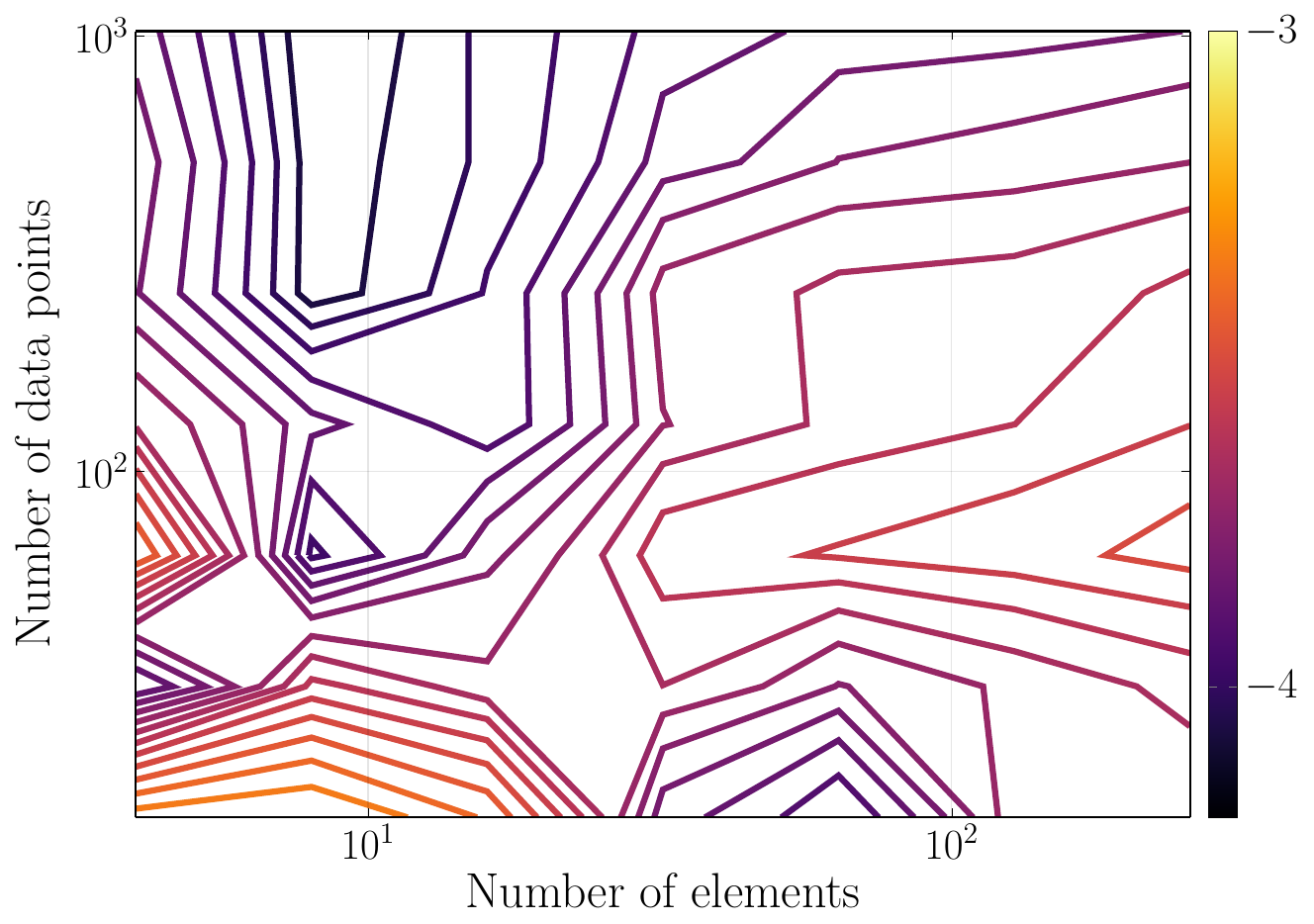}}
      \caption{Convergence of the relative $L^2$ error in the axial displacement of the bar, illustrated in Figure \ref{fig:1dbarGeometry}a, with \textbf{nonlinear strain measures ($\alpha=1.0$)}, using different solvers. The colors correspond to $\log_{10}(\cdot)$ of the errors and the external load is applied in 10 load steps.}\label{fig:1dbarConvNonlin}
    \end{figure}

    \begin{remark}
      Furthermore, we benchmark the case with $u(\arclen) = -\beta \, \sin(\frac{\pi}{L_0} \, \arclen)$, i.e. with a manufactured mirrored with respect to the $x$-axis. We obtain the same results as in Figure \ref{fig:1dbarResults} but mirrored with respect to the $x$-axis, i.e. multiplied by a factor of $-1$. 
      This further confirms the correctness of our implementation of the studied solving strategies for the one-dimensional bar example. 
    \end{remark}

    We then investigate the convergence behavior of the axial displacement obtained with the ADM and GO-ADM solving strategies. 
    Figure \ref{fig:1dbarConvLin} illustrates the convergence of the relative $L^2$-norm error of the axial displacement in the case of linear strain measures. 
    We observe that both solvers lead to the same convergence behavior and accuracy, which is consistent with the observations in the study above. 
    We see that the discrete solution converges to the reference one with increasing number of elements and/or number of data points. 
    Figure \ref{fig:1dbarConvNonlin} illustrates the convergence of the same error in the case of nonlinear strains. 
    We observe that the errors obtained with the ADM-solver converges with increasing number of data points, however, remains almost constant with increasing number of elements. 
    Using the GO-ADM-solver generally leads to lower error levels, i.e. it improves the accuracy overall. 
    However, we do not recognize a clear convergence pattern for this example. 
    For some combinations of mesh resolution and data points, for example with 100 data points and at least 50 elements, we see in Figure \ref{fig:1dbarConvNonlin}b that the relative errors slightly increase with increasing number of elements. 
    To our best knowledge, no error estimate for such nonlinear systems combined with a greedy optimization algorithm is currently available. Hence, we do not have a formal explaination for this non-monotonic convergence behavior. 
    We attribute this observed local error increase to the fact that, for some combinations of mesh resolution and data points, GO-ADM may not reach a sufficiently accurate approximation of the global optima. Therefore, it leads to larger errors in such cases than others when a better approximation is achieved.  
    Furthermore, we note that the chosen maximum number of ``greedy'' searches is the same for all computations. This number, however, is proportional to the number of elements since we count the searches for all elements together (see also Algorithm \ref{alg:goadmsolver}). 
    This necessarily means that using the same maximum number of searches with increasing number of elements might affect the performance of the GO-ADM-solver and hence its convergence behavior.

    \subsubsection{Truss structure}\label{sec:benchmarkTruss}

    \begin{figure}[h]
  \centering
  \includegraphics[width=0.45\textwidth]{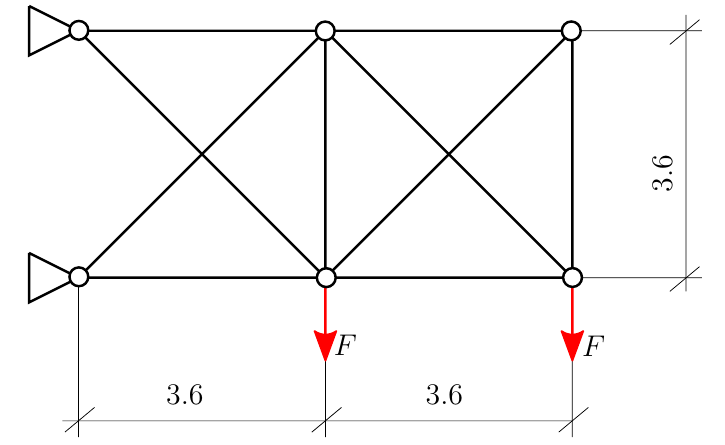}

  \caption{Sketch of a truss structure subjected to vertical nodal forces.}\label{fig:kannoTrussGeometry}
\end{figure}

\begin{figure}[h]
  \centering
  \subfloat[][Deformed configuration]{\includegraphics[width=0.48\textwidth]{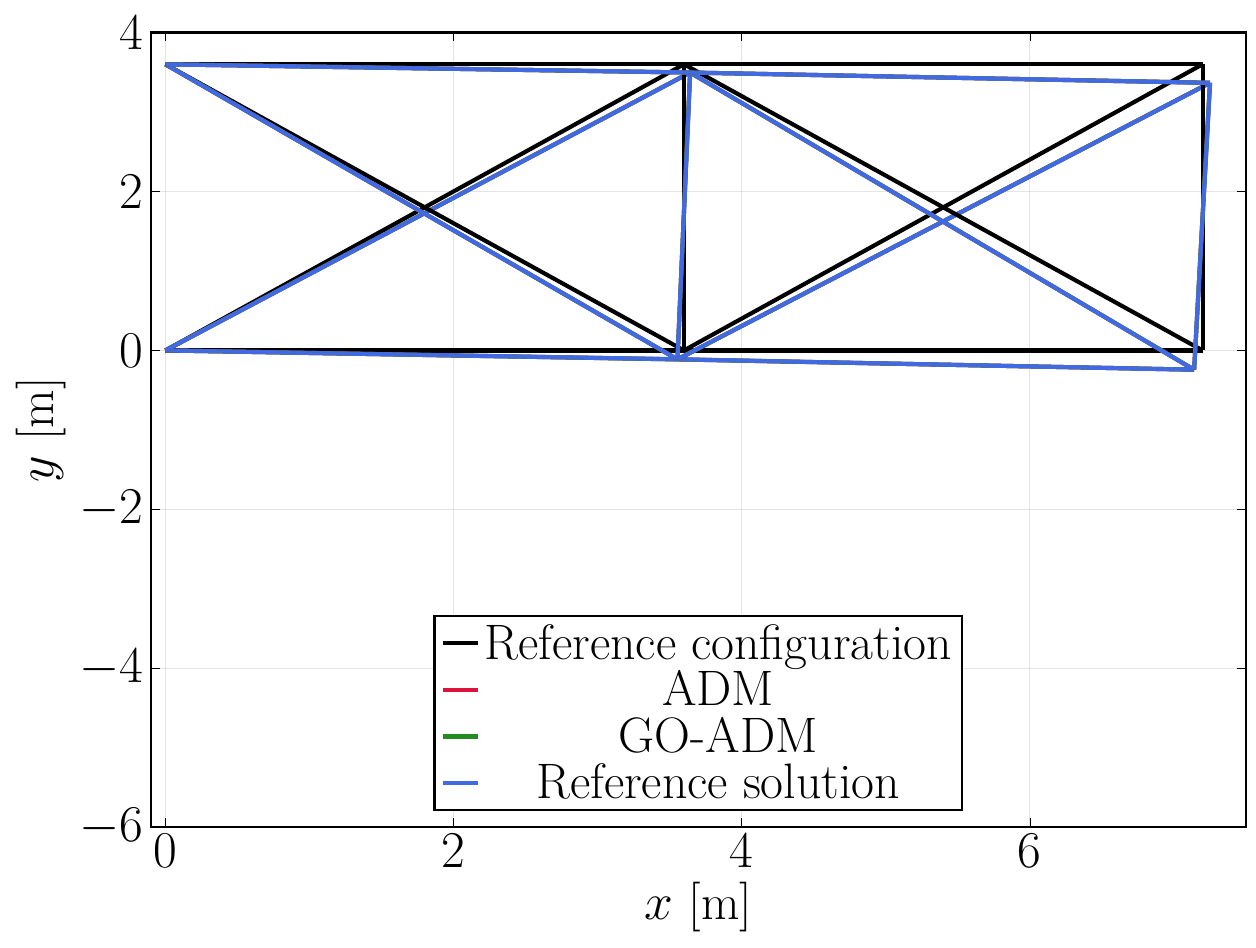}}
  \subfloat[][Axial stress]{\includegraphics[width=0.48\textwidth]{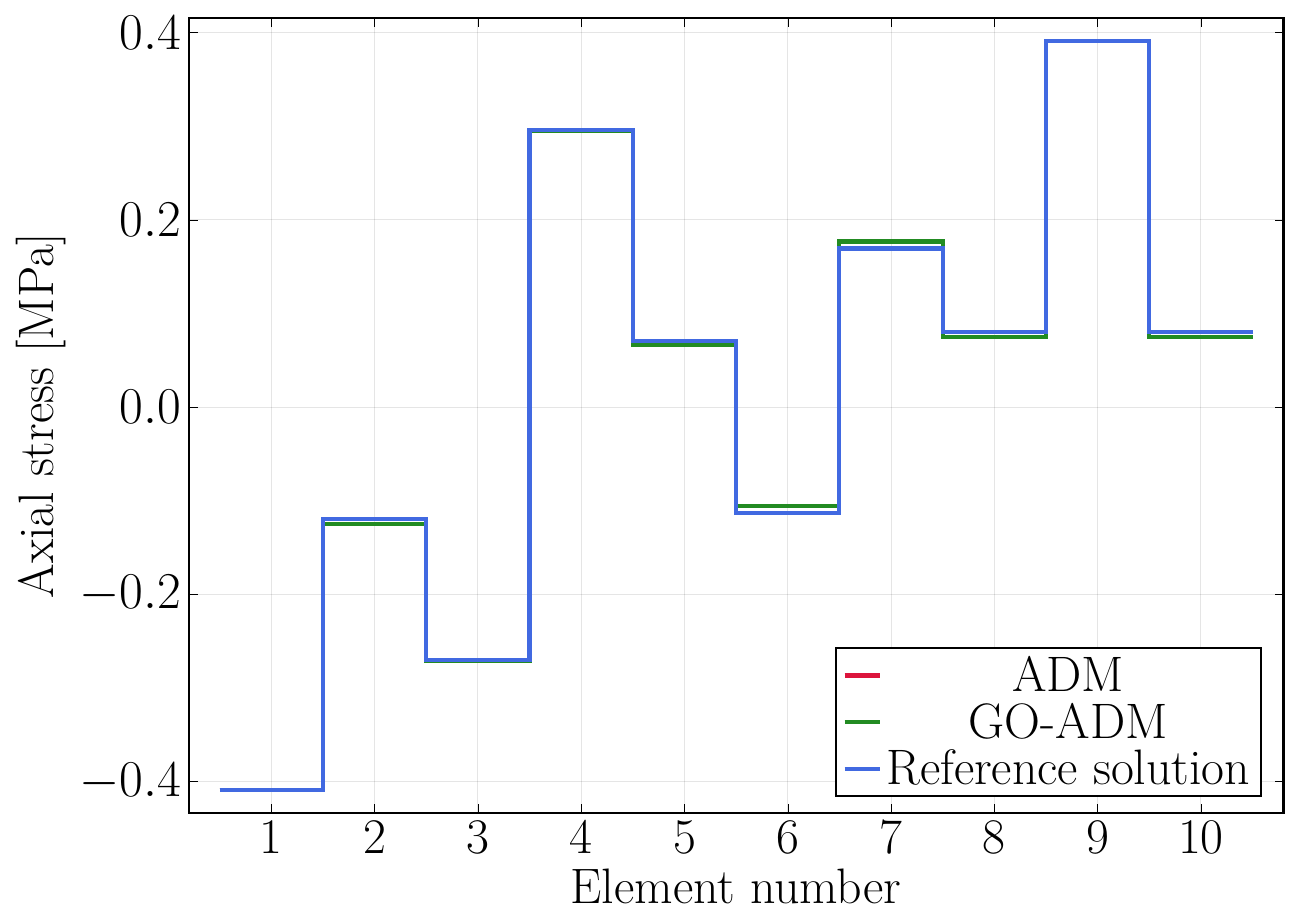}}

  \caption{Deformed configuration with displacements \textbf{scaled by a factor of 50} and axial stresses of the truss structure in Figure \ref{fig:kannoTrussGeometry}, computed with \textbf{linear strain measures} and different solving strategies.}\label{fig:kannoTrussResultsLin}
\end{figure}

\begin{figure}[h]
  \centering
  \subfloat[][Deformed configuration]{\includegraphics[width=0.48\textwidth]{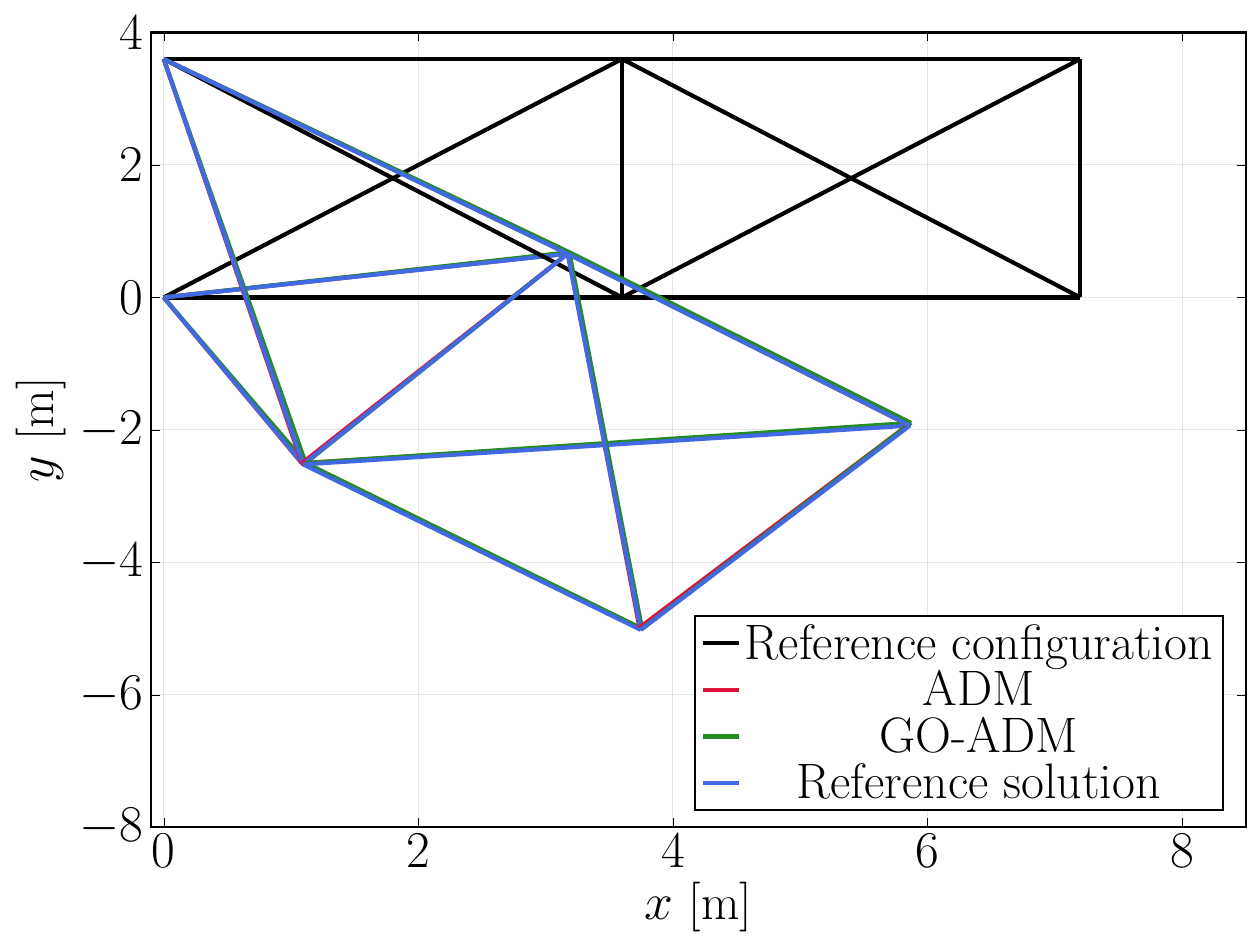}}
  \subfloat[][Axial stress]{\includegraphics[width=0.48\textwidth]{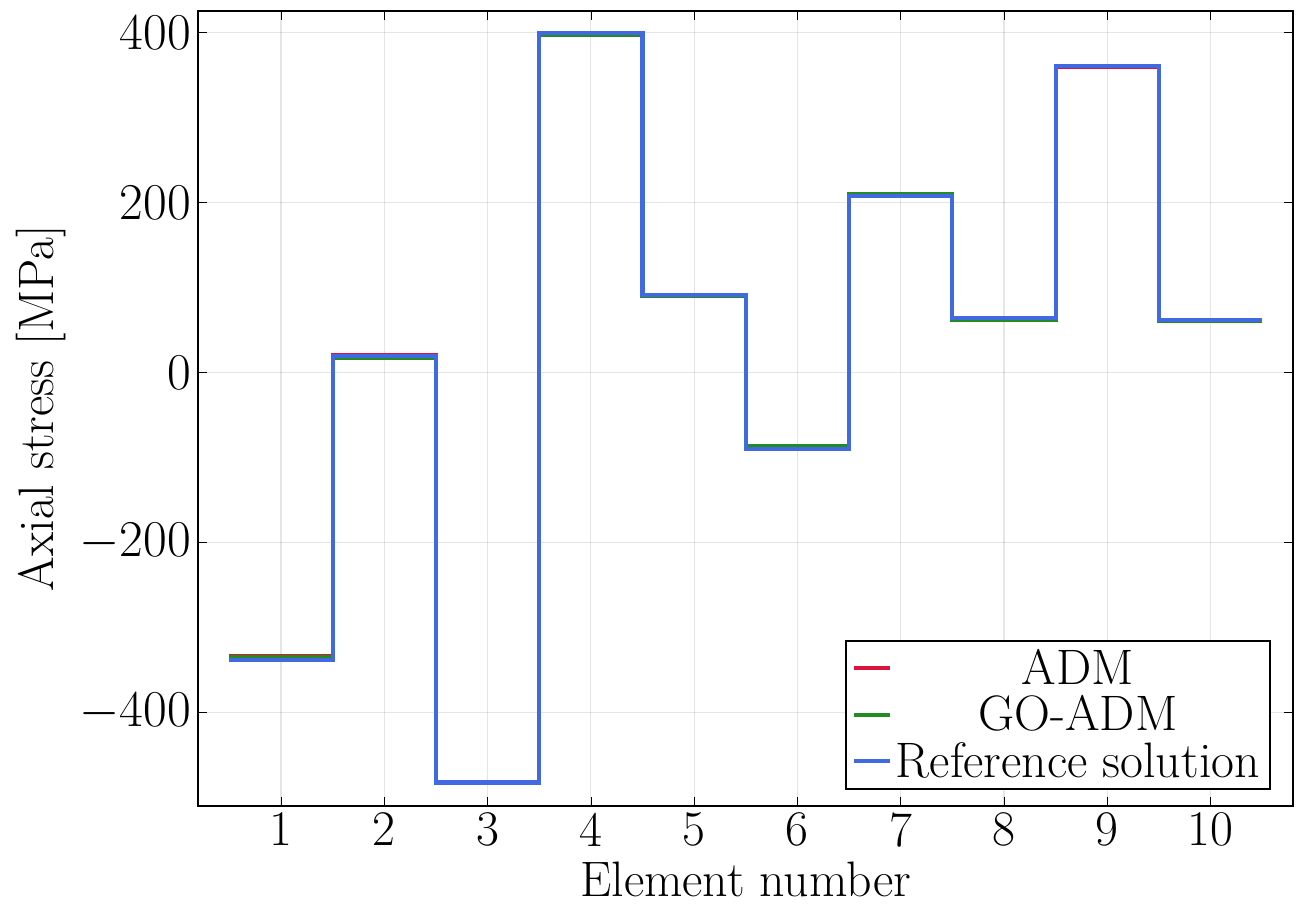}}

  \caption{Deformed configuration and axial stresses of the truss structure in Figure \ref{fig:kannoTrussGeometry}, computed with \textbf{nonlinear strain measures} and different solving strategies.}\label{fig:kannoTrussResultsNonlin}
\end{figure}

\begin{figure}[h]
  \centering
  \includegraphics[width=0.5\textwidth]{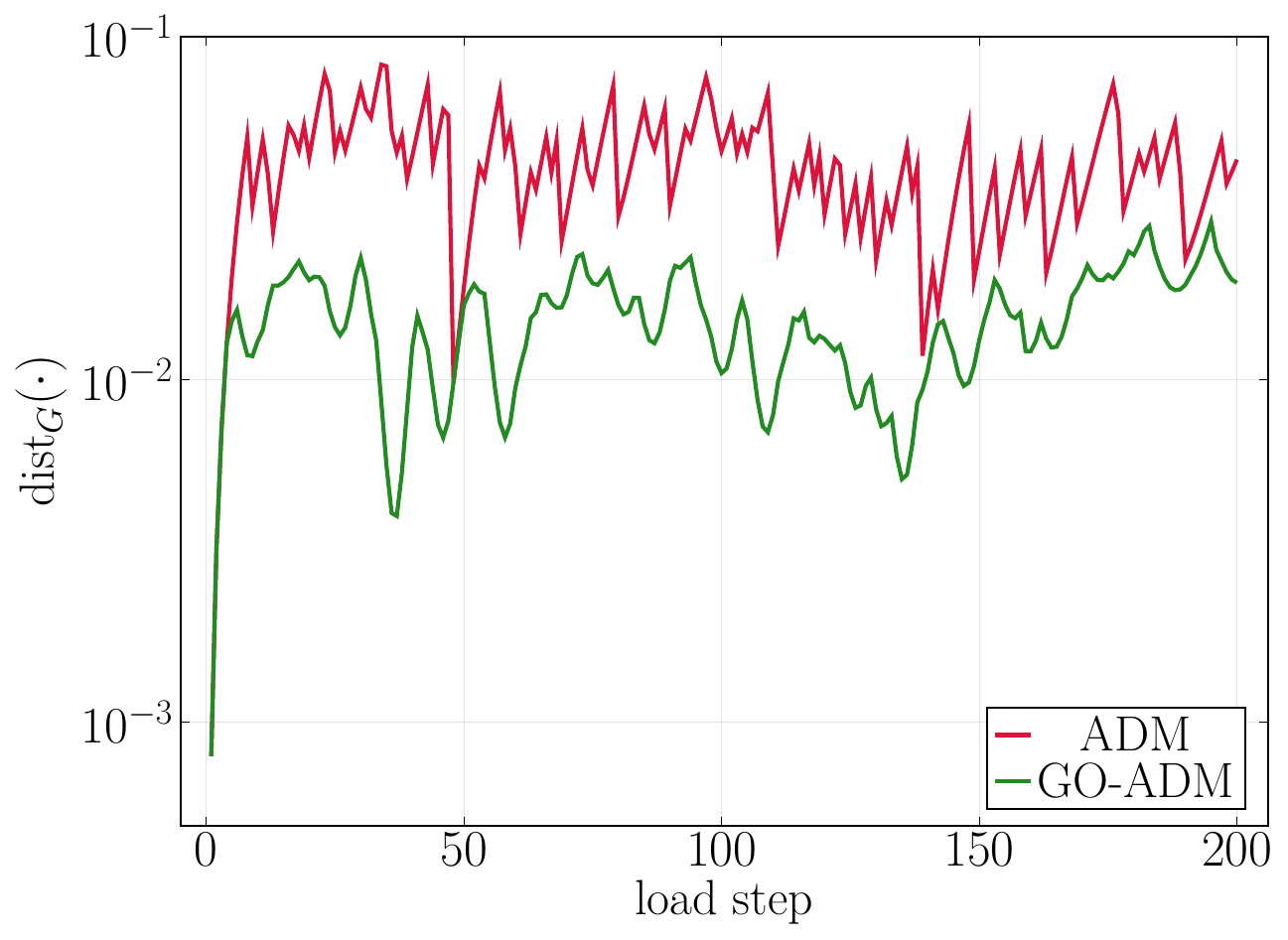}

  \caption{Value of the global objective function over load steps of the truss structure in Figure \ref{fig:kannoTrussGeometry}, computed with \textbf{nonlinear strain measures} and different solving strategies.}\label{fig:kannoTrussResultsNonlinCostFunc}
\end{figure}

\begin{figure}[h]
  \centering
  \subfloat[][Deformed configuration]{\includegraphics[width=0.48\textwidth]{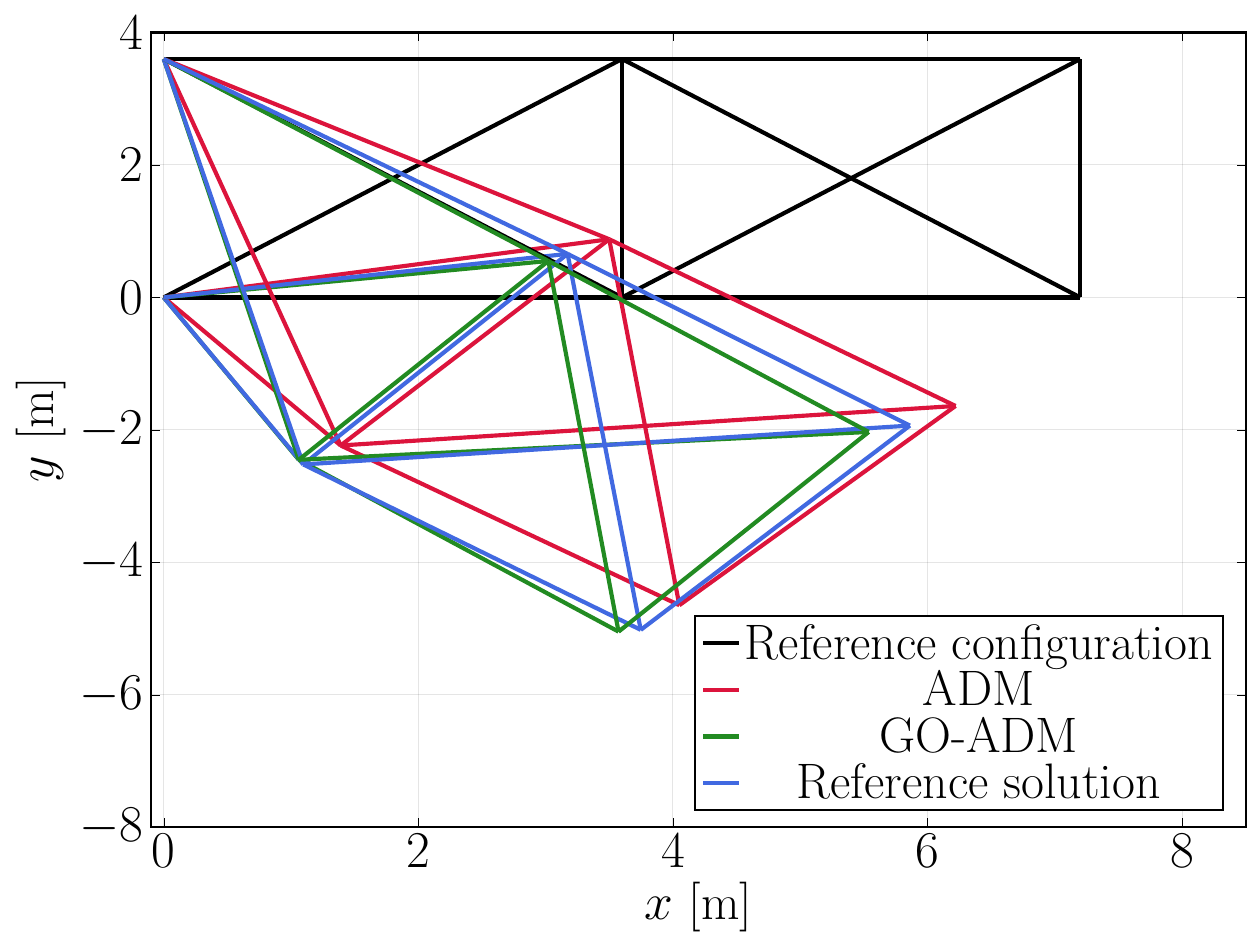}}
  \subfloat[][Axial stress]{\includegraphics[width=0.48\textwidth]{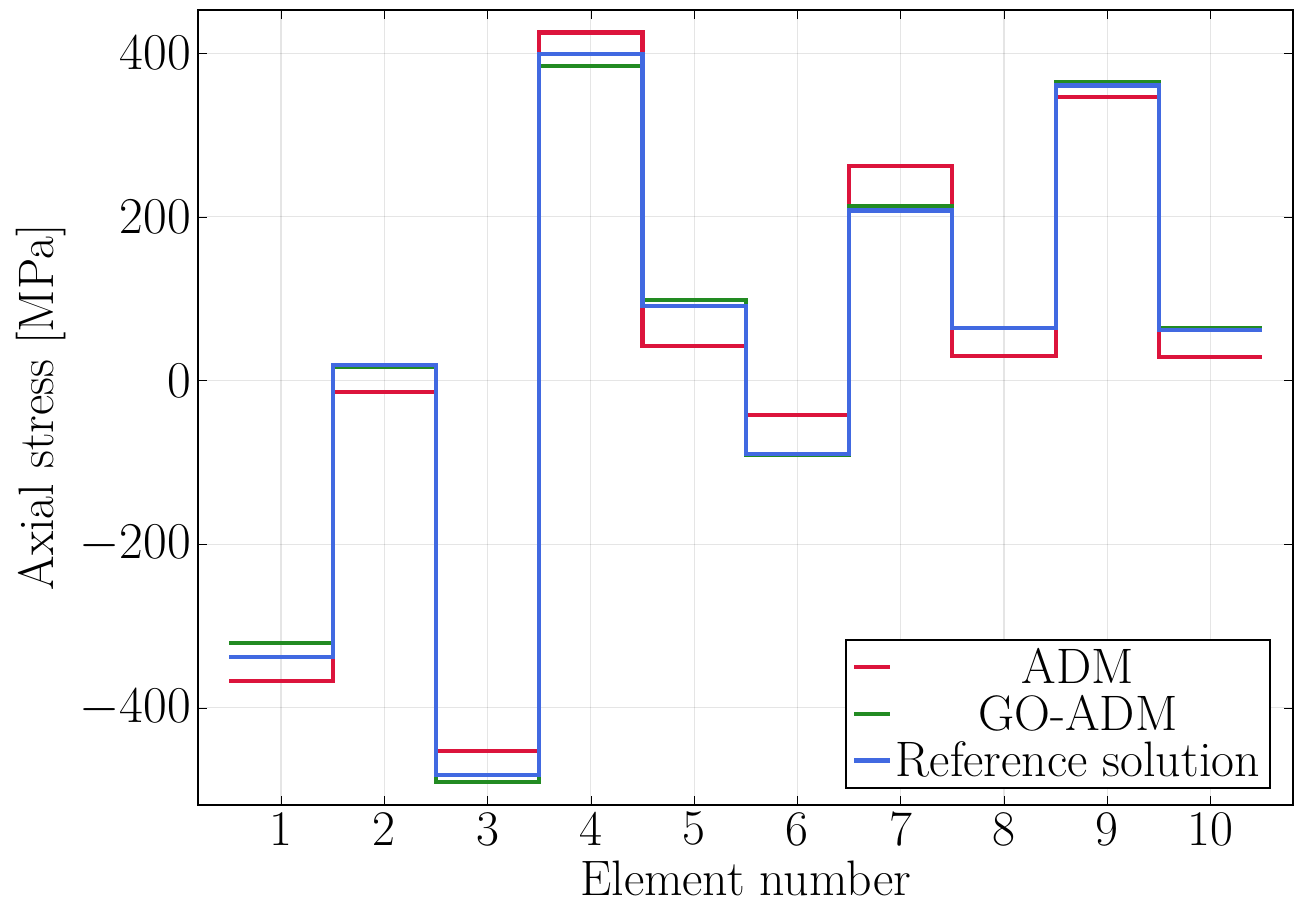}}

  \caption{Deformed configuration and axial stresses of the truss structure in Figure \ref{fig:kannoTrussGeometry}, computed with \textbf{nonlinear strain measures} and different solving strategies, \textbf{using the same number of data points but ten times larger limits}.}\label{fig:kannoTrussResultsNonlinb}
\end{figure}
    
    We consider now a second benchmark of a two-dimensional truss structure illustrated in Figure \ref{fig:kannoTrussGeometry}. 
    This has been employed as a test problem for data-driven solvers, e.g. in \cite{Gebhardtddcmsolution2025}, and also for the MINLP reformulation in \cite{kanno_data_driven_2019}. 
    The cross-sectional area of each truss element is $0.002$ m$^2$ and the structure is subjected to downward nodal forces $F = 400\gamma$ N, where $\gamma$ is an arbitrary load factor (see Figure \ref{fig:kannoTrussGeometry}). 
    We first consider a case with linear strain measures and small deformations. To this end, we choose a load factor of $\gamma =1.0$ and apply the forces in one load step. 
    We obtain the reference solution for this example using an online truss calculator\footnote{Link: \url{https://valdivia.staff.jade-hs.de/fachwerk_en.html}} and generate the dataset with 65 data points including the origin corresponding to the stress-free state. 
    Figure \ref{fig:kannoTrussResultsLin} illustrates the deformed configuration and the axial stresses computed with both the ADM- and the GO-ADM-solvers. 
    For visualization purposes, we plot the former with discrete displacements scaled by a factor of 50. 
    We observe that both solvers lead to the same results that overlap with the reference results. 
    Both solvers require only one ADM iteration and lead to the same value of the global objective function of $1.654 \cdot 10^{-7}$. 
    These results are as expected for this linear system with the structure-specific initialization, as for the previous example of the one-dimensional aluminum bar.

    We then consider the second case with nonlinear strain measures and large deformations. 
    To this end, we choose a load factor of $\gamma =1500$ and apply the forces in 200 steps to obtain large deformations, such that the nonlinear strain contribution becomes sufficiently severe (see also the quadratic term in the strain measures \eqref{eq:strainmeasures}). 
    We compute the reference solution with a standard structural analysis problem using the same nonlinear measures, linear constitutive law that is employed to generate the dataset, and the same discretization (see also Section \ref{sec:discretization}). 
    We note that this is equivalent to an approximate nonlinear optimization problem where the constitutive relation is approximated with a linear function. 
    For more details and discussions on approximate nonlinear optimization problems, we refer to \cite{Gebhardtddcmstatic2020,Gebhardtddcmcontacts2024}. 
    Figure \ref{fig:kannoTrussResultsNonlin} illustrates the deformed configuration and the axial stresses computed with both the ADM- and the GO-ADM-solvers for the studied truss structure. 
    We observe that both solvers approximately lead to the same displacement and stresses that overlap with the reference solution. 
    Based on our numerical experiments, we obtained similar overlapping results with the ADM- and GO-ADM solvers for higher load factors. 
    This necessarily means that for this example, the ADM solver together with the structure-specific initialization approach achieves a very good approximation of the global optima for both small and large deformations. 
    Nevertheless, using the GO-ADM-solver reduces the value of the objective function at all load steps, as illustrated in Figure \ref{fig:kannoTrussResultsNonlinCostFunc}. 
    Furthermore, we investigate the accuracy obtained with the ADM- and GO-ADM solvers when increasing the stress and strain limits of the dataset. 
    We plot again the deformed configuration and the axial stresses for the same load factor and number of data points but ten times the data limits in Figure \ref{fig:kannoTrussResultsNonlinb}. 
    We observe that the ADM solver leads to results (red curves) differing from the reference solution (blue curves), i.e. it does not achieve global optima in this case. 
    This is due to the elementwise search and update of the closest data point, as discussed in Section \ref{sec:goadmalgorithm}. 
    Using our GO-ADM solver achieves a better approximation (green curves) of the global optima, as expected. 
    For the studied truss structure, we also check that the results and dataset are thermomechanically consistent for all cases, as discussed in \cite{Gebhardtddcmhilbert2025}. 
    The presented results confirm that our implementation for this two-dimensional structure.

  \subsection{Applications with real experimental data - Cyclic test of a nylon rope}

\begin{figure}
  \centering
  \subfloat[][Cyclic loads, repeated until $t=55$ min]{\includegraphics[width=0.48\textwidth]{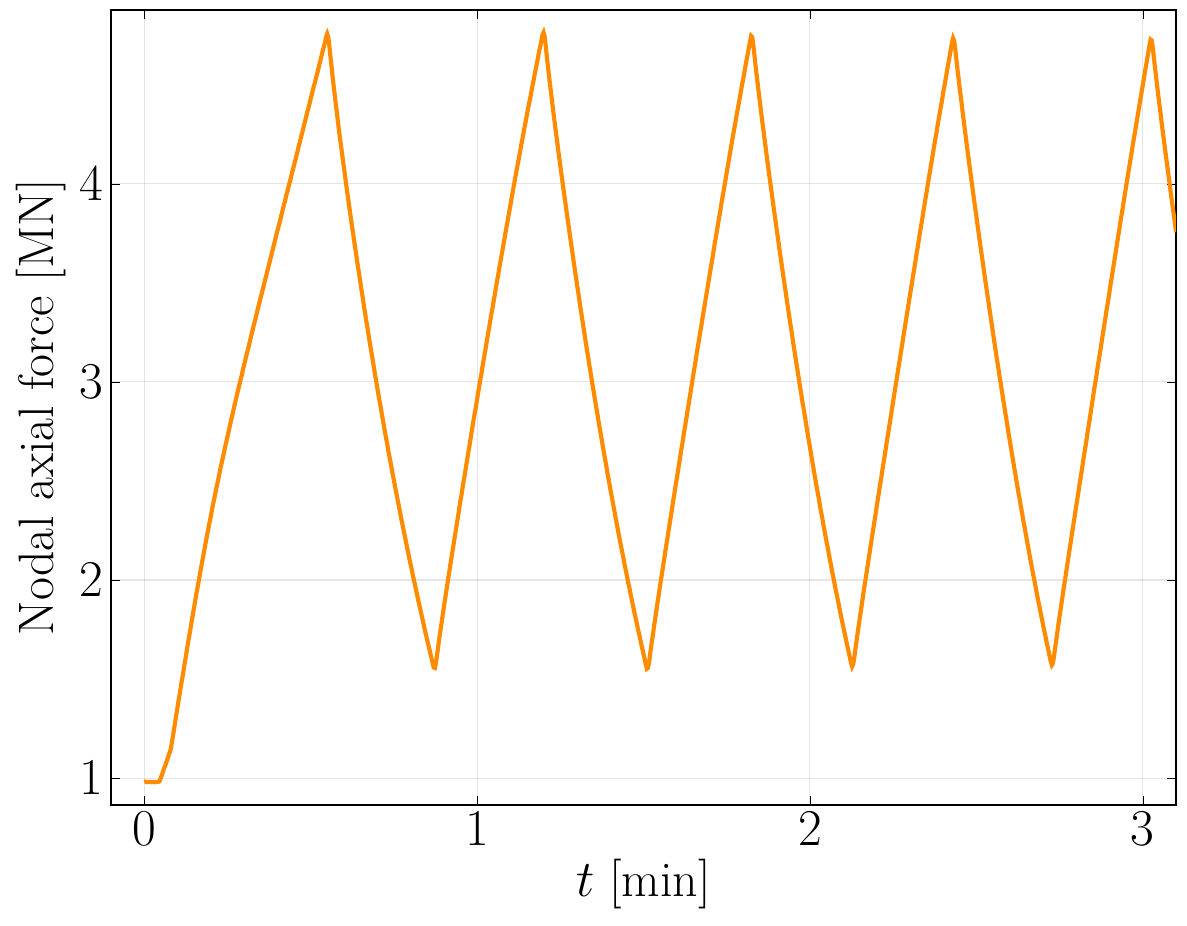}}
  \subfloat[][Provided dataset]{\includegraphics[width=0.51\textwidth]{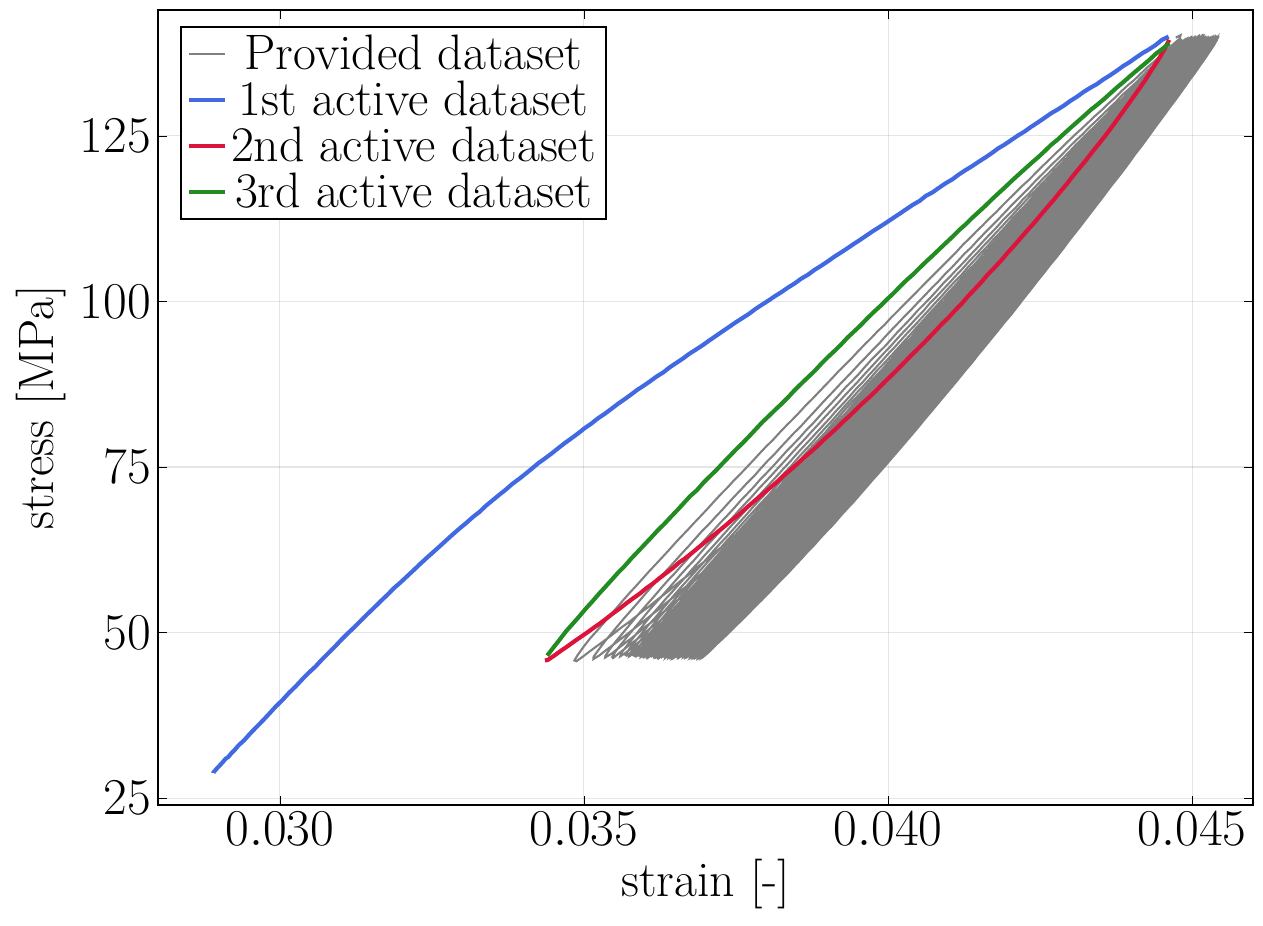}}

  \caption{The provided loading and corresponding constitutive dataset of a cyclic test for a nylon rope. The data consists of more than 16,500 discrete points.}\label{fig:cyclic1DnonlinDataAll}
\end{figure}

  We now apply our solving strategy to real experimental dataset 
  of a cyclic testing of a nylon rope that is commonly employed as mooring lines for offshore structures and tested in industrial testing facilities. 
  For more details on the setup for such industrial tests, we refer to Section 2.3.1 of \cite{tom2025}. 
  With provided loading steps and measured strains, we compute the load-deflection curve consisting of expected hysteresis loop(s). 
  We obtain the loading and strain dataset from an anonymous industry partner for a tested nylon rope of an initial length of $17.010$ m and a cross-sectional diameter of $0.208$ m. 
  During the test, the nylon rope is subjected to a cyclic tensional loading, showed in Figure \ref{fig:cyclic1DnonlinDataAll}a, and its stretched length is measured. 
  The force and measured axial strain at each load step are provided from five repeated tests of the same rope over almost one hour for each test. 
  We obtain the corresponding stress via dividing the applied force by the cross-sectional area. 
  Figure \ref{fig:cyclic1DnonlinDataAll}b illustrates the obtained stresses corresponding to provided strain data from one of the provided tests.

  We note that the provided data consists of more than $16,500$ discrete points of a test over 55 minutes. 
  Here, for visualization purposes, we plot this as a solid curve and illustrate only the first 3 minutes of the cyclic loading (Figure \ref{fig:cyclic1DnonlinDataAll}a), which repeat until $t=55$ minutes. 
  In this work, since we aim at simulating the hysteresis loop(s) in the load-deflection curve for the cyclic testing, we consider only the first cycle of one of the provided tests, which we highlight with different colors in Figure \ref{fig:cyclic1DnonlinDataAll}b. 
  We note that the measurement begins with non-zero load and hence the dataset does not include the origin corresponding to the stress-free state, which can be artificially added. 
  Following the discussions in \cite{Gebhardtddcmhilbert2025}, we check that the provided dataset is thermomechanically consistent, i.e. the dataset obeys fundamental laws of physics, particularly thermodynamics. 
  For our computations, we divide the original dataset in three sub-datasets, following the first loading, first unloading, and second loading path, which is illustrated with blue, red, and green colors in Figure \ref{fig:cyclic1DnonlinDataAll}b, respectively. 
  This necessarily means that we consider relevant, or rather active, dataset for each phase of the test in separated computations. 
  We note that in realistic scenarios, the splitting of experimental data is not always known since experiments do not always follow paths. 
  To preserve the sequential effect between the sub-datasets, from the second sub-dataset, we start the computation using the sub-dataset with an initial state that is the solution of the last computation using the previous sub-dataset. 
  We are aware that this division is a manual improvement of the dataset quality overall, nevertheless, is a common choice in computational practice to reduce the total number of data points. 

\begin{figure}
  \centering
  \includegraphics[width=0.44\textwidth]{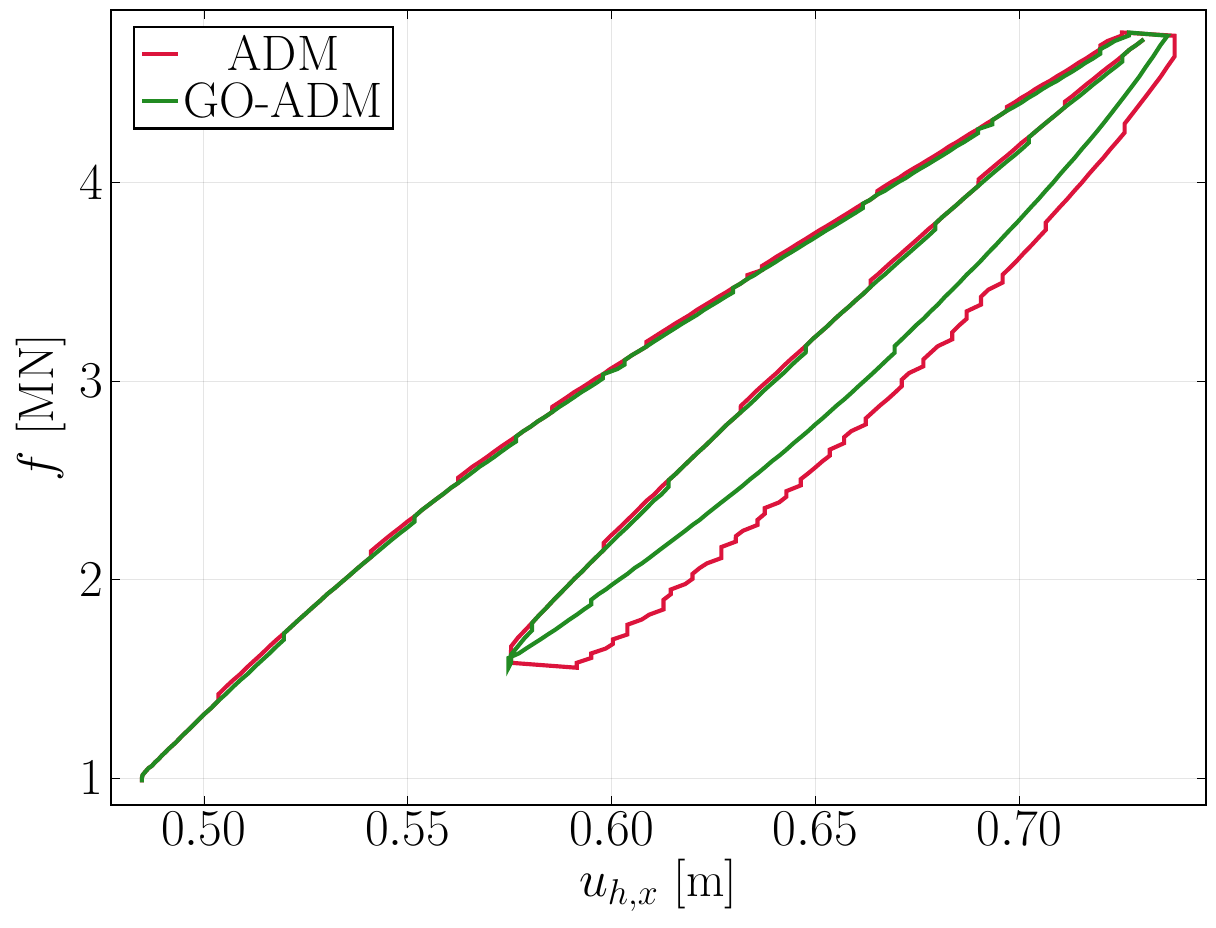}

  \caption{The load-deflection curve of a nylon rope under a cyclic testing. The results are computed with different solvers and \textbf{nonlinear strain measures}.}\label{fig:cyclic1DnonlinLoadDeflecCurv}

\vspace{0.3cm}

  \centering
  \subfloat[][Dataset no. 1]{\includegraphics[width=0.44\textwidth]{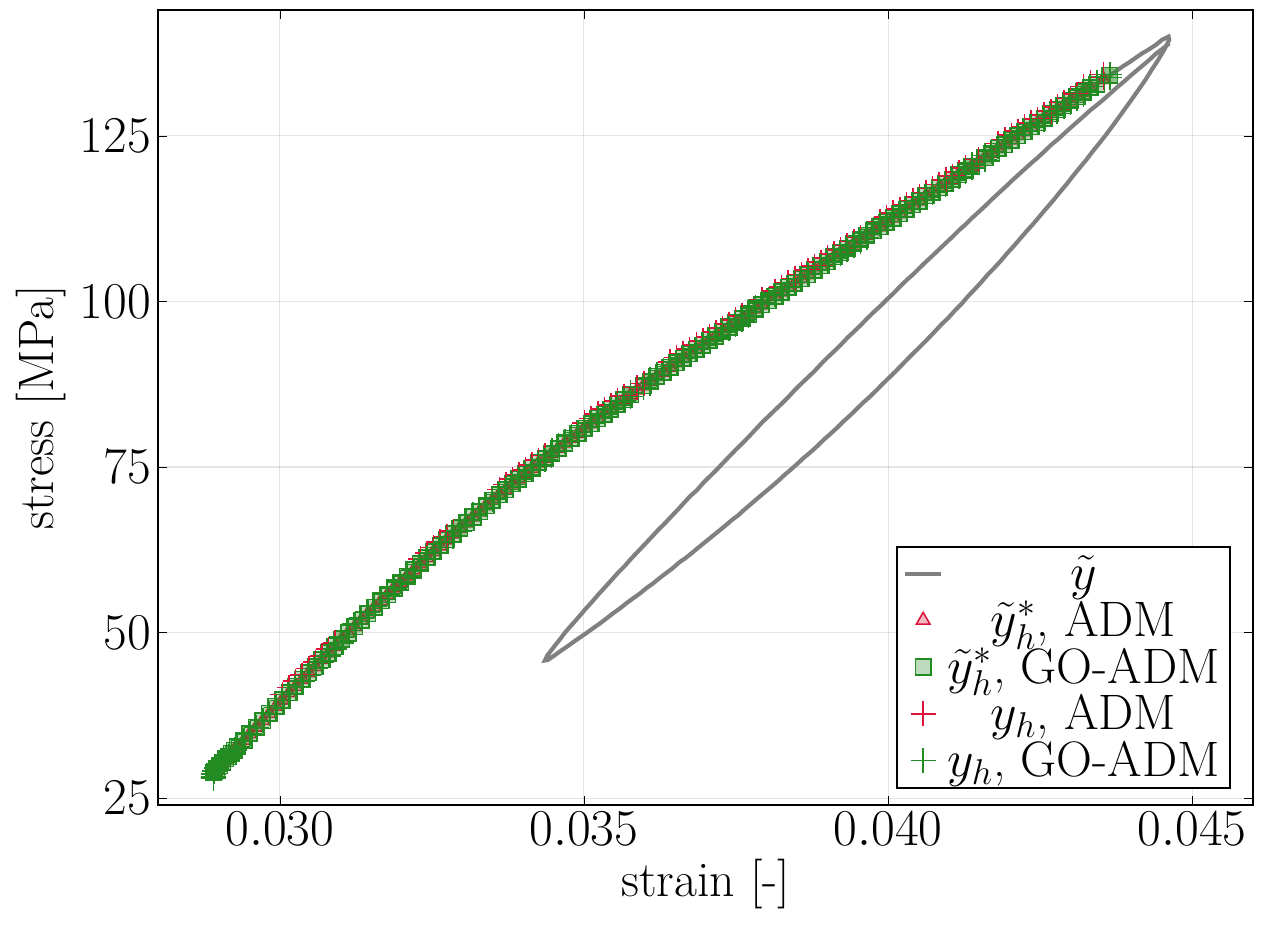}}
  \subfloat[][Dataset no. 2]{\includegraphics[width=0.44\textwidth]{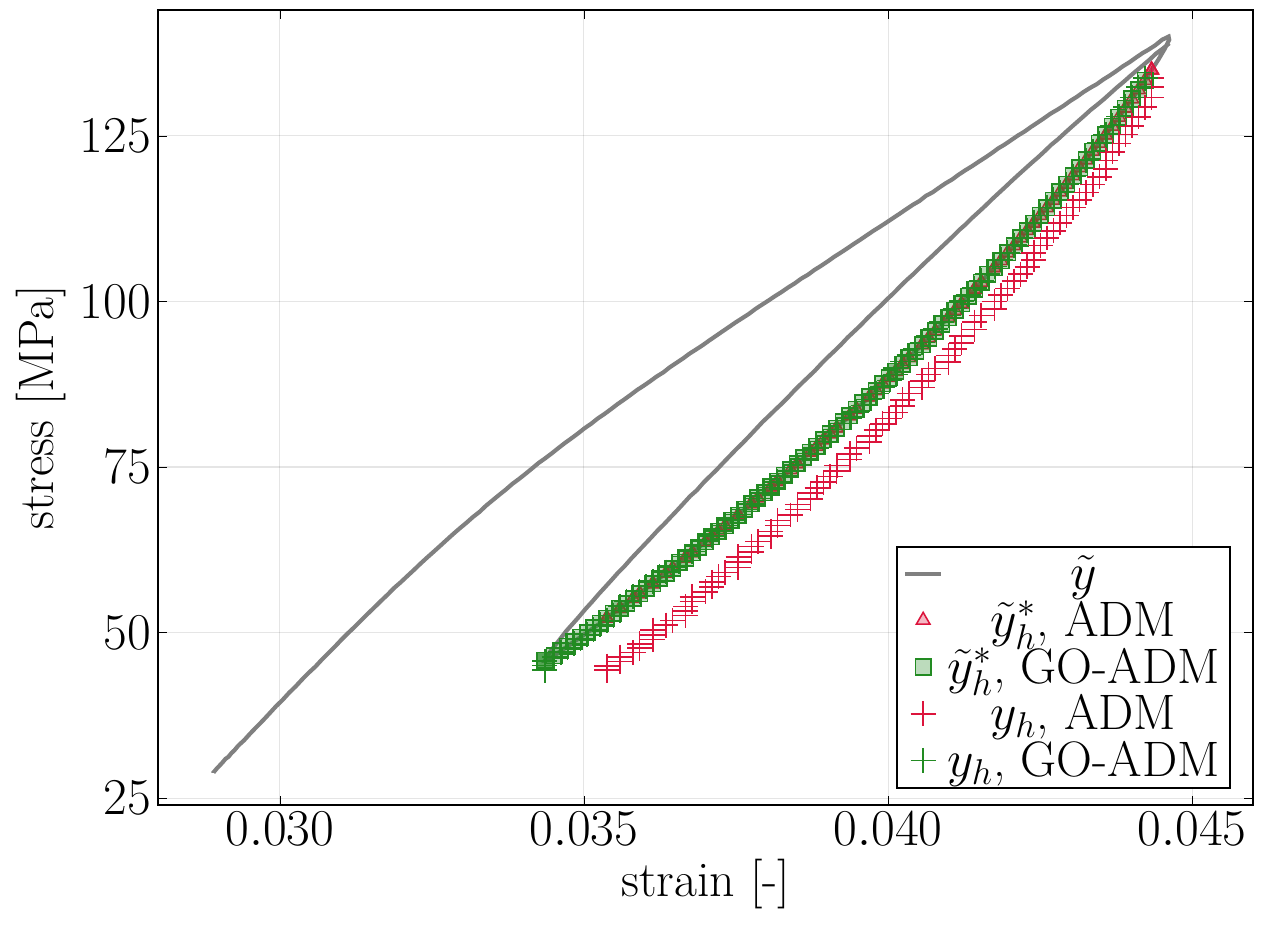}}

  \subfloat[][Dataset no. 3]{\includegraphics[width=0.44\textwidth]{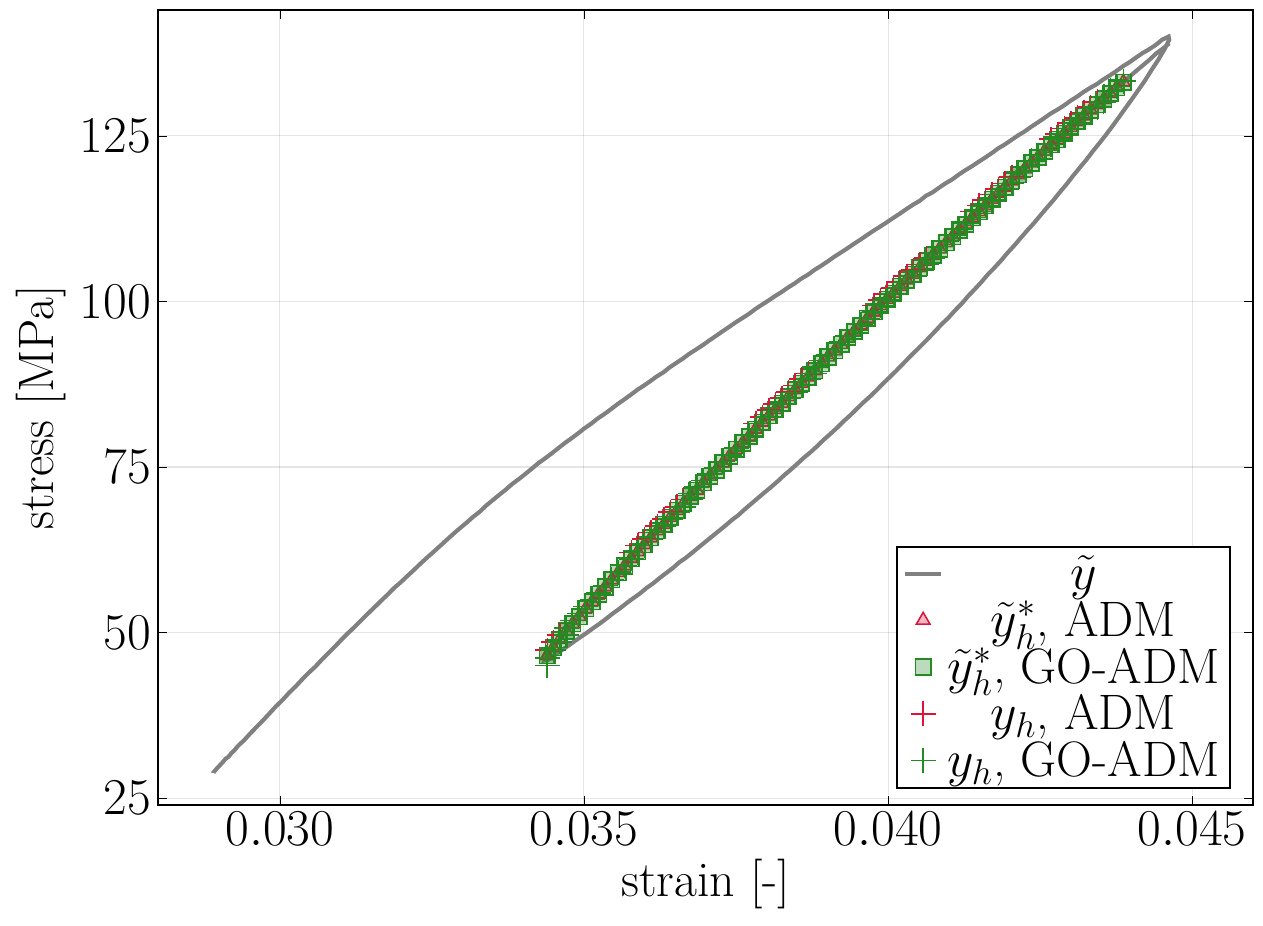}}
  \subfloat[][dist$_G(\cdot)$]{\includegraphics[width=0.44\textwidth]{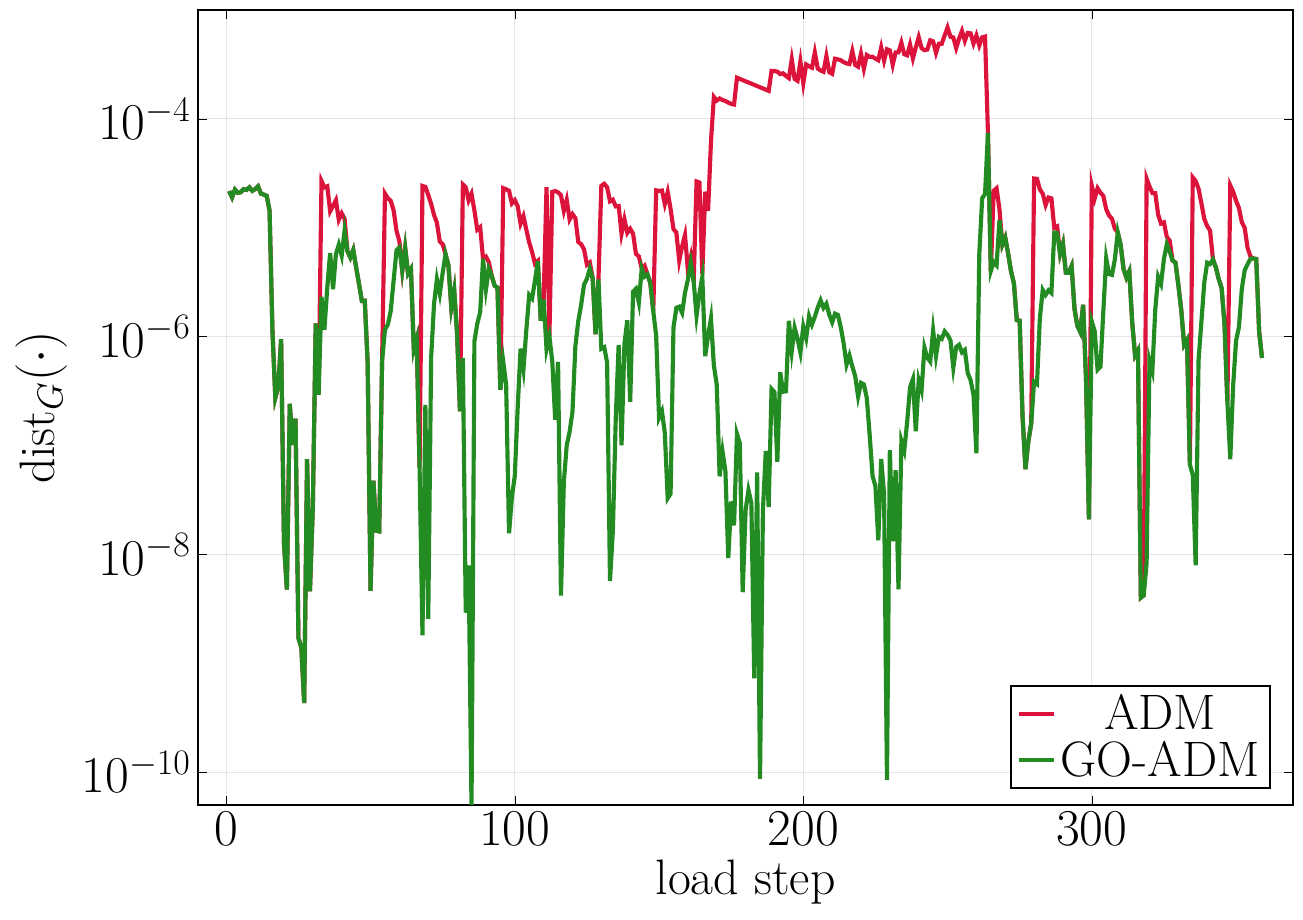}}

  \caption{Datasets including the computed stress-strain pairs and converged minimizer (a-c) and the value of the global objective function at each load step of a nylon rope under a cyclic testing. The results are computed with different solvers and \textbf{nonlinear strain measures}.}\label{fig:cyclic1DnonlinDataNcost}
\end{figure}

  \begin{remark}
    We note that splitting the datasets and computations is one approach to deal with history-dependent problems such as the cyclic tests studied in this work, given that the history of the loading and data generation is known. 
    In general, such problems, particularly when dealing with history-dependent materials, require advanced approaches to capture this dependency during data searching procedure, for instance, introduced in \cite{Ladeveze2019,Bartel2023}. 
    Integration of such approaches to our GO-ADM solving strategy is out of scope of this work and is considered for future work.
  \end{remark}

  We model the one-dimensional tested nylon rope as a one-dimensional bar structure that is constrained on the left boundary and subjected to an axial nodal force at the right boundary. 
  We apply the same force values and steps as the provided loading data. 
  For our computations, we employ the nonlinear strain measures, merely for the sake of generality, and discretize the bar with 16 elements which lead to an essentially good discretization based on numerical experiments. 
  Since the nylon rope exhibits small deformations under the cyclic testing, the quadratic term of the strain measures \eqref{eq:strainmeasures} is relatively insignificant compared to the linear one. 
  Figure \ref{fig:cyclic1DnonlinLoadDeflecCurv} illustrates the load-deflection curve obtained with the ADM- and GO-ADM-solvers for the first loading cycle.  
  We observe that using any of these two solvers we can simulate the hysteresis loop in the load-deflection curve.   
  We also see that both solvers approximately lead to the same curve following the first and second loading paths. 
  Focusing on the unloading path, using the ADM-solver leads to larger axial displacement at the same load step than the GO-ADM-solver. 
  To gain better insights, we illustrate the computed stress-strain pairs and minimizers corresponding to each loading phase in Figures \ref{fig:cyclic1DnonlinDataNcost}a-c. Here, we include the provided dataset (solid curve) as a reference solution. 
  We observe that both solvers achieve a very good match following the first and second loading paths. 
  Following the unloading path, the ADM-solver, however, leads to slightly larger axial strains than the GO-ADM-solver and also the measured values. 
  This is consistent with the observations in the load-deflection curve discussed earlier. 
  This necessarily means that using the GO-ADM-solver improves the results and achieves a better approximation of the optima, which is also reflected in the value of the global objective function illustrated in 
  Figure \ref{fig:cyclic1DnonlinDataNcost}d. 
  We observe that the objective function obtained with the ADM solving strategy shows a jump to larger values during the unloading path. 
  Using our GO-ADM solving strategy generally reduces the values of the objective function, particularly during this unloading phase. 
  Moreover, following the discussions in \cite{Gebhardtddcmhilbert2025}, we also check that the obtained results are thermomechanically consistent for all cases.

  \subsection{Global optimality and robustness}

  We now numerically illustrate favorable properties with respect to the global optimality and robustness of our solving strategy in the case of nonlinear constitutive relation. 
  In particular, we show via two-dimensional truss structures that our solving strategy achieves a better approximation of the globally optimal solution, compared to the standard solving strategy based on ADM. 
  This, however, comes at the expense of increased computational cost, reflecting a trade-off between solution quality and efficiency. 
  We also briefly investigate and discuss the effect of the data initialization and of unsymmetrically distributed and noisy datasets. 
  We numerically illustrate that for the studied truss structure, our solving strategy generally improves the robustness. 
  Particularly, it reduces the effect of the data initialization and of the dataset on the value of the global objective function, 
  and reduces the effect of an unsymmetrically distributed dataset on the accuracy of the discrete solution. 
  We also show that, however, an unsymmetrical data distribution and noisy data increase 
  the computational cost of our solving strategy more significantly than the ADM solving strategy, i.e. its computational cost is less robust with respect to the data distribution.

  \subsubsection{Optimal solution versus computational cost}

  \begin{figure}[h]
    \centering
    \includegraphics[width=0.3\textwidth]{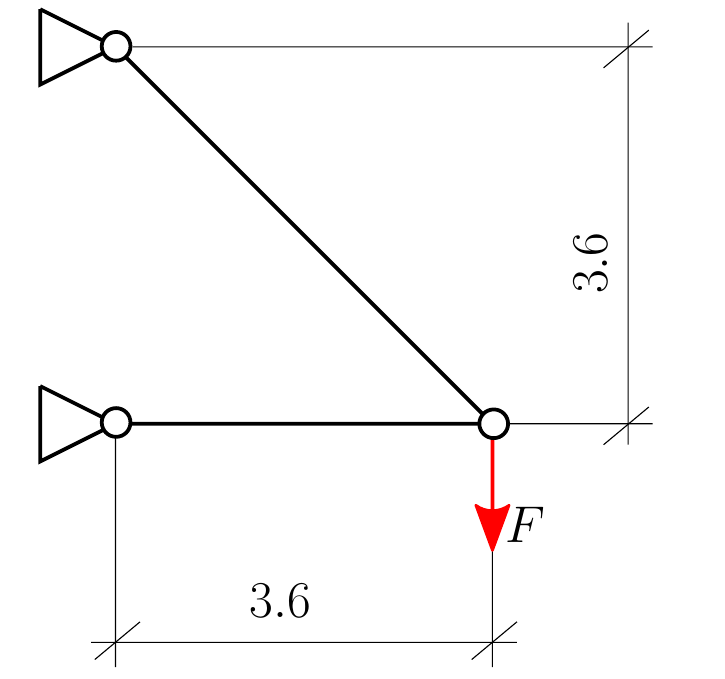}

    \caption{Sketch of a simplified truss structure.}\label{fig:kannoTrussGeometrySimp}
  \end{figure}

  \begin{figure}
  \centering
  \subfloat[][$\vect{\curconfig}_h$ with $\vect{u}_h \times 50$, $\alpha=0$, $\gamma=1$]{\includegraphics[width=0.47\textwidth]{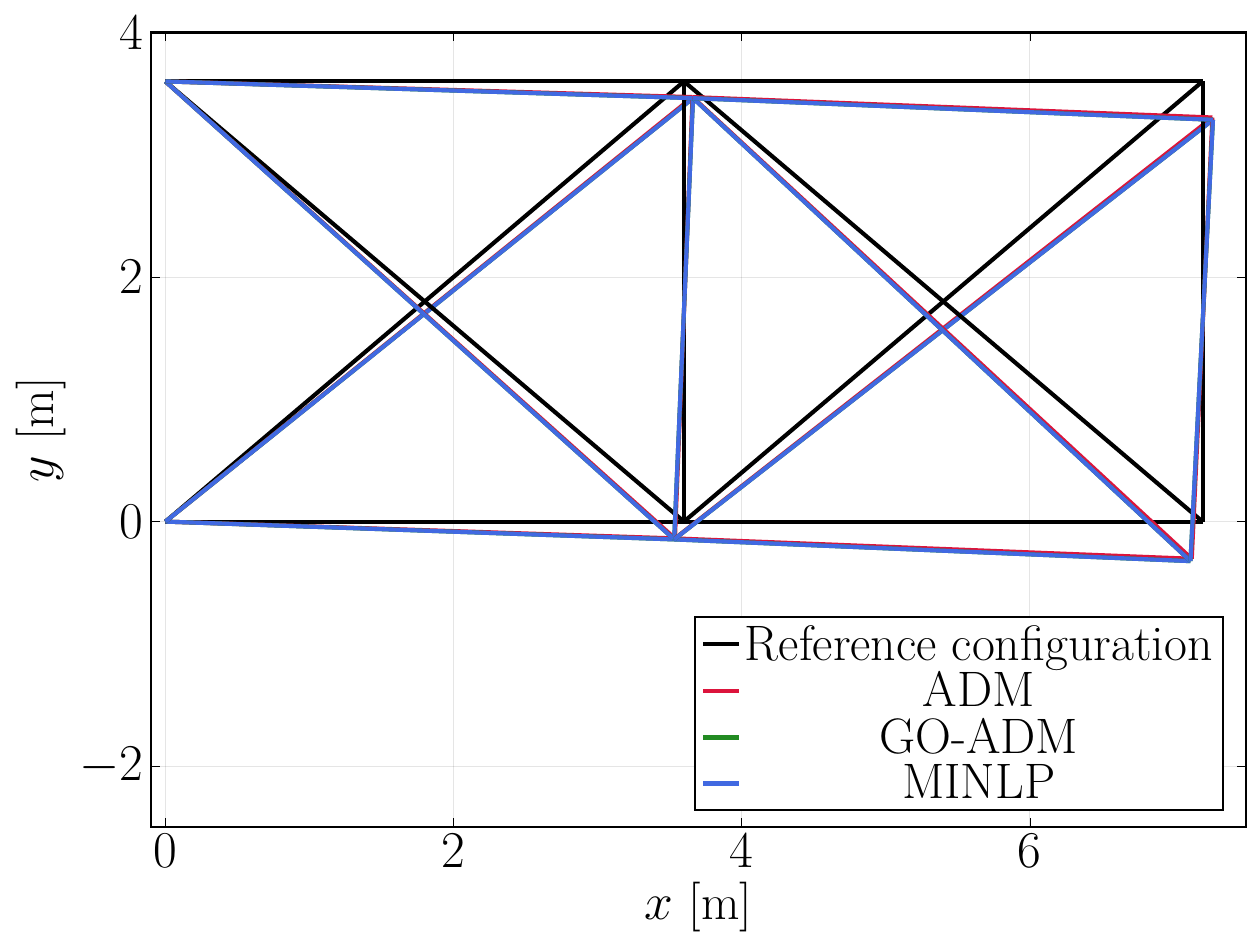}}
  \subfloat[][$\vect{\curconfig}_h$, $\alpha=1$, $\gamma=100$]{\includegraphics[width=0.47\textwidth]{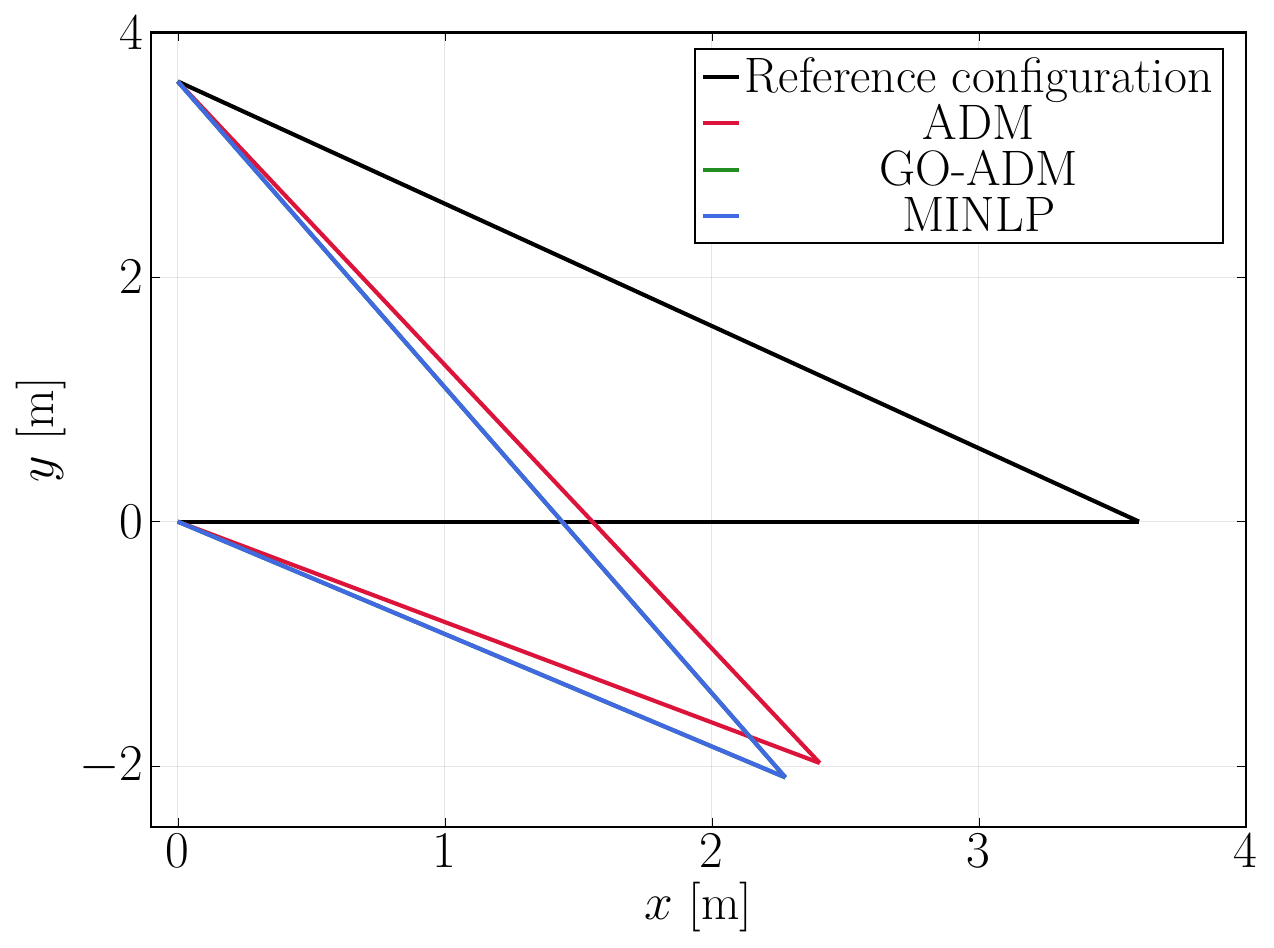}}

  \subfloat[][$s_h$, $\alpha=0$, $\gamma=1$]{\includegraphics[width=0.47\textwidth]{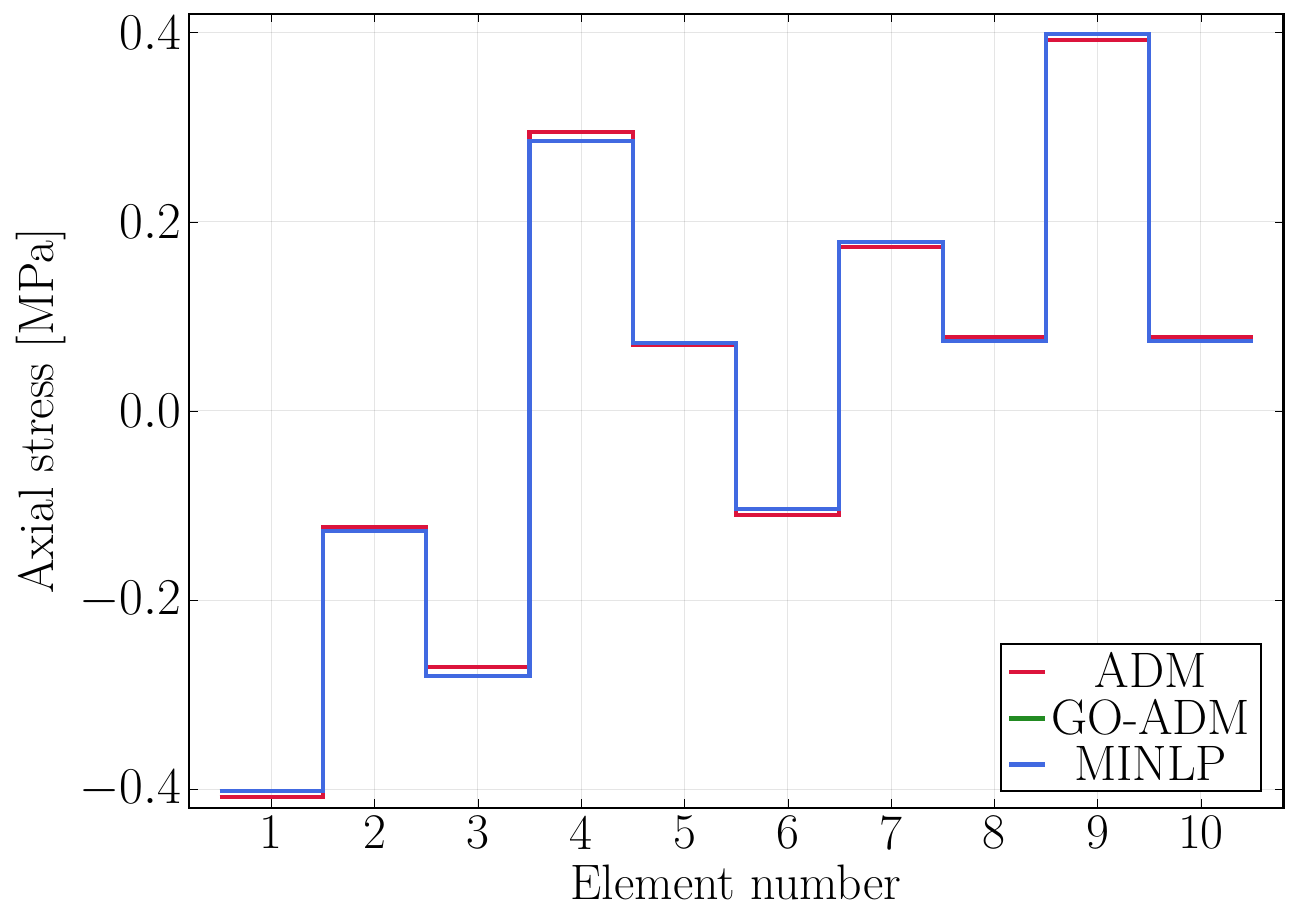}}
  \subfloat[][$s_h$, $\alpha=1$, $\gamma=100$]{\includegraphics[width=0.47\textwidth]{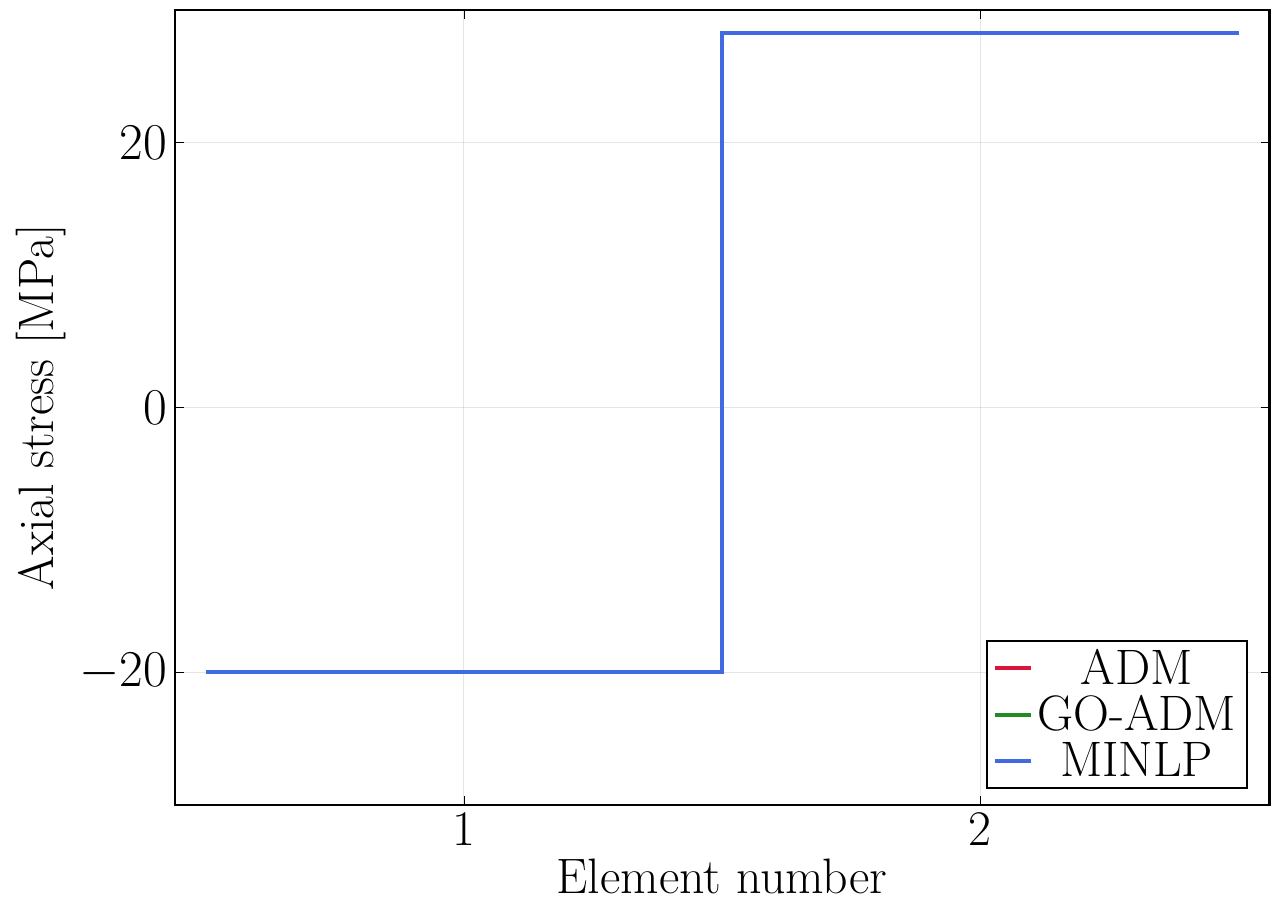}}

  \subfloat[][$\dataset$, $\alpha=0.0$, $\gamma=1$]{\includegraphics[width=0.47\textwidth]{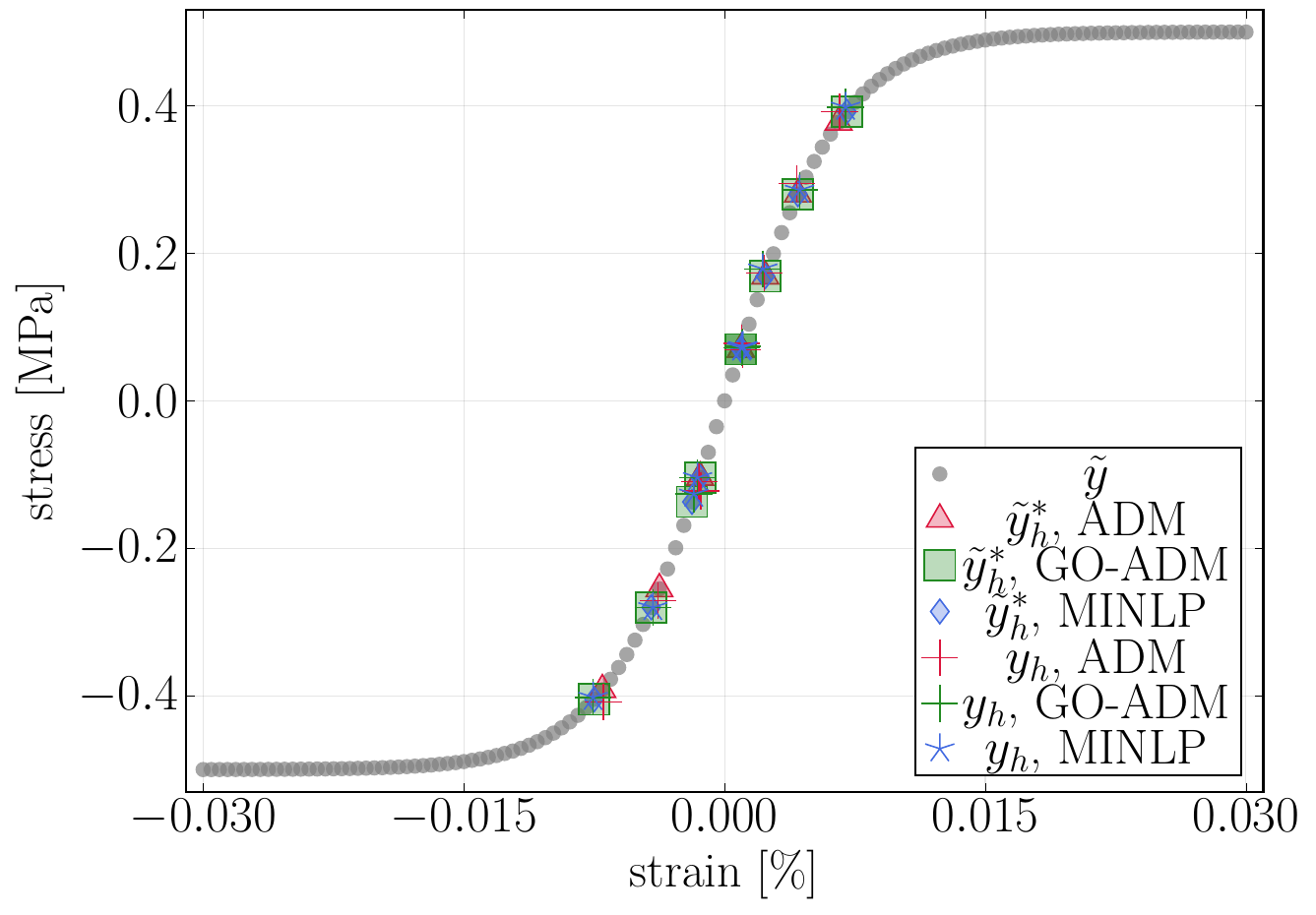}}
  \subfloat[][$\dataset$, $\alpha=1.0$, $\gamma=100$]{\includegraphics[width=0.47\textwidth]{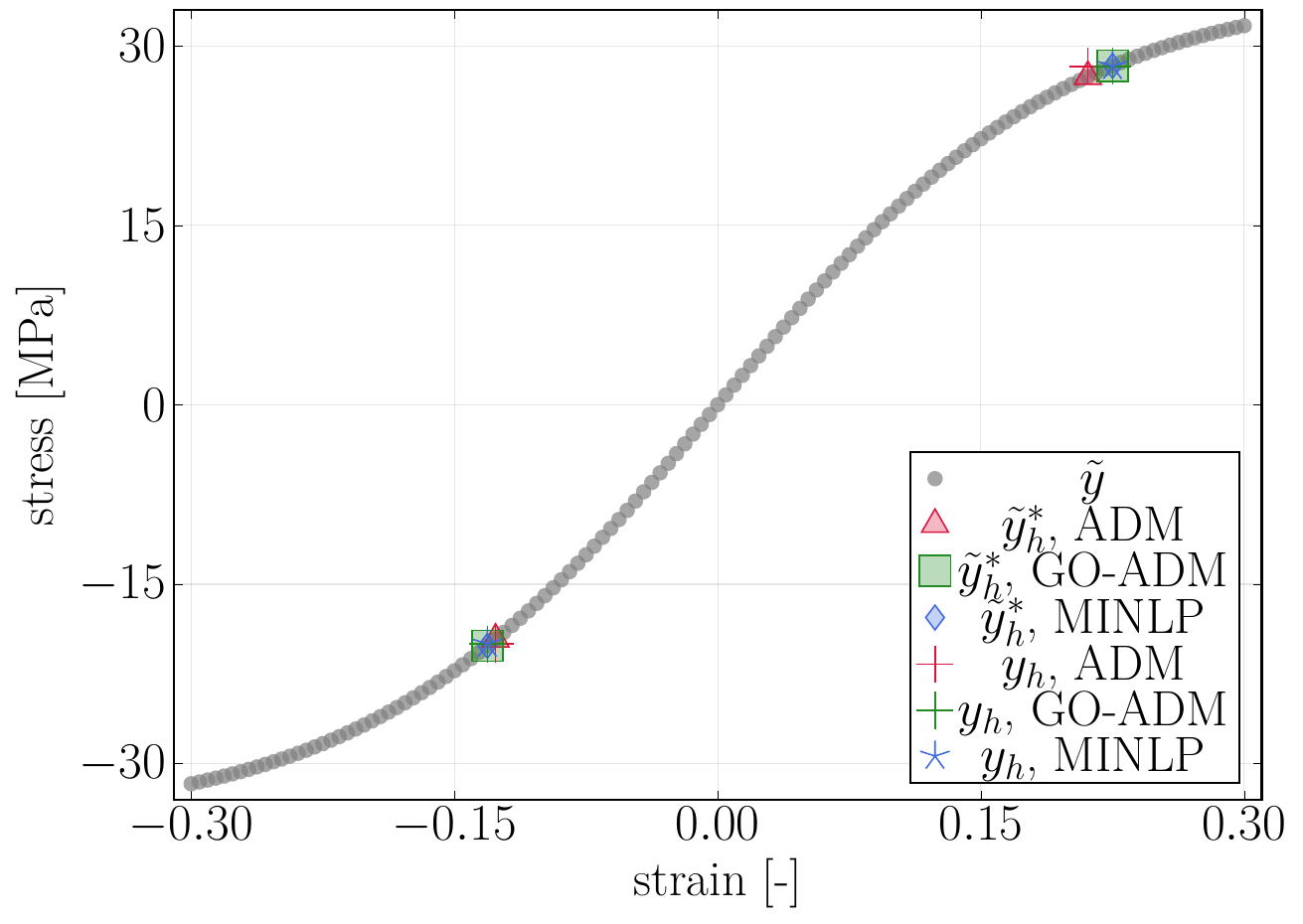}}

  \caption{Deformed configuration and axial stress of the truss structure in Figures \ref{fig:kannoTrussGeometry} and \ref{fig:kannoTrussGeometrySimp}, and the dataset including computed phase state, computed with linear ($\alpha=0.0$) and nonlinear ($\alpha=1.0$) strain measures. 
  The computation is performed with \textbf{129 data points} and the \textbf{structure-specific} initialization approach.}\label{fig:kannoTrussNonlinEnonlinDataNullspace}
\end{figure}

\begin{figure}
  \centering
  \subfloat[][$\alpha = 0.0$]{\includegraphics[width=0.48\textwidth]{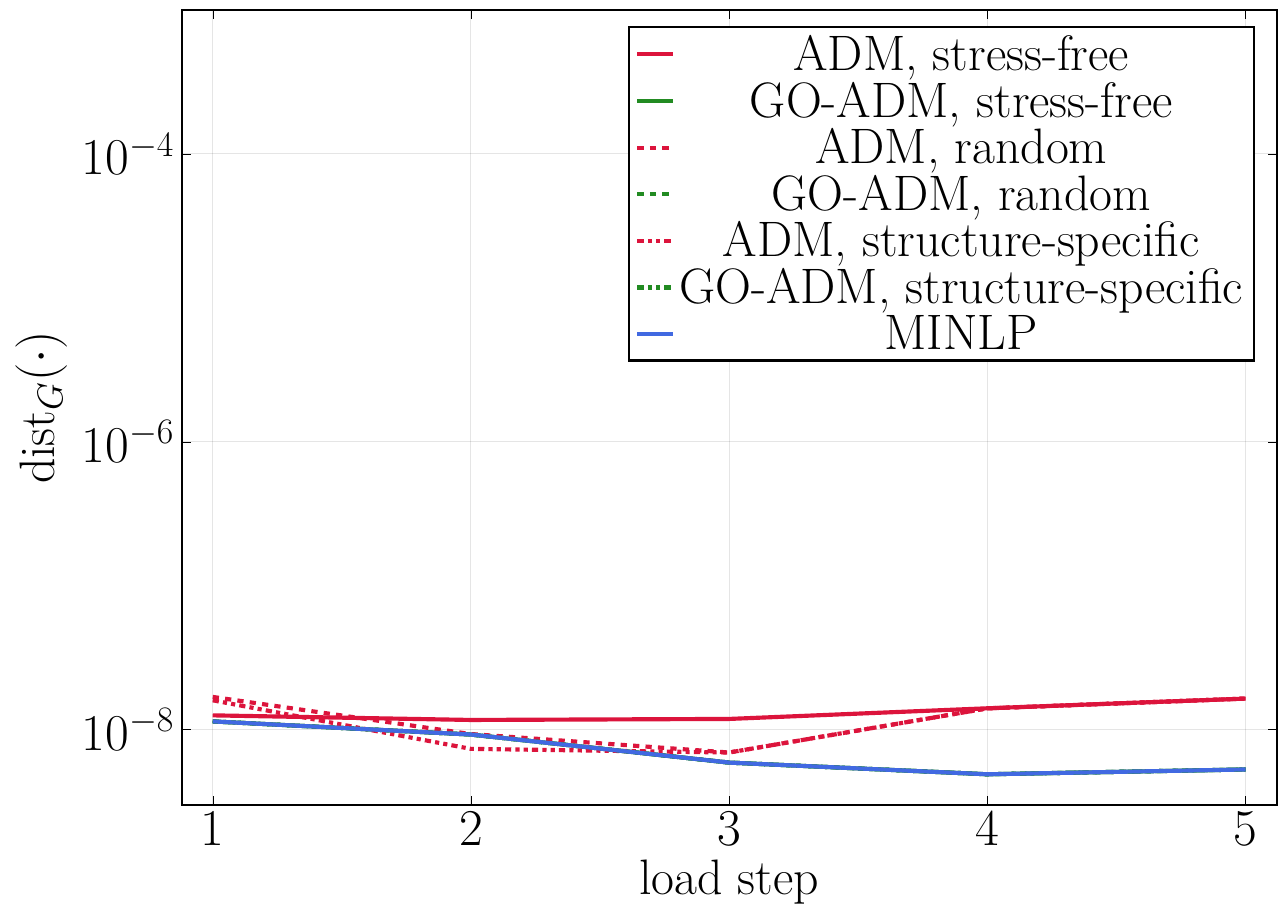}}
  \subfloat[][$\alpha = 1.0$]{\includegraphics[width=0.48\textwidth]{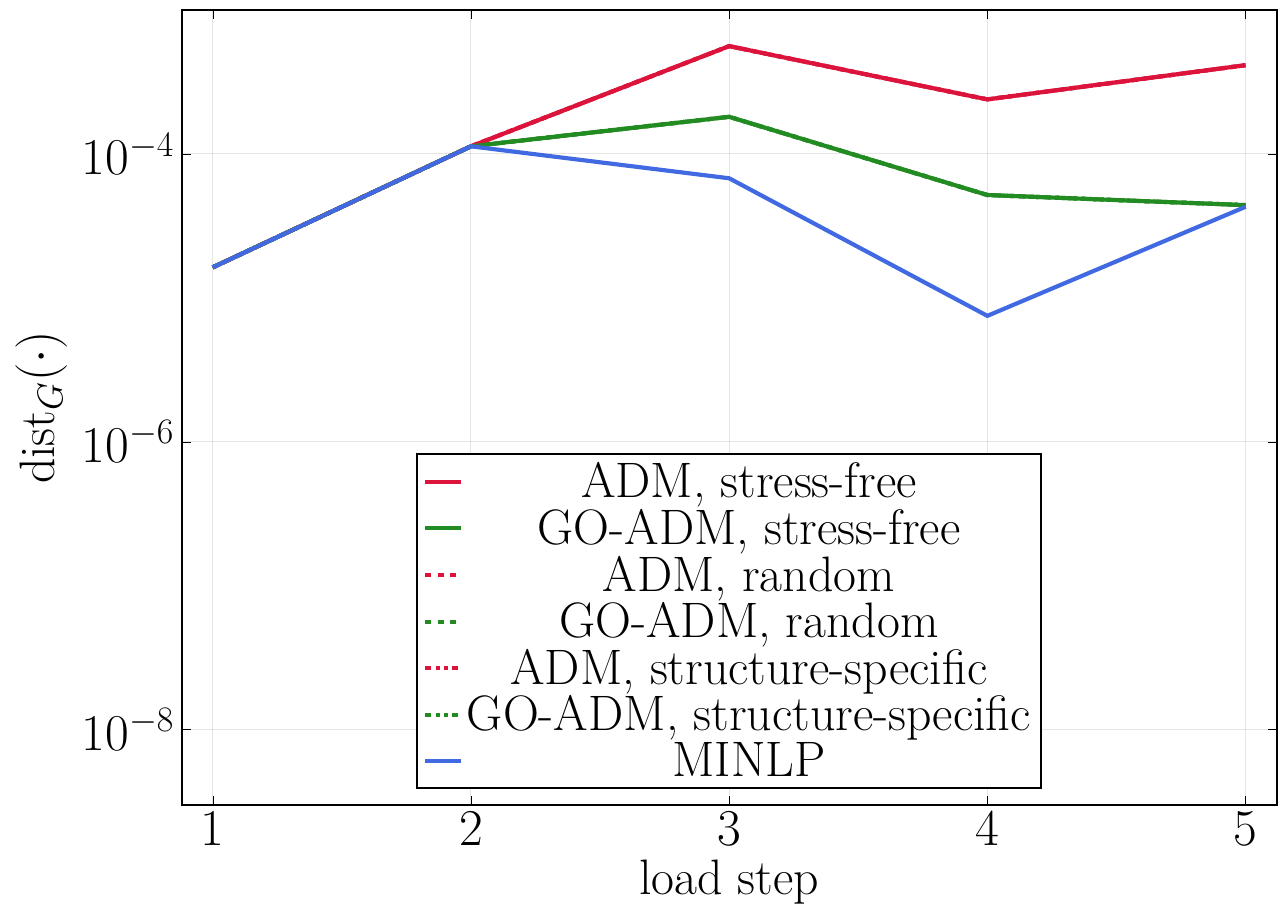}}

  \caption{Global objective function of the truss structure examples associated with Figure \ref{fig:kannoTrussNonlinEnonlinDataNullspace}, using different initialization approaches with a dataset of \textbf{129 data points}.}\label{fig:kannoTrussNonlinEnonlinDataCostFunc}
\end{figure}

\begin{table}[h]
  \centering
  \begingroup
  \setlength{\extrarowheight}{.2em} 
  \setlength{\tabcolsep}{6pt}       

  \begin{tabularx}{\textwidth}{|l|*{6}{>{\centering\arraybackslash}X|}}
    \hline
    Solving strategy
      & \multicolumn{3}{c|}{$\alpha = 0$ (5 load steps)}
      & \multicolumn{3}{c|}{$\alpha = 1$ (5 load steps)} \\
    \cline{2-7}
      &
      stress-free & random & structure-specific
      & stress-free & random & structure-specific \\
    \hline

    \multicolumn{7}{|l|}{\textbf{ADM}} \\
    \hline
    Total Newton--Raphson & 36 & 42 & 32 & 69 & 90 & 63 \\
    iterations            & & & & & & \\
    \hline
    Number of ADM         & 18 & 21 & 16 & 18 & 22 & 16 \\
    iterations            & & & & & & \\
    \hline
    Computing time [s]    & 0.00514 & 0.00664 & 0.00486 & 0.00921 & 0.01155 & 0.00773 \\
    \hline

    \multicolumn{7}{|l|}{\textbf{GO-ADM}} \\
    \hline
    Total Newton--Raphson & 60 & 66 & 56 & 132 & 150 & 126 \\
    iterations            & & & & & & \\
    \hline
    Total ADM iterations  & 30 & 33 & 28 & 30 & 33 & 28 \\
    \hline
    Number of ``greedy''    & 505 & 505 & 505 & 505 & 505 & 505 \\
    searches                & & & & & & \\
    \hline
    Computing time [s]      & 0.00907 & 0.01616 & 0.00922 & 0.01616 & 0.02456 & 0.01564 \\
    \hline

    \multicolumn{7}{|l|}{\textbf{MINLP}} \\
    \hline
    Computing time [s]               & \multicolumn{3}{|c|}{0.05411} & \multicolumn{3}{|c|}{0.59617} \\
    \hline

  \end{tabularx}
  \endgroup
  \caption{Summary of iterations and computing time using different data initialization approaches and solving strategies for truss structures exhibiting \textbf{linear ($\alpha=0.0$) and nonlinear strains ($\alpha=1.0$)}. The metrics corresponding to the GO-ADM-solver are obtained for all searches and load steps.}
  \label{tab:kannoTrussNonlinDataCompCost}
\end{table}

  We now consider datasets associated with nonlinear constitutive relations. 
  We illustrate via two truss structures with such datasets that our solving strategy also achieves 
  a better approximation of the globally optimal solution, however, comes at the cost of computational effort. 
  To this end, we define a nonlinear constitutive relation based on the illustration of the dataset in \cite{kanno_data_driven_2019}, using the sigmoid function as follows: 
  \begin{equation}\label{eq:sigmoidDataFunc}
    \tilde{s}(\tilde{e}) =  \tilde{S} \,  \left(\frac{2}{1 + e^{-\tilde{e}}}- 1 \right) \,,
  \end{equation}
  where $\tilde{S}$ is the maximum stress value included in the dataset. 
  For our computations in this subsection, we generate the dataset with 129 discrete points using this constitutive function, for which we check its thermomechanical consistency based on the discussions in \cite{Gebhardtddcmhilbert2025}. 
  For this example, we obtain the globally optimal solution as the reference employing the mixed-integer quadratic programming solver with an objective function using the $L^2$-norm, 
  as discussed in Section \ref{sec:minlpForm}. 
  We refer to this as the solution of a mixed-integer nonlinear optimization problem (MINLP), which is more general than the mixed-integer quadratic programming problem.

  We compare the discrete solution, the converged minimizer, the value of the global objective function, and the computational cost in terms of the total number of iterations and computing time when using the ADM, GO-ADM solving strategies, and the MINLP-solver. 
  We give the details of the computer specifications for our computations in \ref{sec:computerSpecific}. 
  To gain better insights on different nonlinearity sources and their effect, we investigate two cases with a linear and nonlinear strain measures. 
  For the former, we consider the same truss structure 
  studied in Section \ref{sec:benchmarkTruss} (see also Figure \ref{fig:kannoTrussGeometry}) and again   
  choose a unit load factor of $\gamma = 1.0$ for small deformations. We now apply the nodal forces in 5 load steps to gain better insights in the development of the objective function over load steps. 
  For the case with nonlinear strain measures, 
  due to the enormous computational cost when using the MINLP solver, we consider a simplified structure, illustrated in Figure \ref{fig:kannoTrussGeometrySimp}, that has the same cross-section and material properties as the one showed in Figure \ref{fig:kannoTrussGeometry} and is subjected to a downward nodal force, $F = 400\gamma$ N, at its single unconstrained node. 
  To obtain large deformations and hence sufficient nonlinear strain contribution, we choose a load factor of $\gamma = 100$ and also apply the nodal force in 5 steps. 
  For both study cases, we choose a maximum number of ``greedy'' searches of 100 when using the GO-ADM-solver.

  \begin{remark}
    The simplified truss structure, illustrated in Figure \ref{fig:kannoTrussGeometrySimp}, has an unstable solution. This occurs when the unconstrained node aligns with the other two on the vertical axis. 
    To avoid this scenario, we choose a load factor $\gamma = 100$ and generate the dataset with strains ranging in $[-0.2,\,0.2]$ (see also Figure \ref{fig:kannoTrussNonlinEnonlinDataNullspace}f). 
  \end{remark}

  We first focus on the accuracy of the discrete solution obtained with our and the ADM solving strategy and hence employ the structure-specific initialization approach. 
  This choice is based on its favorable property of guaranteeing global optima for the case with linear strains and certain symmetry, as well as the convergence of the Newton-Raphson scheme for the case with nonlinear strains, as discussed in Section \ref{sec:datainitialization}. 
  Figure \ref{fig:kannoTrussNonlinEnonlinDataNullspace} illustrates the deformed configuration, the discrete axial stresses, the computed stress-strain pairs, and the converged minimizer obtained with both solvers for linear (Figures \ref{fig:kannoTrussNonlinEnonlinDataNullspace}a, c, e) and nonlinear strains (Figures \ref{fig:kannoTrussNonlinEnonlinDataNullspace}b, d, f). 
  For the former case, we have the same observations as the example with linear constitutive relation studied in Section \ref{sec:benchmarkTruss}: both solvers lead to the same discrete solution as the reference obtained with the MINLP-solver. The solved stress-strain pairs and the obtained minimizers, however, slightly differ when using the ADM-solver but remains approximately the same when using either the GO-ADM- or the MINLP-solver. 
  For the case with nonlinear strains, we observe that using our solving strategy leads to larger deformations and the same solution as the globally optimal solution obtained with the MINLP-solver (see Figures \ref{fig:kannoTrussNonlinEnonlinDataNullspace}b, d). The discrete solution and the minimizers obtained with the ADM-solver show slightly larger difference than the case with linear strains but remain approximately the same when using ours and the MINLP-solver. 
  Comparing to the discrete solution obtained for the case with linear constitutive relation studied in Section \ref{sec:datainitialization} (see also Figure \ref{fig:kannoTrussResultsNonlin}), 
  we see that the material nonlinearity reduces the accuracy achieved with the ADM-solver despite the employed structure-specific initialization. 
  As discussed in Section \ref{sec:datainitialization}, the main reason is that this initialization approach is based on a linear system and hence does not guarantee global optimality for nonlinear cases.

  Focusing on the value of the global objective function $\globObjFunc(\cdot,\cdot)$, illustrated in Figure \ref{fig:kannoTrussNonlinEnonlinDataCostFunc}, 
  we observe that for the case with linear strain measures, all three solvers lead to approximately the same values of $\globObjFunc(\cdot,\cdot)$ at all load steps. 
  For the case with nonlinear strains, using the GO-ADM-solver slightly reduces $\globObjFunc(\cdot,\cdot)$ at all load steps, as expected, and achieves the same values as the MINLP-solver at the first two and last load steps. 
  These observations are consistent with the difference observed in the solved stress-strain pair and obtained minimizer showed in Figures \ref{fig:kannoTrussNonlinEnonlinDataNullspace}e and f. 
  The difference in the objective function between our and the MINLP-solver illustrates that ours does not achieve the global optimal at all load steps but a good approximation, leading to approximately the same discrete solution, i.e. accuracy. 
  Focusing on the total number of iterations and computing times when using our and the ADM-solver, showed in Table \ref{tab:kannoTrussNonlinDataCompCost}, 
  we observe that for both linear and nonlinear strain measures, using the GO-ADM-solver requires approximately 1.5 times to twice as many iterations for the Newton-Raphson scheme and updating the stress-strain data pairs (ADM iterations) compared to the ADM-solver, as well as twice the computing time. 
  This increased cost is caused by the recalculation in each ``greedy'' search, as discussed in Section \ref{sec:goadmalgorithm}. 
  Using the MINLP-solver, on the other hand, requires about 5 times and one order of magnitude more time for the linear and nonlinear strains, respectively. 
  Notably, for the case with nonlinear strains, the MINLP-solver requires one order of magnitude more time to finish the computations than the case with linear strains, despite our simplifications, which illustrates its sensitivity with respect to nonlinearity.

  We then numerically investigate the effect of different initialization approaches on the value of the objective function $\globObjFunc(\cdot,\cdot)$ and the computational cost for the studied truss structure. 
  We note that we do not focus on their effect on the accuracy in this work since it has been discussed in \cite{Gebhardtddcmsolution2025} and Section \ref{sec:datainitialization}. 
  Figure \ref{fig:kannoTrussNonlinEnonlinDataCostFunc} illustrates the values of $\globObjFunc(\cdot,\cdot)$ at each load step for the case with linear and nonlinear strains, obtained with different initialization approaches.  
  We observe that for the case with linear strains, using the ADM-solver with the stress-free initialization leads to slightly different $\globObjFunc(\cdot,\cdot)$, while using our solver leads to the same objective function for all initialization approaches. 
  For the case of nonlinear strains, all three approaches lead to the same objective function at all load steps when using either our or the ADM-solver, except the first load step when using a random initialization. 
  We see that when using the ADM-solver, a random initialization leads to slightly higher values of $\globObjFunc(\cdot,\cdot)$ at the first load steps, which is already reduced at earlier load steps when using our solver. 
  We note that this is not conclusive based on only one computation with a random initialization but requires a statistical study with multiple replicates. 
  Such a study is out of scope of this work and is considered for future work. 
  We conclude that for the studied truss structure and dataset, the data initialization does not affect the objective function.   
  A random initialization slightly increases the computational cost, which can be reduced to the same effort using either the stress-free or structure-specific initialization.

  \subsubsection{Effects of data distribution}

  \begin{figure}
  \centering
  \subfloat[][$\vect{\curconfig}_h$, unsymmetrical dataset]{\includegraphics[width=0.47\textwidth]{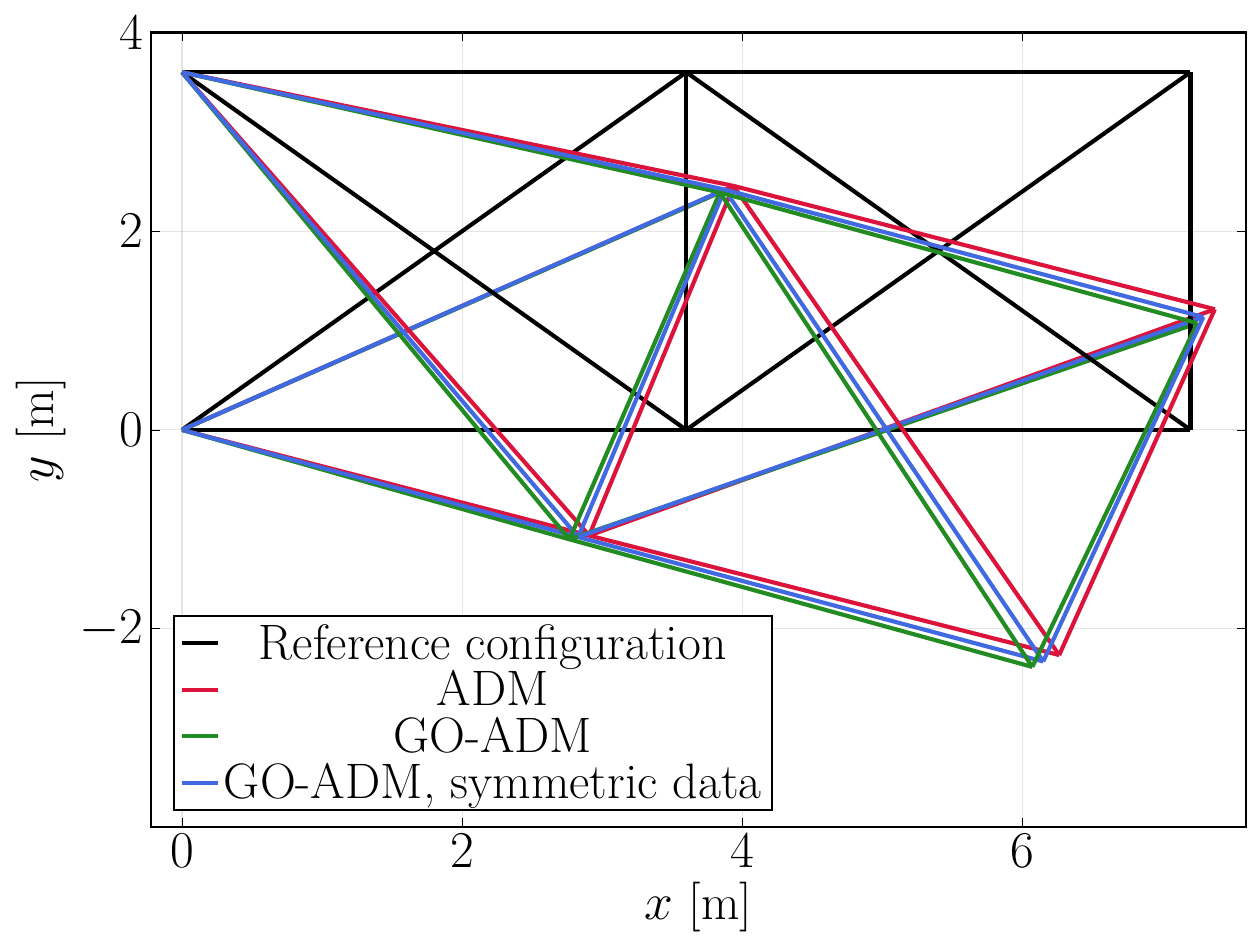}}
  \subfloat[][$\vect{\curconfig}_h$, noisy dataset]{\includegraphics[width=0.47\textwidth]{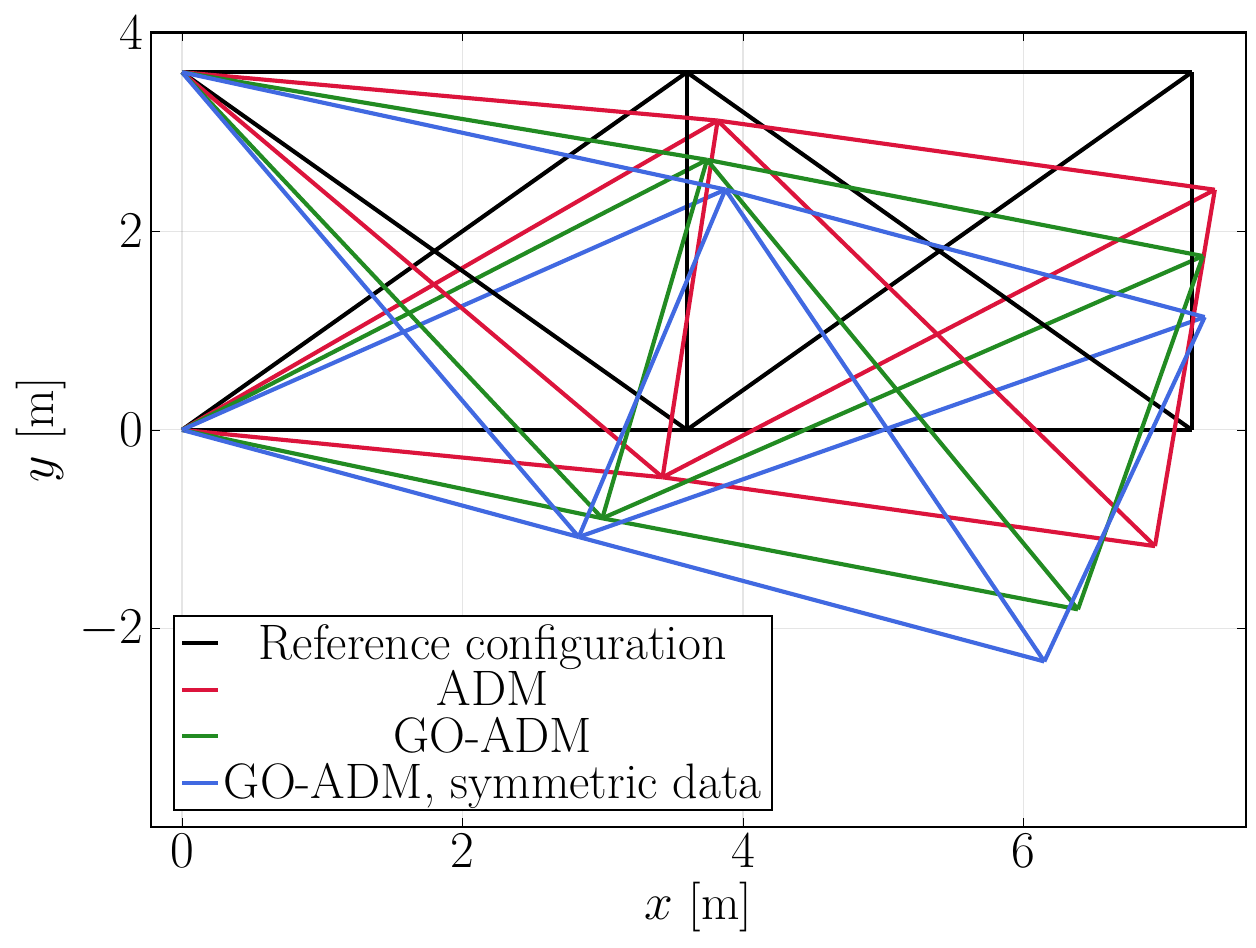}}

  \subfloat[][$s_h$, unsymmetrical dataset]{\includegraphics[width=0.47\textwidth]{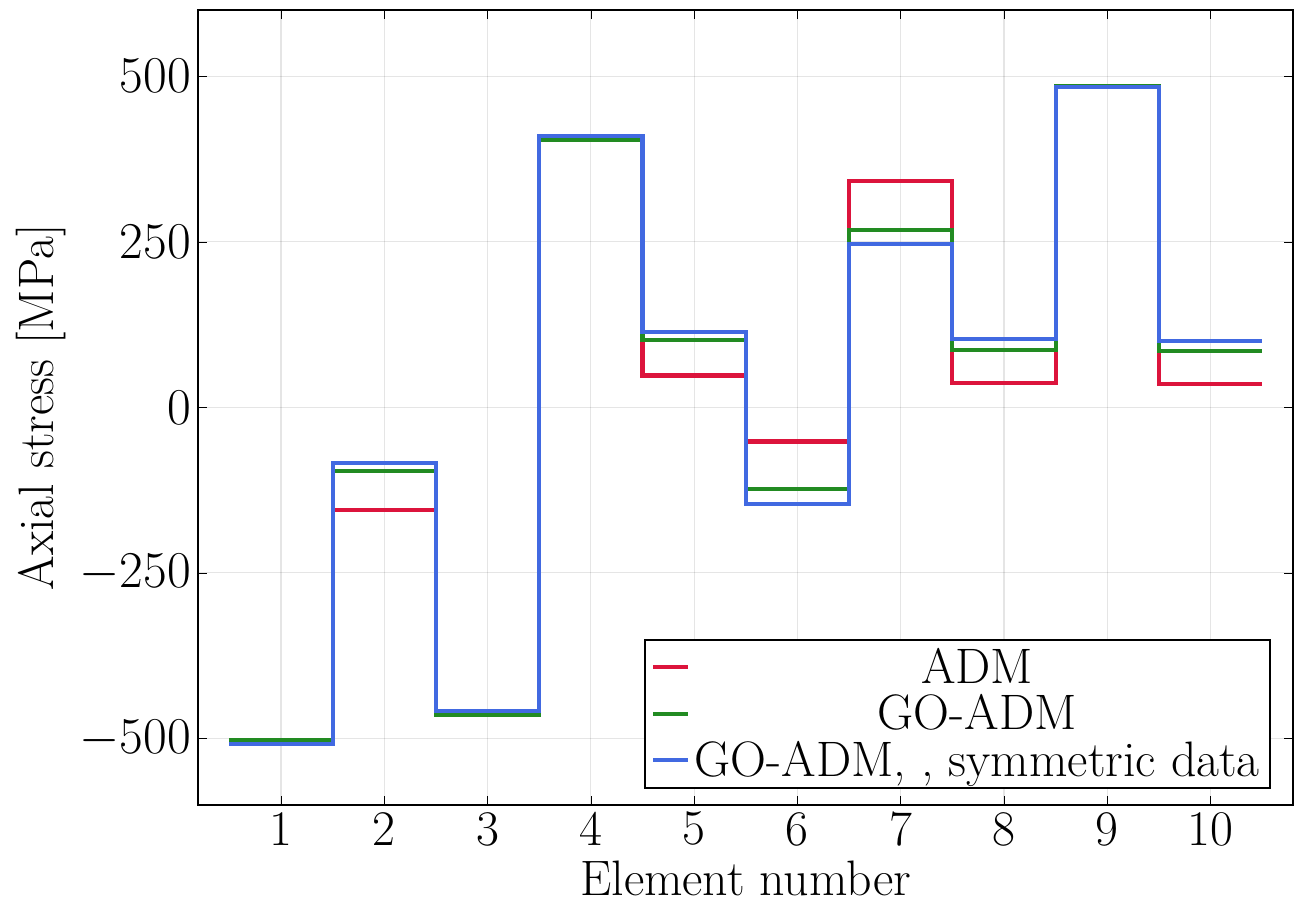}}
  \subfloat[][$s_h$, noisy dataset]{\includegraphics[width=0.47\textwidth]{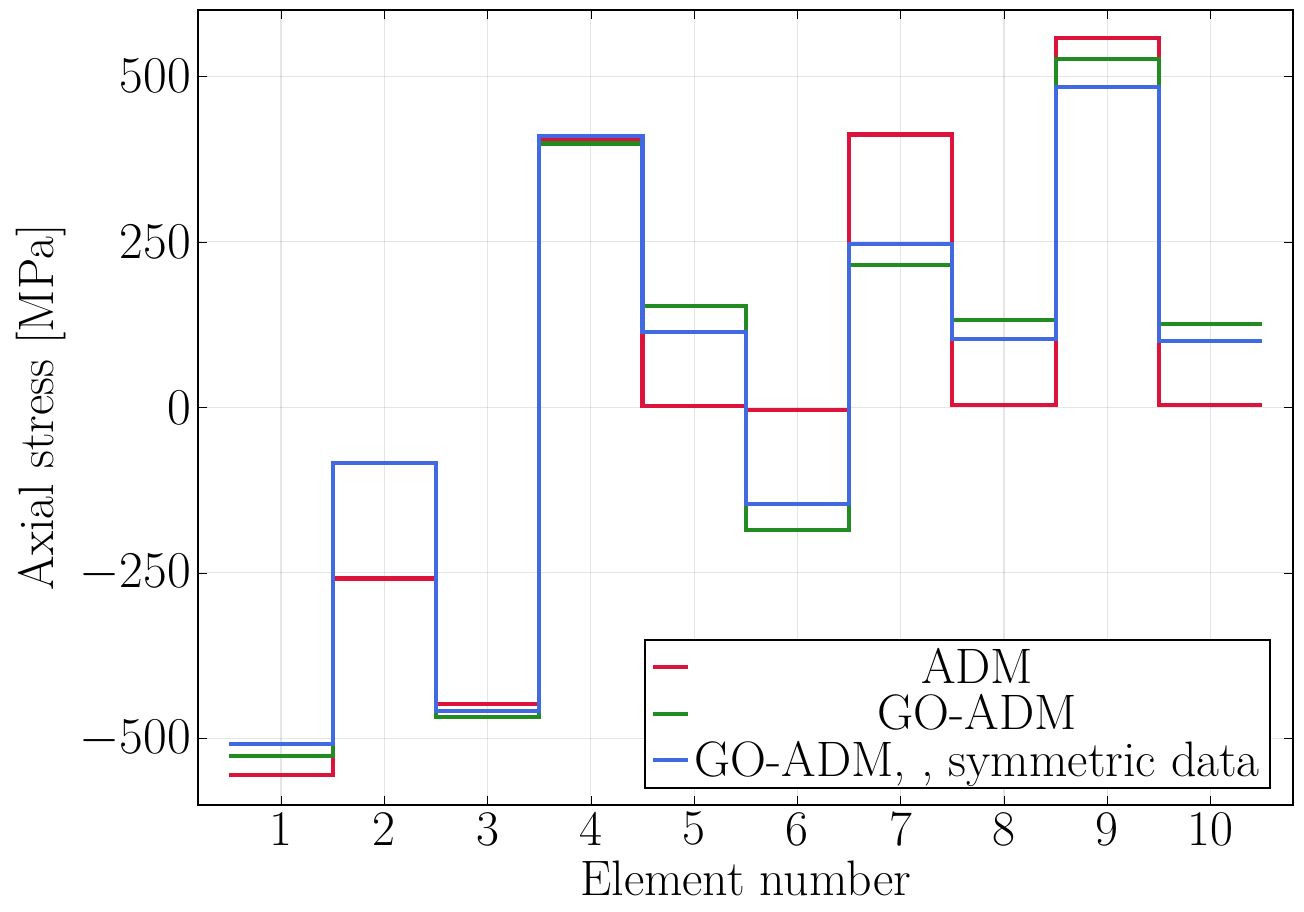}}

  \subfloat[][Unsymmetrical $\dataset$]{\includegraphics[width=0.47\textwidth]{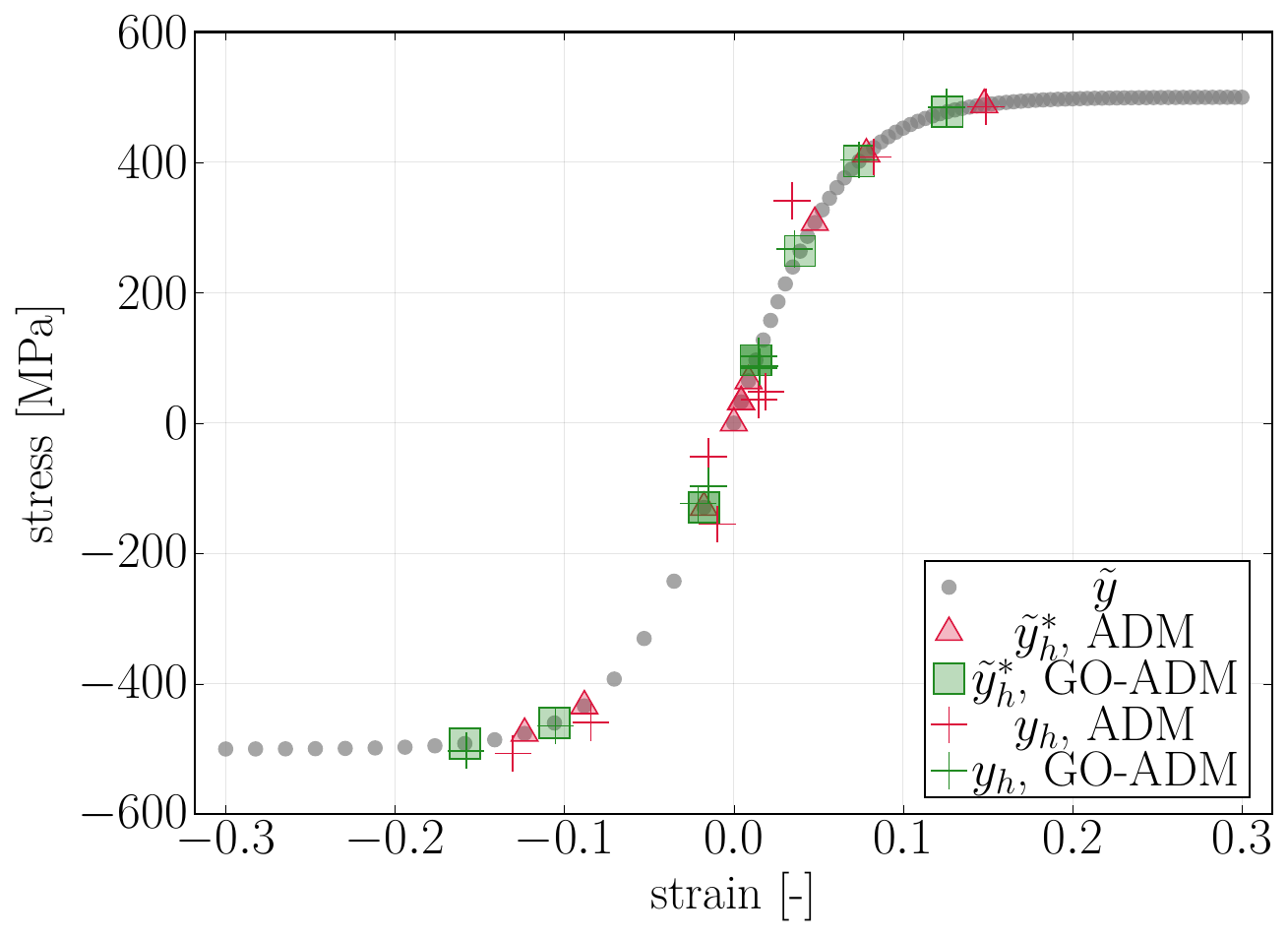}}
  \subfloat[][Noisy $\dataset$]{\includegraphics[width=0.47\textwidth]{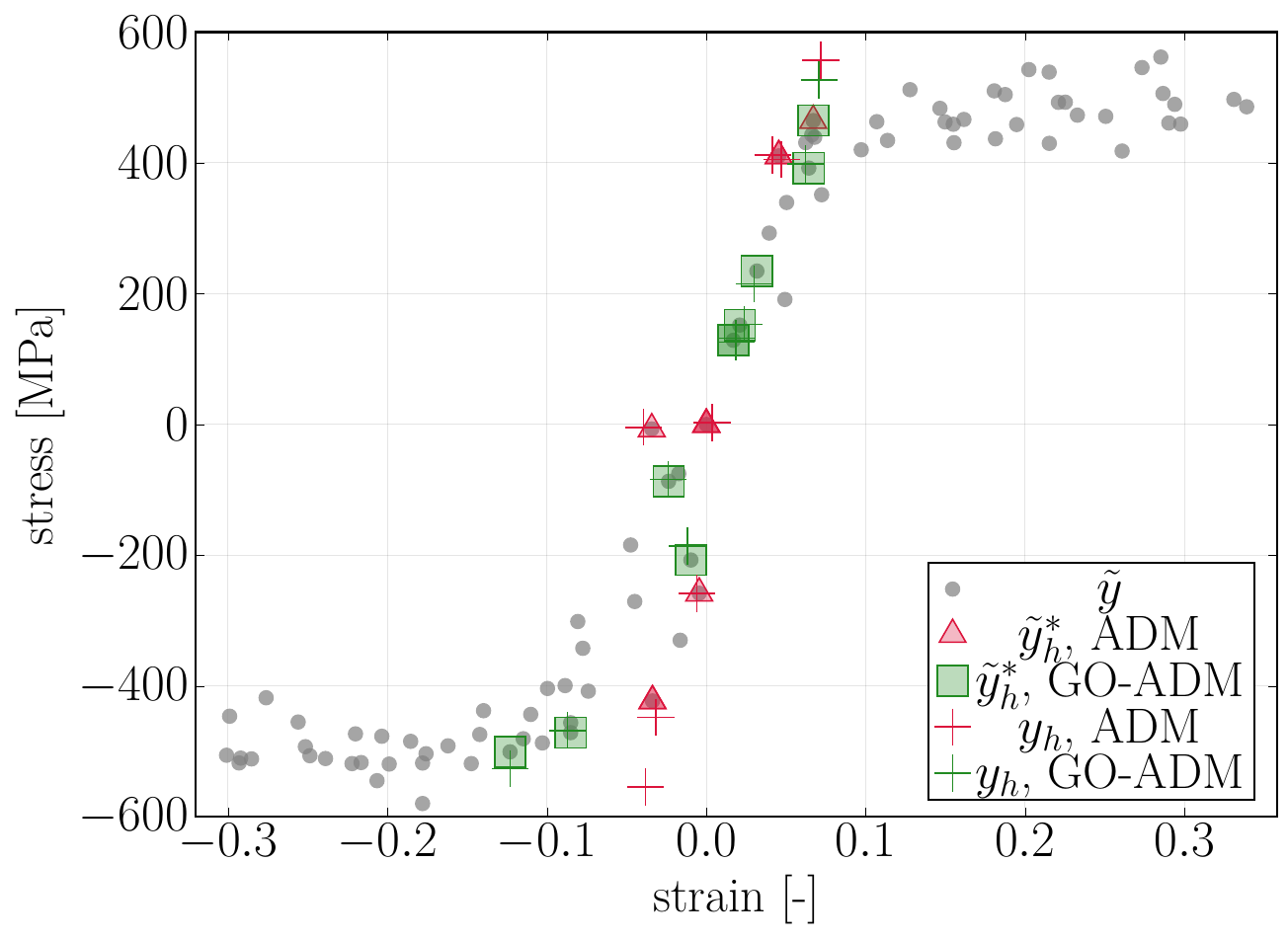}}

  \caption{Deformed configuration, axial stress, and the computed stress-strain pairs of the truss structure in Figure \ref{fig:kannoTrussGeometry}, computed with a \textbf{unsymmetrical (left column)} and \textbf{noisy (right column)} dataset, using \textbf{nonlinear strain measures} and the \textbf{structure-specific} initialization approach.}\label{fig:kannoTrussNonlinEdataEffectResultsNullspace}
\end{figure}

\begin{figure}
  \centering
  \subfloat[][Unsymmetrical dataset]{\includegraphics[width=0.48\textwidth]{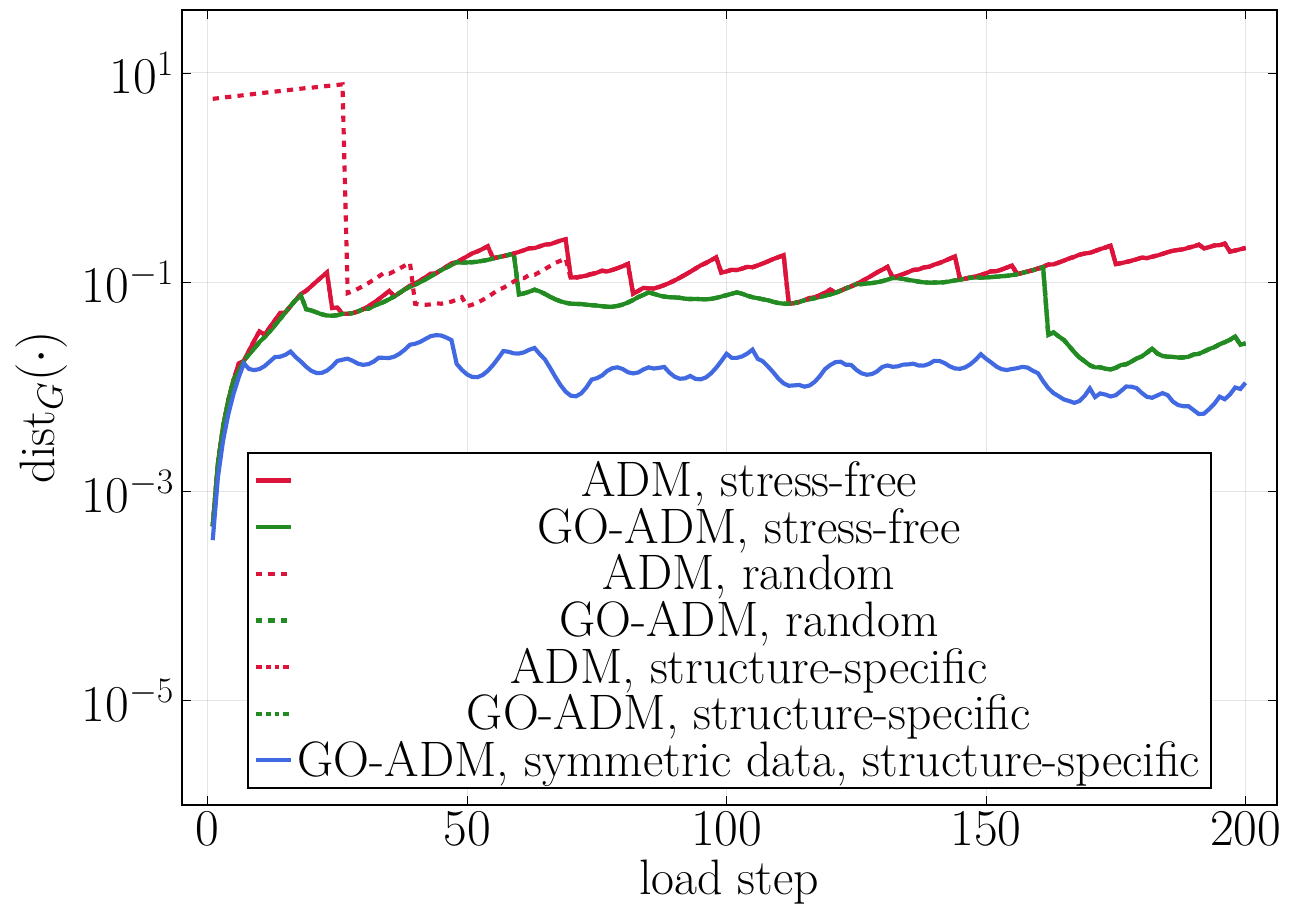}}
  \subfloat[][Noisy dataset]{\includegraphics[width=0.48\textwidth]{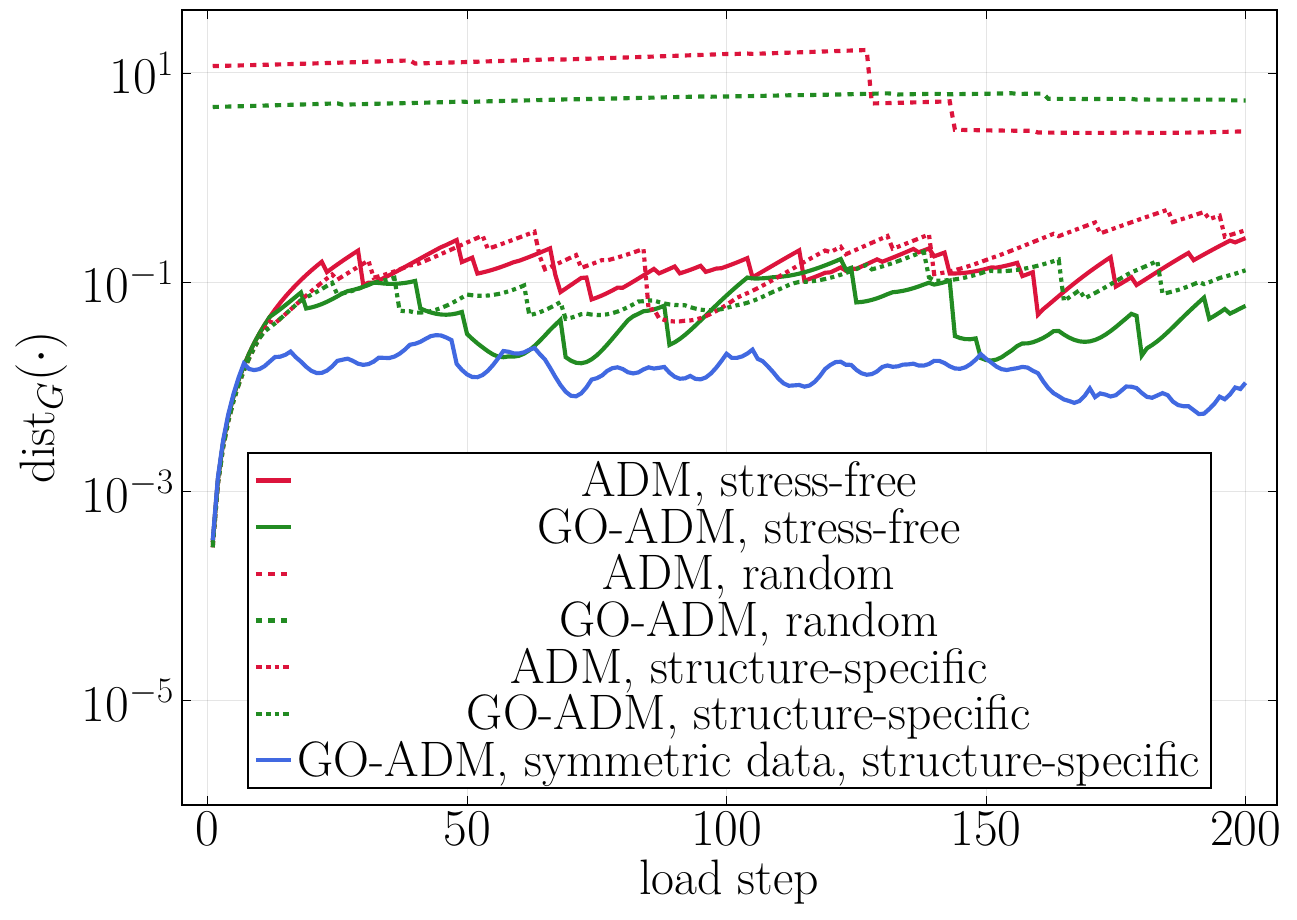}}

  \caption{Values of the global objective function over load steps, obtained when using unsymmetrical and noisy datasets and different data initialization approaches for the computations of the truss structure in Figure \ref{fig:kannoTrussGeometry} with \textbf{nonlinear strain measures}.}\label{fig:kannoTrussNonlinEdataEffectCostFunc}
\end{figure}

  We now numerically illustrate for the truss structure studied in Section \ref{sec:benchmarkTruss} (see also Figure \ref{fig:kannoTrussGeometry}) with a nonlinear constitutive relation that our GO-ADM solving strategy improves the robustness with respect to the data distribution and initialization. 
  In particular, it reduces the effect of an unsymmetrical distribution and noises in the dataset on the global objective function $\globObjFunc(\cdot,\cdot)$. 
  Moreover, our solving strategy removes the dependency of $\globObjFunc(\cdot,\cdot)$ on the initialization for the case of an unsymmetrically distributed dataset, and  
  reduces the effect of noisy dataset on the accuracy of the discrete solution. 
  We also numerically show that the computational cost of our solver is less robust with respect to the data distribution than the ADM-solver. 
  Particularly, an unsymmetrical data distribution and noisy data increase its computational effort more significantly than the ADM-solver.

  For our numerical studies, we generate the dataset using the same nonlinear constitutive relation \eqref{eq:sigmoidDataFunc} as in the previous subsection. 
  Since we now focus on the effect of the data distribution, we consider more data points of 87 points in all cases, which requires significantly more computing time when using the MINLP solver for the studied truss structure. 
  Hence, we do not include the results obtained with the MINLP-solver but the GO-ADM-solver using a symmetric distributed dataset as reference results. 
  We note that the latter generally does not guarantee global optima, however, achieves a good approximation and hence is sufficient for our comparison purposes to investigate the effect of an unsymmetrically distributed and noisy dataset in this subsection. 
  For the following computations, we consider nonlinear strain measures, large deformations with a load factor of $\gamma = 1500$, and apply the nodal forces in 200 steps.

  We first investigate the effect of an unsymmetrically distributed dataset on the accuracy and the global objective function $\globObjFunc(\cdot,\cdot)$. 
  We generate the dataset with 80$\%$ of data points distributed in the first (positive stresses and strains) and the remaining 20$\%$ in the fourth quadrant (negative stresses and strains) (see also Figure \ref{fig:kannoTrussNonlinEdataEffectResultsNullspace}e). 
  We also check for this dataset and the obtained results that they are thermomechanically consistent as discussed in \cite{Gebhardtddcmhilbert2025}. 
  Figures \ref{fig:kannoTrussNonlinEdataEffectResultsNullspace}a, c, and e illustrate the deformed configuration, the discrete axial stresses, the dataset including the solved stress-strain pair and the converged minimizer, obtained with both solvers, respectively. 
  Here, we include the deformed configuration and discrete stress field obtained with the GO-ADM-solver using a symmetrically distributed dataset as a reference (blue curve in Figures \ref{fig:kannoTrussNonlinEdataEffectResultsNullspace}a and c). 
  We observe that our solving strategy achieves a better approximation of the discrete solution computed with a symmetric data distribution than the ADM-solver. 
  This means that our solver reduces the negative effect of the unsymmetrical data distribution on the accuracy. 
  Focusing on $\globObjFunc(\cdot,\cdot)$ evaluated at each load step, illustrated in Figure \ref{fig:kannoTrussNonlinEdataEffectCostFunc}a, we see that both solvers lead to $\globObjFunc(\cdot,\cdot)$ at approximately the same order of magnitude, except the last 50 load steps where our solver reduces $\globObjFunc(\cdot,\cdot)$ approximately by one order of magnitude. 
  Moreover, we see that $\globObjFunc(\cdot,\cdot)$ obtained with an unsymmetrical data distribution is of one order of magnitude larger than that obtained with a symmetric distribution (blue curve), illustrating the negative effect of this unsymmetry on $\globObjFunc(\cdot,\cdot)$.

  We then briefly study the effect of a noisy dataset on the accuracy of the discrete solution and the global objective function $\globObjFunc(\cdot,\cdot)$ and consider random noises based on a normal distribution with the same zero mean value and standard deviation of $0.06$. 
  We first normalized the symmetrically distributed dataset such that both strain and stress values are in the interval $[-1,1]$. 
  We then add random noises to both strains and stresses before scaling them back to their original range, 
  such that the noises are distributed over the whole dataset and is not scaled with either stress or strain value. 
  Figure \ref{fig:kannoTrussNonlinEdataEffectResultsNullspace}f shows the generated noisy dataset, for which we check its thermomechanical consistency, following discussions in \cite{Gebhardtddcmhilbert2025}. 
  Note that random noises possibly pollute and lead to thermomechanically inconsistent data points. 
  In this work, we simply replace the thermomechanically inconsistent data points with their original counterparts, given that the original ones are thermomechanically consistent. 
  This approach is not to denoise the dataset but only to recover the thermomechanical consistency of some data points since we want to include noise in the dataset for our investigation. 
  In general, data denoising is not trivial, particularly for non-synthetic datasets. Addressing this issue would require extending the GO-ADM solving strategy with specific techniques, such as the max-ent \cite{Kirchdoerfer2017ddcm} or locally convex recontruction (LCR) approaches \cite{He2020ddcmnoise}. Such extensions are beyond the scope of the present work and are therefore deferred to future research.

  Figures \ref{fig:kannoTrussNonlinEdataEffectResultsNullspace}b, d, and f illustrate the deformed configuration, the discrete axial stresses, the dataset including the solved stress-strain pair and the converged minimizer, obtained with both solvers, respectively. 
  We check that the obtained results are also thermomechanically consistent. 
  We observe that both solvers lead to smaller deformations than those obtained with dataset without noises (compare to the green and blue curves in Figure \ref{fig:kannoTrussNonlinEdataEffectResultsNullspace}a), illustrating the negative effect of noisy data on the accuracy. 
  We note that due to random noises, the results are not conclusive without comprehensive statistical studies. 
  Nevertheless, compared to the ADM-solver, our solver improves the accuracy for this example and hence is more robust with respect to noisy data. 
  Focusing on the values of the global objective function $\globObjFunc(\cdot,\cdot)$ over load steps, illustrated in Figure \ref{fig:kannoTrussNonlinEdataEffectCostFunc}b, we see that our solver reduces $\globObjFunc(\cdot,\cdot)$ at all load steps. 
  Both solvers lead to an objective function that has approximately the same order of magnitude as the case using dataset without noises (compare to Figure \ref{fig:kannoTrussNonlinEdataEffectCostFunc}a). 
  This necessarily means that the random noises in this example do not significantly affect $\globObjFunc(\cdot,\cdot)$. 
  In Table \ref{tab:kannoTrussNonlinEdataEffectCompCost}, we report the total number of iterations for the Newton-Raphson scheme and updating the stress-strain data pairs (ADM iterations) and the computing time, required for our and the ADM-solver, when using an unsymmetrically distributed and noisy dataset.  
  We observe that using GO-ADM-solver requires more iterations and computing time of one order of magnitude than the ADM-solver since they are scaled by the number of ``greedy'' searches, similar to our observations in the examples of the previous subsection. 
  We see that the unsymmetrical data distribution and noises significantly increase the number of Newton-Raphson and ADM iterations, approximately one and two orders of magnitude when using the ADM- and our solver, respectively. 
  This is also reflected in the computing time. 
  The different scale when using the ADM- and our solver illustrates that the computational cost of our solving strategy strongly depends on the data distribution and hence is less robust than the ADM-solver in this regard.

  \begin{table}[htbp]
  \centering
  \begingroup
  \setlength{\extrarowheight}{.2em} 
  \setlength{\tabcolsep}{6pt}       

  \begin{tabularx}{\textwidth}{|l|*{6}{>{\centering\arraybackslash}X|}}
    \hline
    Solving strategy
      & \multicolumn{3}{c|}{Unsymmetrical dataset}
      & \multicolumn{3}{c|}{Noisy dataset} \\
    \cline{2-7}
      &
      stress-free & random & structure-specific
      & stress-free & random & structure-specific \\
    \hline

    \multicolumn{7}{|l|}{\textbf{ADM}} \\
    \hline
    Total Newton--Raphson & 1099 & 1139 & 1099 & 820 & 814 & 820 \\ 
    iterations            & & & & & & \\
    \hline
    Number of ADM         & 272 & 280 & 272 & 205 & 202 & 205 \\
    iterations            & & & & & & \\
    \hline
    Computing time [s]    & 0.50135 & 0.51621 & 0.48184 & 0.47264 & 0.45601 & 0.45023 \\
    \hline

    \multicolumn{7}{|l|}{\textbf{GO-ADM}} \\
    \hline
    Total Newton--Raphson & 30069 & 30233 & 30069 & 46339 & 46714 & 46339 \\ 
    iterations            & & & & & & \\
    \hline
    Total ADM iterations  & 7263 & 7292 & 7263 & 11363 & 11449 & 11363 \\
    \hline
    Number of ``greedy''    & 20209 & 20209 & 20209 & 40200 & 40200 & 40200 \\
    searches                & & & & & & \\
    \hline
    Computing time [s]      & 11.92655 & 12.06396 & 11.97260 & 18.26092 & 18.21942 & 18.91328 \\
    \hline

  \end{tabularx}
  \endgroup
  \caption{Summary of iterations and computing time using different data initialization approaches and solving strategies with a \textbf{unsymmetrical} and \textbf{noisy} dataset, using \textbf{nonlinear strain measures}. The metrics corresponding to the GO-ADM-solver are obtained for all searches and load steps.}
  \label{tab:kannoTrussNonlinEdataEffectCompCost}
\end{table}

  Furthermore, we investigate the effect of three considered initialization approaches on the global objective function $\globObjFunc(\cdot,\cdot)$ and the computational cost when using an unsymmetrically distributed and noisy dataset. 
  Figure \ref{fig:kannoTrussNonlinEdataEffectCostFunc} also includes $\globObjFunc(\cdot,\cdot)$ evaluated at each load step when using different initialization approaches. 
  We observe that in the case of an unsymmetrical data distribution, the stress-free and structure-specific initialization lead to the same objective function when using either our or the ADM-solver. 
  In the case of noisy dataset, the stress-free initialization achieves slightly smaller values of $\globObjFunc(\cdot,\cdot)$. 
  Using a random initialization and the ADM-solver, on the other hand, leads to higher values approximately of two orders of magnitude, irrespective of the employed solver. 
  One reason could be a non-converging ADM scheme when updating the stress-strain data pairs.  
  Using our solver removes dependency of $\globObjFunc(\cdot,\cdot)$ on the initialization for an unsymmetrical data distribution but not for noisy dataset. 
  We note that the results obtained with a random initialization and dataset with random noises are not conclusive without comprehensive statistical studies. 
  Focusing on the required total number of iterations and computing time required for each solver when using different initialization, reported in Table \ref{tab:kannoTrussNonlinEdataEffectCompCost}, 
  we have the same observations as in the case study with symmetric data distribution in the previous subsection. 
  In all cases, using either a stress-free or structure-specific initialization requires the same total number of iterations and computing time for this example, while a random initialization slightly increases computational cost. 
  We conclude that for the studied truss structure with nonlinear strains and datasets, the stress-free and structure-specific initialization lead to the same objective function and approximately the same computational cost. 
  A random initialization increases the objective function and computational cost, which can be further amplified by noisy data.

\section{Summary and conclusions}\label{sec:conclusions}

In this work, we extended and generalized our solving strategy, introduced in \cite{viljar2025}, based on the greedy optimization algorithm and the alternating direction method, for nonlinear systems. 
The introduced solving algorithm combines the direct data-driven solver based on the alternating direction method (ADM), introduced in \cite{ortiz_ddcm_2016}, and the Newton-Raphson scheme (see also \cite{Keip_ddcm_nonlinearbar}) at each load step, while ``greedy'' searching for better alternatives of the stress-strain data pairs to reduce the global objective function. 
We briefly discussed the computational cost of our solving strategy that is scaled with the number of searches, compared to the standard ADM-solver. 
We integrated three common initialization approaches of the data pairs: a random, a stress-free, and a structure-specific initialization, in our solving strategy. 
Particularly, we initialize the data pairs with one of these approaches for the computations at the first load step, and employ the minimizer obtained from the preceding load step as initial data pairs at the next load step. 
For nonlinear systems, the three studied initialization approaches do not guarantee globally optimal solution. 
A random initialization might lead to a diverging Newton-Raphson scheme and hence does not guarantee numerical stability. 
Using either a structure-specific or stress-free initialization improves this stability.

We numerically illustrated via one-dimensional bar and two-dimensional truss structures exhibiting nonlinear strains with different constitutive datasets 
that our solving strategy achieves a better approximation of the global optima. 
This, however, comes at the expense of higher computational cost in terms of the number of iterations and computing time, which is scaled by the number of ``greedy'' searches. 
Using our solving strategy, we also computed the hysteresis loop in the load-deflection curve of a nylon rope during an industrial cyclic testing. 
We also numerically show for truss structures that our solving strategy generally improves the robustness. 
Particularly, it reduces the effect of the initialization approaches, an unsymmetrical data distribution, and a noisy data on the global objective function. 
It also reduces the effect of an unsymmetrical data distribution on the accuracy of the discrete structural solution. 
The computational cost of our solving strategy, however, strongly depends on the data distribution and is increased more significantly by the asymmetry or noises in the data distribution, compared to the standard solving strategy based on ADM.

Our results open various potential directions for future research. 
One aspect is to employ accelerating approaches for the direct data-driven solver and/or the greedy optimization algorithm, such as the 
approximate nearest-neighbor algorithms \cite{Eggersmann2021ddcm}, 
tree-based search algorithms \cite{Bentley1975,Zheng2020,Bahmani2021}, or 
using adaptive hyperparameters for the distance-minimizing method \cite{Nguyen2022ddcm}, 
momentum-based acceleration techniques \cite{Montaut2022greedy,Montaut2024greedy}, or 
randomization and multi-armed bandit approaches \cite{Rakotomamonjy2015Greedy}. 
These would accelerate the material data searches within both solving strategy, without affecting the accuracy of the solution. 
Along this line of research, it is particularly interesting to investigate and systematically compare the performance of the proposed GO-ADM solving strategy with the defective restarting approach \cite{Rocha2025}, in order to gain insights into effective acceleration mechanisms for GO-ADM. 
Another potential approach is the Dijkstra's algorithm \cite{dijkstra1959} and its extended version, the A* 
algorithms \cite{hart1968}, for pathfinding, which would improve the ``greedy'' searching step. 
A second aspect is to improve and/or develop new initialization approaches for nonlinear systems. 
One can employ the solution and minimizer of the corresponding linear system using the standard direct data-driven solver as initial stress-strain data pairs and solution guess. 
An alternative is to solve an approximate nonlinear optimization problem based on an approximated constitutive relation for the nonlinear system. 
This would improve the accuracy,  
aiming at achieving global optima, however, at the expense of higher computational cost for the initialization step. 
A third aspect is to extend the GO-ADM solving strategy when dealing with non-convex optimization problems such as snap-through or buckling. One can integrate the framework introduced in \cite{Kuang2023ddcmsnapthrough} for snap-through problems into the GO-ADM solver or consider the data of the load-deflection curve for the optimization problem as introduced in \cite{Romero2026ddcm}. 
When dealing with history-dependent materials or loading, one can integrate the proposed solving strategy with data searching strategies that capture the history dependency, 
for instance \cite{Ladeveze2019,Bartel2023}. 
Another challenging scenarios is noisy and sparse datasets, for which one can extend the GO-ADM solving strategy with promising techniques such as the max-ent \cite{Kirchdoerfer2017ddcm} or locally convex recontruction (LCR) approaches \cite{He2020ddcmnoise}. 
Another aspect for future work is to extend the application of the introduced solving strategy 
and corroborate its potential for finite element formulations of more complex structural models, in particular, geometrically exact beams, shells, and solid elements.

\section*{Acknowledgments}

C.G.\ Gebhardt, B.A. Roccia and T.-H. Nguyen gratefully acknowledge the financial support from the European Research Council through the ERC Consolidator Grant “DATA-DRIVEN OFFSHORE” (Project ID 101083157).

\appendix

\section{Linearization}\label{sec:linearization}

    In this section, we illustrate the intermediate steps regarding the 
    linearization of the nonlinear Equation \eqref{eq:1stOptCondEq} 
    in Section \ref{sec:continuous_form}.  
    For the sake of readability, we repeat Equation \eqref{eq:1stOptCondEq} here:
    \begin{equation}
        \begin{aligned}
                0 = \, \delta\left( \globObjFunc \left(\yF ,\, \ytilde\right)
            + \map\left( \lF,\, \xF; \, \vect{f} \right)
            \right) 
            = \, & \langle \delta \vect{u}, \, \mathcal{B}^T \mu + \linB \vect{\lambda} \, s \rangle_{L^2(\domain)} \\
            + \, & \langle \delta e, \, c \, (e - \tilde{e}) - \mu \rangle_{L^2(\domain)} \\
            + \, & \langle \delta s, \, \frac{1}{c} (s - \tilde{s}) + \mathcal{B} \vect{\lambda} \rangle_{L^2(\domain)} \\
            + \, & \langle \delta \mu, \, \epsilon(\vect{u}) - e \rangle_{L^2(\domain)} \\
            + \, & \langle \delta \vect{\lambda}, \, \mathcal{B}^T s - \vect{f} \rangle_{L^2(\domain)}
            = \, \vect{g}(\qV) \,,
            \end{aligned}
    \end{equation}
    where $\qV^T = \left[\vect{u}^T \; e \; s \; \mu \; \vect{\lambda}^T \right]^T$. 
    The operator $\mathcal{B}_1$ results from the derivative $\frac{\partial}{\partial \vect{u}} \left(\mathcal{B} \vect{\lambda} \right) \delta \vect{u}$. 
    We recall the strain-displacement operator and compute this derivative as follows:
    \begin{align}
        & \mathcal{B}(\cdot) = \vect{\refconfig}^\prime \cdot (\cdot)^\prime + \alpha \, \vect{u}^\prime \cdot (\cdot)^\prime \,, \\
        & \frac{\partial}{\partial \vect{u}} \left(\mathcal{B} \vect{\lambda} \right) \delta \vect{u} 
            = \alpha \, \delta \vect{u}^\prime \cdot \vect{\lambda}^\prime 
            = \delta \vect{u} \cdot \alpha \, (\cdot)^\prime \mat{I} \vect{\lambda}^\prime 
            = \delta \vect{u} \cdot \underbrace{\alpha \,(\cdot)^\prime (\cdot)^\prime \mat{I}}_{\mathcal{B}_1} \vect{\lambda} 
            = \delta \vect{u} \cdot \mathcal{B}_1 \vect{\lambda} \,.
    \end{align}
    We note that the operator $\mathcal{B}_1$ is symmetrical, i.e. $\mathcal{B}_1^T = \mathcal{B}_1$. 
    The linearized Equations \eqref{eq:linearization} centers on the computation of 
    the derivative of $\vect{g}(\qV)$ with respect to $\qV$ that is:
    \begin{equation}
        \begin{aligned}
            \frac{\partial}{\partial \qV} \vect{g}(\qV) \, \Delta \qV
            \, 
            =\, & \langle \delta \vect{u}, \, 
                \mu \, \mathcal{B}_1 \Delta \vect{u} + 
                \mathcal{B}^T \Delta \mu + 
                s \, \mathcal{B}_1 \Delta \vect{\lambda} + 
                \mathcal{B}_1 \vect{\lambda} \, \Delta s \rangle_{L^2(\domain)} \\
            +\, & \langle \delta e, \, c \Delta e - \Delta \mu \rangle_{L^2(\domain)} \\
            +\, & \langle \delta s, \, \frac{1}{c} \Delta s + 
                \vect{\lambda}^T \mathcal{B}_1 \Delta \vect{u} + 
                \mathcal{B} \Delta \vect{\lambda} \rangle_{L^2(\domain)} \\
            +\, & \langle \delta \mu, \, \mathcal{B} \Delta \vect{u} - \Delta e \rangle_{L^2(\domain)} \\
            +\, & \langle \delta \vect{\lambda}, \, 
                s \, \mathcal{B}_1 \Delta \vect{u} +
                \mathcal{B}^T \Delta s \rangle_{L^2(\domain)} \,,
        \end{aligned}
    \end{equation}
    where the terms $\mu \, \mathcal{B}_1 \Delta \vect{u}$ and $s \, \mathcal{B}_1 \Delta \vect{u}$, or in general $a \, \mathcal{B}_1 \Delta \vect{u}$, 
    where $a$ is a scalar-valued function, 
    result from the derivative of $\delta \vect{u}^T \mathcal{B}^T \, a$ with respect to $\vect{u}$, that is:
    \begin{equation}
        \frac{\partial}{\partial \vect{u}} \left(\delta \vect{u}^T \, \mathcal{B}^T a \right) \Delta \vect{u} = 
            \frac{\partial}{\partial \vect{u}} \left( a \, \left(\vect{\refconfig}^\prime \cdot \delta \vect{u}^\prime + \alpha \, \vect{u}^\prime \cdot \delta \vect{u}^\prime \right)^T \right) \Delta \vect{u}
            = a \, \alpha \, \delta \vect{u}^\prime \cdot \Delta \vect{u}^\prime
        = \delta \vect{u} \cdot a \, \mathcal{B}_1 \Delta \vect{u} \,.
    \end{equation}
    This forms the linearized Equations \eqref{eq:linearization}.

\section{Scaling of the linearized matrix equation}\label{sec:scalingEq}

    To reduce the condition number of the system matrix $\mat{A}$ in the linearized Equation \eqref{eq:kkt_matrix_sys} (the KKT system), we scale the first two and the last matrix equations by a factor $\scaleFac$ and employ the identity $1 = \scaleFac \, \frac{1}{\scaleFac}$ for the second and first terms on the left- and right-hand sides of the third matrix equations, respectively. 
    For readability, we first repeat the explicit expression of the original matrix equations \eqref{eq:kkt_matrix_sys}, 
    $\mat{A} \, \Delta \qhat \,=\, \vect{b}$, here:
    \begin{equation}
        \begin{aligned}        
        & \int_\domain \, \begin{bmatrix}
            \alpha \, \mu_h \, \mat{N}^{\prime\,T} \mat{N}^\prime & \mathbf{0} & \alpha \, \mat{N}^{\prime\,T} \vect{\lambda}_h^\prime \mat{R} & \mat{N}^{\prime\,T} \vect{a} \, \mat{R} & \alpha \, s_h \, \mat{N}^{\prime\,T} \mat{N}^\prime \\
            \mathbf{0} & c\,\mat{R}^T \mat{R} & \mathbf{0} & -\mat{R}^T \mat{R} & \mathbf{0} \\
            \alpha \,\mat{R}^T \, \vect{\lambda}_h^{\prime \, T} \mat{N}^\prime & \mathbf{0} & \frac{1}{c} \,\mat{R}^T \mat{R} & \mathbf{0} & \mat{R}^T \vect{a}^T \mat{N}^\prime \\
            \mat{R}^T \vect{a}^T \mat{N}^\prime & -\mat{R}^T \mat{R} & \mathbf{0} & \mathbf{0} & \mathbf{0} \\
            \alpha \, s_h \, \mat{N}^{\prime\,T} \mat{N}^\prime & \mathbf{0} & \mat{N}^{\prime\,T} \vect{a} \, \mat{R} & \mathbf{0} & \mathbf{0}
        \end{bmatrix} \, \begin{bmatrix}
            \Delta \vect{\hat{u}} \\ \Delta \vect{\hat{e}} \\ \Delta \vect{\hat{s}} \\ \Delta \vect{\hat{\mu}} \\ \Delta \vect{\hat{\lambda}}
        \end{bmatrix} \, \mathrm{d} \arclen \nonumber \\
        & \qquad      
        =\, - \int_\domain \, \begin{bmatrix}
            \mat{N}^{\prime\,T} \vect{b} \\
            c \, \mat{R}^T \mat{R} \, \left(\vect{\hat{e}} - \vect{\tilde{e}}\right) - \mat{R}^T \mu_h  \\
            \frac{1}{c} \, \mat{R}^T \mat{R} \, \left(\vect{\hat{s}} - \vect{\tilde{s}}\right) + \mat{R}^T \vect{a} \cdot \, \vect{\lambda}_h \\
            \mat{R}^T \left( \epsilon_h(\vect{u}_h) - e_h \right) \\
            \mat{N}^{\prime\,T} \vect{a} \, s_h - \mat{N}^T \vect{f}
        \end{bmatrix} \, \mathrm{d} \arclen \,, \\
        & \text{where } \vect{a} = \vect{\refconfig}_h^\prime + \alpha \, \vect{u}_h^\prime \,, \quad \text{and } 
        \vect{b} = \mu_h \, \vect{\refconfig}_h^\prime + \alpha \, \mu_h \, \vect{u}_h^\prime + \alpha \, s_h \, \vect{\lambda}_h^\prime \,.
        \end{aligned}
    \end{equation}

    \noindent
    The modified equations, with $\scaleFac$ highlighted in blue, are then:
    \begin{equation}
        \begin{aligned}        
        & \int_\domain \, \begin{bmatrix}
            \alpha \, \scaleFacHi \, \mu_h \, \mat{N}^{\prime\,T} \mat{N}^\prime & \mathbf{0} & \alpha \, \mat{N}^{\prime\,T} \vect{\lambda}_h^\prime \mat{R} & \mat{N}^{\prime\,T} \vect{a} \, \mat{R} & \alpha \, \scaleFacHi\, s_h \, \mat{N}^{\prime\,T} \mat{N}^\prime \\
            \mathbf{0} & c\,\scaleFacHi\, \mat{R}^T \mat{R} & \mathbf{0} & -\mat{R}^T \mat{R} & \mathbf{0} \\
            \alpha \,\mat{R}^T \, \vect{\lambda}_h^{\prime \, T} \mat{N}^\prime & \mathbf{0} & \frac{1}{\scaleFacHi\,c} \,\mat{R}^T \mat{R} & \mathbf{0} & \mat{R}^T \vect{a}^T \mat{N}^\prime \\
            \mat{R}^T \vect{a}^T \mat{N}^\prime & -\mat{R}^T \mat{R} & \mathbf{0} & \mathbf{0} & \mathbf{0} \\
            \alpha \, \scaleFacHi\, s_h \, \mat{N}^{\prime\,T} \mat{N}^\prime & \mathbf{0} & \mat{N}^{\prime\,T} \vect{a} \, \mat{R} & \mathbf{0} & \mathbf{0}
        \end{bmatrix} \, \begin{bmatrix}
            \Delta \vect{\hat{u}} \\ \Delta \vect{\hat{e}} \\ \scaleFacHi\,\Delta \vect{\hat{s}} \\ \scaleFacHi\,\Delta \vect{\hat{\mu}} \\ \Delta \vect{\hat{\lambda}}
        \end{bmatrix} \, \mathrm{d} \arclen \nonumber \\
        & \qquad      
        =\, - \int_\domain \, \begin{bmatrix}
            \mat{N}^{\prime\,T} \scaleFacHi\,\vect{b} \\
            \scaleFacHi\,c \, \mat{R}^T \mat{R} \, \left(\vect{\hat{e}} - \vect{\tilde{e}}\right) - \mat{R}^T \scaleFacHi\,\mu_h \\
            \frac{1}{\scaleFacHi\,c} \, \mat{R}^T \mat{R} \, \left(\scaleFacHi\,\vect{\hat{s}} - \scaleFacHi\,\vect{\tilde{s}}\right) + \mat{R}^T \vect{a} \cdot \, \vect{\lambda}_h \\
            \mat{R}^T \left( \epsilon_h(\vect{u}_h) - e_h \right) \\
            \mat{N}^{\prime\,T} \vect{a} \, \scaleFacHi\,s_h - \mat{N}^T \scaleFacHi\,\vect{f}
        \end{bmatrix} \, \mathrm{d} \arclen \,. \\
        & \text{Note that } \scaleFacHi\,\vect{b} = 
    \scaleFacHi\,\mu_h \, \vect{\refconfig}_h^\prime + \alpha \, \scaleFacHi\,\mu_h \, \vect{u}_h^\prime + \alpha \, \scaleFacHi\,s_h \, \vect{\lambda}_h^\prime \,.
        \end{aligned}
    \end{equation}
    Solving this modified system of equations is equivalent to solving a system with scaled variables fields, $\scaleFac\,s_h$ and $\scaleFac\,\mu_h$, and using scaled weighting factor, $\scaleFac\,c$, for the objective function and scaled external force vector, $\scaleFac\,\vect{f}$.

\section{Newton-Raphson iteration scheme}\label{sec:Newtonraphsonalgo}

    \begin{algorithm}
\textbf{Input}:
	solution guess $\qhat^{(0)}$, 
    selected data $\ytilde$,
    external force vector $\vect{f}$ \\
\textbf{Output}: $\qhat$
\begin{algorithmic}[1]
    \State $k = 1$      \Comment{Number of iterations}
    \While{$\Delta\,\qhat \,\geq\, \delta$}        \Comment{Convergence tolerance $\delta$}
        \State Assemble $\mat{A} = \mat{A}\left(\qhat^{(k-1)}\right)$ and $\vect{b} = \vect{b}\left(\qhat^{(k-1)},\,\ytilde,\, \vect{f}\right)$     \Comment{See Equation \eqref{eq:kkt_matrices}}
        \State Solve $\mat{A} \, \Delta \qhat = \vect{b}$
        \State $\qhat^{(k)} \,=\, \qhat^{(k-1)} + \Delta\,\qhat$
        \State k\,+=\,1
    \EndWhile    
\caption{Newton-Raphson scheme for the ADM-solver in Algorithm \ref{alg:admsolver}.}\label{alg:NRscheme}
\end{algorithmic}
\end{algorithm}

    We describe the standard Newton-Raphson scheme integrated in the solving strategies discussed in Section \ref{sec:solvingstrategy}. 
    It centers on solving the linearized Equation \eqref{eq:kkt_matrix_sys} for the solution increment, which is employed to update the solution until convergence. 
    The required inputs include a solution guess, which is commonly the solution obtained with the previous load step, selected value for the stress-strain data pairs $\ytilde$, and the external force vector. 
    These inputs and the updated solution are employed to assemble the system matrix $\mat{A}$ and the right-hand side $\vect{b}$ every iteration. 
    We note that the Newton-Raphson scheme is called each time the data pairs $\ytilde$ are updated within the ADM-solver (see also Algorithm \ref{alg:admsolver}).

\section{Computer specifications and codes}\label{sec:computerSpecific}

    \begin{table}[htbp]
  \centering
  \begingroup
  \setlength{\extrarowheight}{.3em}
  \setlength{\tabcolsep}{9pt}

  \begin{tabularx}{\textwidth}{ >{\hsize=0.5\hsize\raggedright\arraybackslash}X >{\hsize=1.5\hsize\arraybackslash}X }
    \toprule
    \textbf{Specification} & \textbf{Details} \\
    \midrule
    CPU & AMD Ryzen 7 5800X, 8 cores / 16 threads, 3.8 GHz base clock \\    
    RAM & 32 GB DDR4 \\    
    Storage & 1 TB NVMe SSD (WDS100T3X0C-00SJG0) \\   
    Operating System & Windows 11 Pro 64-bit, Version 26100 \\    
    Julia Version & 1.11.3+0.x64.w64.mingw32 \\    
    Gurobi Version & 12.0.0 build v12.0.0rc1 (win64 – Windows 11.0 (26100.2)) \\
    \bottomrule
  \end{tabularx}
  \endgroup
\end{table}

    We describe here the computer specifications for our computations in Section \ref{sec:results}, particularly, for the runtime comparison between the solving strategies. 
    Moreover, our implementations using the \texttt{Julia} programming language     
    are available on \href{https://github.com/viljargjerde/Datasolver.jl}{\textcolor{blue}{Github}} (\url{https://github.com/viljargjerde/Datasolver.jl}).

\bibliographystyle{elsarticle-num}
\bibliography{sections/refs}

\end{document}